\documentclass[aps,preprint,showpacs,preprintnumbers,superscriptaddress]{revtex4-1}
\usepackage{multirow}
\usepackage{diagbox}
\usepackage{booktabs}
\usepackage{array}
\usepackage{natbib}
\usepackage{amsmath,amssymb,graphicx,bm,float,siunitx,color}
\usepackage{bm,float,siunitx,graphicx}
\usepackage[figuresright]{rotating}
\usepackage{color}
\usepackage{epstopdf}
\usepackage{mathrsfs}
\usepackage{subcaption}

\usepackage{mathrsfs,amsmath,amsthm,latexsym,amssymb,amsfonts,epsfig,cancel,enumerate,graphicx,subcaption,txfonts,overpic}
\usepackage[justification=raggedright,singlelinecheck=false]{caption}

\newcommand{\be}{\begin{equation}}
\newcommand{\ee}{\end{equation}}
\newcommand{\1}{\left}
\newcommand{\2}{\right}
\newcommand{\dif}{\,\mathrm{d}}

\newcommand{\n}{\nu}

\renewcommand{\th}{\theta}

\begin{document}
\title{\boldmath 
Images of shadow and thin accretion disk around Bardeen black hole surrounded by perfect fluid dark matter}

\author{Haiyuan Feng \footnote{Corresponding author}}
\email{Email address:  fenghaiyuan@sxnu.edu.cn}
\affiliation{School of Physics and Electronic Engineering, Shanxi Normal University, Taiyuan 030031, China}

\author{Ziqiang Cai}
\email{Email address:gs.zqcai24@gzu.edu.cn}
\affiliation{School of Physics, Guizhou University, Guiyang 550025, China}

\author{Hao-Peng Yan}
\email{Email address: yanhaopeng@tyut.edu.cn}
\affiliation{College of Physics, Taiyuan University of Technology, Taiyuan 030024, China}

\author{Rong-Jia Yang }
\email{Email address: yangrongjia@tsinghua.org.cn}
\affiliation{College of Physical Science and Technology, Hebei University, Baoding 071002, China}

\author{Jinjun Zhang \footnote{Corresponding author}}
\email{Email address: zhangjinjun@sxnu.edu.cn }
\affiliation{School of Physics and Electronic Engineering, Shanxi Normal University, Taiyuan 030031, China}

\begin{abstract}
We investigate the shadow and optical appearance of Bardeen black hole (BH) immersed in perfect fluid dark matter (PFDM). Using EHT observations of M87* and Sgr A*, we constrain the DM parameter to a narrow range \(b/M \sim \mathcal{O}(10^{-1}-10^{-2})\) for M87* and to \(\mathcal{O}(10^{-2}-10^{-3})\) for Sgr A*. From these constraints we derive a rough  prediction for the PFDM density near the shadow scale (\(R_{\mathrm{sh}}\sim5M\)): \(\rho_{\mathrm{PFDM}} \sim 0.27\)–\(2.67\,\mathrm{g/cm^3}\) for Sgr A*, dropping to \(\sim10^{-24}\)–\(10^{-25}\,\mathrm{g/cm^3}\) at 100 pc. 
Moreover, increasing \(b\) substantially enlarges the photon sphere, impact parameter, shadow radius, and suppresses the observed disk brightness, while the magnetic charge \(g\) produces only negligible corrections completely masked by PFDM on macroscopic scales.  Subsequently, we investigate the primary/secondary images, flux, and redshift profiles for the PFDM‑Bardeen BH using the Novikov–Thorne disk model, and compare these quantities with those of NFW, Dehnen-type and Moore DM BHs. The four BH types exhibit distinct densities at the shadow radius and at 100 pc, offering a potential distinguishing signature. Furthermore, for all DM BH models, blueshift appears in the primary image as inclination increases, while the secondary image remains redshift dominated even at high inclinations. Hence, if significant blueshifted emission were detected at low
inclination, the predictions of these four DM models would
be seriously challenged.

\end{abstract}

\maketitle

\section{Introduction}
The pioneering observations by the Event Horizon Telescope (EHT) collaboration have ushered in a new era of black hole (BH) imaging and strong-field gravity studies. The release of the first image of the supermassive BH M87* in 2019 \cite{EventHorizonTelescope:2019dse,EventHorizonTelescope:2019uob,EventHorizonTelescope:2019jan,EventHorizonTelescope:2019ths,EventHorizonTelescope:2019pgp,EventHorizonTelescope:2019ggy} marked a major milestone, sparking extensive investigations into BH shadows as powerful probes for testing gravitational theories and exploring the electromagnetic characteristics of compact objects \cite{Lima:2021las,Volkel:2020xlc,Afrin:2021imp,Guerrero:2021ues,Badia:2020pnh,Cunha:2019ikd,Tsukamoto:2017fxq,Cunha:2017qtt,Cunha:2016bjh,Abdujabbarov:2016hnw,Wu:2025hcu,Cai:2025rst,Feng:2024iqj,Liu:2024brf}. More recently, in May 2022, the EHT collaboration presented the image of Sgr A*, the supermassive BH at the center of the Milky Way, revealing an asymmetric emission ring \cite{EventHorizonTelescope:2022wkp,EventHorizonTelescope:2022apq,EventHorizonTelescope:2022wok,EventHorizonTelescope:2022exc,EventHorizonTelescope:2022urf,EventHorizonTelescope:2022xqj}. These successive advancements have greatly enhanced the statistical robustness of BH shadow observations, offering a promising avenue to place stringent constraints on alternative BHs that deviate from the standard Kerr paradigm \cite{Herdeiro:2022yle}. Motivated by these observational breakthroughs, it has become essential to understand how the intrinsic properties of BHs and their surrounding environments influence the formation and morphology of the observed shadow.

The BH shadows represent one of the most striking observational signatures of strong gravitational fields. It appears as a dark region surrounded by a bright emission ring, formed by the gravitational lensing and photon capture effects near the event horizon. For a non-rotating BH, such as the Schwarzschild BH (Sch BH), the shadow exhibits a perfectly circular boundary. This property was originally analyzed by Synge and later extended by Luminet, who explored how the presence of thin accretion disk modifies the observed shadow \cite{Synge:1966okc,Luminet:1979nyg}. In contrast, the shadow produced by a rotating BH, such as the Kerr BH, becomes asymmetric and deformed as a result of spin-induced frame-dragging effects \cite{Hioki:2009na,Bardeen:1973tla}. The morphology of the shadow is determined by the specific spacetime geometry surrounding BH. In recent decades, a substantial body of research has focused on analyzing shadow characteristics across diverse BHs, both within the framework of GR and in various modified gravity theories \cite{1,2,3,4,5,6,7,8,9,10,11,12,13,14,15,16,17,18,19,20,Cai:2025pan,Cai:2025rst}. These studies highlight that the remarkable success of the EHT in imaging BH shadows has inaugurated a new frontier in observational relativity.

Moreover, imaging techniques for thin accretion disks-ranging from semi-analytic approaches to advanced ray-tracing methods coupled with radiative transfer—serve as a powerful tool for testing gravity in the strong-field regime. Building upon the classical Shakura–Sunyaev model for geometrically thin and optically thick disks and its relativistic generalization by Novikov and Thorne \cite{Shakura:1972te,Page:1974he}, semi-analytic analyses such as that of Luminet have provided direct and secondary disk images along with analytical predictions for radiation flux, while modern numerical ray-tracing simulations have enabled realistic modeling of accretion-disk images in a variety of BHs \cite{1993A&A...272..355V,1995CoPhC..88..109S,1997PASJ...49..159F,Broderick:2005my,Dexter:2016cdk,Vincent:2011wz,Cunha:2016bjh,Hou:2022eev,Zhang:2024lsf,Gyulchev:2019tvk,Shaikh:2019hbm,Bambi:2019tjh,Johannsen:2016uoh,PhysRevD.102.104041,Okyay:2021nnh,He:2024amh}. Although the geometric outline of the shadow is primarily determined by the spacetime structure rather than the detailed accretion dynamics, the observed luminosity distribution, ring morphology, and photon ring substructures are highly sensitive to the emission of thin accretion disks and radiative processes \cite{21,22,23,24,25,26,27,28}. Therefore, by combining physically consistent disk models with realistic radiative transfer and high-precision ray-tracing techniques, accretion-disk imaging also provides a robust observational framework to  discriminate between GR and its possible modifications.

Recent observations have yet to definitively exclude deviations from GR, leaving open the possibility of testing alternative theories of gravity. Although GR has achieved remarkable success in describing gravitational phenomena from Solar-System scales to BH astrophysics, it inevitably predicts the formation of spacetime singularities inside BHs, where curvature invariants diverge and the classical theory loses predictive power \cite{t1974one,Deser:1974cz,Deser:1974xq,Hawking:1973uf,tHooft:1974toh,Cognola:2013fva,Zhang:2019dgi,Penrose:1964wq,Hawking:1976ra,Christodoulou:1991yfa}. Such singularities are generally regarded as indications of the incompleteness of GR in the strong-field regime and have motivated the search for regular (nonsingular) BHs. To resolve this issue, regular BH models have been proposed, among which the Bardeen BH stands as the  most paradigmatic example \cite{1968qtr..conf...87B,Ayon-Beato:2000mjt}. The BH was later reinterpreted by Ayon‑Beato and García as an exact solution of GR coupled to nonlinear electrodynamics (NLED), corresponding to the gravitational field of a magnetic monopole $g$ \cite{Hayward:2005gi,Ayon-Beato:1998hmi,Berej:2006cc}. This feature makes the Bardeen geometry not merely a mathematical construction, but a theoretically motivated spacetime connected to classical gravity and NLED effects. Moreover, it provides a concrete framework to test fundamental physics questions: 
whether magnetic monopoles–predicted by Dirac’s quantization condition and grand unified theories-can exist in nature. Observational signatures from the shadow and accretion disk, as constrained by the EHT, also offer a unique opportunity to probe the magnetic monopole indirectly, given that direct searches for monopoles have so far yielded null results.



Additionally, observations of galaxy clusters and galactic rotation curves have long suggested the existence of an unseen mass component, giving rise to the dark matter (DM) hypothesis first proposed by Zwicky and later supported by numerous astrophysical and cosmological observations \cite{1933AcHPh...6..110Z,1980ApJ...238..471R,1937AnLun...6....1H,1936ApJ....83...23S}. Although DM has not been directly detected electromagnetically, strong evidence from cluster dynamics, flat rotation curves, and cosmic microwave background anisotropies firmly supports its existence \cite{1980ApJ...238..471R,Planck:2013pxb}. Nevertheless, the fundamental nature of DM remains unknown, and a variety of DM models have been proposed in the literature, including cold dark matter (CDM) \cite{Blumenthal:1982mv,Frenk:1983zz}, warm dark matter (WDM) \cite{Hogan:2000bv,Dalcanton:2000hn}, self-interacting dark matter (SIDM) \cite{Spergel:1999mh}, and axion-like DM models \cite{Gavrilik:2023wti}. Different DM scenarios may lead to distinct density profiles and dynamical effects around compact objects. Moreover, given that BHs are typically located at the centers of galaxies, they are expected to be deeply embedded within DM halos. This implies that both the spacetime geometry and the dynamics of nearby matter are inevitably affected by the gravitational potential of the surrounding DM distribution. A variety of empirical density profiles have been introduced to describe the spatial distribution of DM halos, such as the  Navarro–Frenk–White (NFW) \cite{Navarro:1995iw}, Dehnen \cite{Dehnen:1993uh}, Burkert \cite{Burkert:1995yz}, and Moore \cite{Moore:1994yx} models. These profiles differ in their central cusp or core structures and asymptotic fall-off rates, thereby producing distinct modifications of the gravitational field around BHs. As one of the DM candidates, the perfect fluid dark matter (PFDM) was proposed by Kiselev \cite{Kiselev:2002dx,Kiselev:2003ah} and further studied in subsequent works \cite{Li:2012zx}, offering a reasonable explanation for the asymptotically flat rotation curves of spiral galaxies \cite{Haroon:2018ryd,Hou:2018avu}. This phenomenological model parameterizes the gravitational effects of DM through a simple energy-momentum tensor, with a density profile \(\rho \sim b/r^3\), thereby introducing only a single parameter \(b\) to characterize its influence on BH. Such simplicity makes PFDM particularly suitable for quantitative constraints using EHT observations of BH shadows and accretion images, allowing us to probe the interplay between PFDM and regular BHs without relying on detailed microscopic DM models. Accordingly, we study regular Bardeen BH surrounded by PFDM (PFDM-Bardeen BH), focusing on the DM imprints in their accretion images.


The structure of this paper is outlined as follows. Section 2 introduces the Bardeen BH surrounded by DM and discusses the theoretical constraints governing the existence of inner and outer horizons. Section 3 employs EHT observations to place bounds on the model parameters, with particular emphasis on the range and magnitude of the DM parameter. In Section 4, we classify photon trajectories into direct emission, lensing rings, and photon rings, and examine how the model parameters influence various associated physical quantities, including the transfer function and intensity distribution. Section 5 investigates the primary and secondary images of the accretion disk, focusing on the behavior of the observed radiation flux and redshift, and analyzes how the parameters shape the optical appearance of the disk. The final section provides a summary of our main conclusion.

\section{Bardeen black hole surrounded by  perfect fluid dark matter}

The Bardeen BH stands as a representative example of a regular, singularity-free solution. First proposed by Bardeen as a theoretical counterexample to the notion that gravitational collapse inevitably leads to a singularity concealed by an event horizon \cite{PhysRevD.55.7615,1968qtr..conf...87B}, it challenged the classical understanding of BH interiors. Decades later, Ayón-Beato and García provided a physical foundation for this model by demonstrating that it arises as an exact solution to Einstein’s equations coupled with nonlinear electrodynamics (NLED), corresponding to the gravitational field generated by a magnetic monopole \cite{Ayon-Beato:2000mjt}. This reinterpretation transformed the Bardeen metric from a purely phenomenological construct into a physically consistent spacetime endowed with well defined matter source.

The geometry of Bardeen BH is characterized by the following metric 
\be
\label{1}
ds^2 = -\left(1 - \frac{2Mr^2}{(r^2 + g^2)^{3/2}}\right) dt^2+\left(1 - \frac{2Mr^2}{(r^2 + g^2)^{3/2}}\right)^{-1} dr^2 + r^2 d\Omega^2,
\ee
here the parameters $M$ and $g$ correspond to the ADM mass and the magnetic charge, respectively. The angular part of the metric, \( d\Omega^2 = d\theta^2 + \sin^2\theta d\phi^2 \), represents the standard line element on a unit two-sphere. Unlike the Reissner–Nordström case, where the charge contribution falls off as \( 1/r^2 \), the Bardeen modification introduces \( 1/r^3 \) dependence. This higher-order decay ensures that all curvature invariants—including \( R \), \( R_{\mu\nu}R^{\mu\nu} \), and \( R_{\mu\nu\alpha\beta}R^{\mu\nu\alpha\beta} \)—remain finite even at the origin \( r = 0 \). As a result, the Bardeen metric successfully eliminates the curvature singularity while still admitting the presence of event horizons. Solving the equation governing the existence of inner and outer horizons yields a critical magnetic charge of \( g/M = 0.7698 \), at which the two horizons merge at \( r_+/M = r_-/M = r_\ast/M = 1.08866 \) \cite{Cui:2025qxu}. Consequently, the spacetime admits two distinct horizons when \( g/M \in (0, 0.7698) \), whereas for values of \(g/M\) exceeding this limit, no horizon is present.

From physical perspective, the regular nature of Bardeen BH arises from the interaction between gravity and a nonlinear electromagnetic field, which is described by the Lagrangian \( \mathcal{L}(F) \)
\be
\label{2}
\mathcal{L}(F)=\frac{3 M}{|g|^3}\left(\frac{\sqrt{2 g^2 F}}{1+\sqrt{2 g^2 F}}\right)^{\frac{5}{2}},
\ee
which satisfies \( F_{\mu\nu} = \nabla_{\mu}A_{\nu} - \nabla_{\nu}A_{\mu} \), with \( \mathcal{L}(F) \) being a function of the electromagnetic invariant \( \tfrac{1}{4}F_{\mu\nu}F^{\mu\nu} \). Incorporating PFDM into the Bardeen BH provides a natural framework to examine the interplay between DM and regular BHs. As PFDM captures the large scale DM distribution influencing galactic and BH environments, its inclusion enables an assessment of how DM modifies the horizon structure of the Bardeen BH. Assuming the BH is embedded in PFDM background, Zhang et al. \cite{Zhang:2020mxi} derived the corresponding PFDM-Bardeen BH by coupling gravity with nonlinear electromagnetic field, for which the energy momentum tensor is given by
\be
\label{3}
T^{\mu}_{\n} = \text{diag}(-\rho, p_r, p_\theta, p_\phi),
\ee
with
\be
\label{4}
\rho=-p_r=\frac{b}{r^3}, \quad p_\theta=p_\phi=\frac{b}{2 r^3},
\ee
where \( \rho \), \( p_r \), and \( p_\theta = p_\phi \) denote the energy density, radial pressure, and tangential pressures of PFDM, respectively \cite{Kiselev:2002dx,Li:2012zx}. The weak energy condition \( T_{tt} \ge 0 \) requires \( b \ge 0 \) \cite{Zhang:2020mxi}. The parameter $b$ characterizes the DM density and its dynamical effect on the spacetime geometry near BH, rather than the global cosmological DM content. Observationally, \( b \) can be constrained by comparing theoretical predictions—such as galaxy rotation curves or BH shadow sizes—with astrophysical data. In the following section, we constrain these parameters using the observational data from EHT.

Since the PFDM-Bardeen BH under consideration is static and spherically symmetric, the corresponding metric can be represented as
\be
\label{5}
d s^2=-f(r) d t^2+f(r)^{-1} d r^2+r^2 d \Omega^2,
\ee
and
\be
\label{6}
f(r)=1-\frac{2 M r^2}{\left(r^2+g^2\right)^{\frac{3}{2}}}-\frac{b}{r} \ln \frac{r}{|b|}.
\ee

It is clear from the above expression that setting both the parameter \( b = 0 \) and the magnetic charge \( g = 0 \) recovers the Sch BH. When PFDM is absent (\( b = 0 \)), the metric reduces to the standard Bardeen BH, whereas vanishing magnetic charge (\( g = 0 \)) corresponds to the Sch BH immersed PFDM (PFDM-Sch BH). It is worth noting that the PFDM‑Bardeen BH can also be derived from modified gravity models such as \(f(R)\) gravity coupled with the NLED Lagrangian Eq.\eqref{2}. The additional effective energy–momentum tensor arising from the \(f(R)\) sector may act as PFDM, thus yielding the metric solution \eqref{6} \cite{Nojiri:2024qgx,Feng:2023riq,Zheng:2018fyn}. This reflects the well‑known degeneracy between DM and modified gravity in explaining rotation curves. In this paper, we adopt the PFDM interpretation to investigate the BH.

\begin{figure}[H]
\centering
\begin{minipage}{0.5\textwidth}
\centering
\includegraphics[scale=1,angle=0]{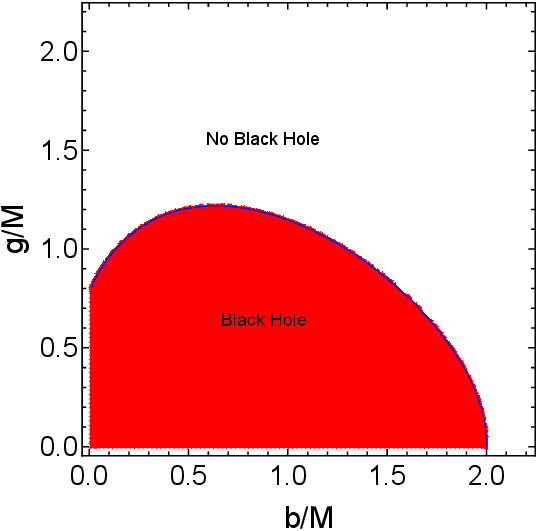}
\end{minipage}
\caption{\label{fig.1}{On the horizontal axis, we have dimensionless parameter $b/M$, while the vertical axis represents the dimensionless magnetic charge $g/M$. The red area corresponds to the BH region, whereas the white area denotes the naked singularity.}}
\end{figure}

Figure \ref{fig.1} illustrates the regions where the inner and outer horizons of the PFDM-Bardeen BH exist. The red area denotes the presence of both horizons, the white area corresponds to naked singularity, and the blue boundary represents the extremal BH. In this context, the extremal condition is governed by the equations \( f(r_\ast, g, b) = 0 \) and \( f'(r_\ast, g, b) = 0 \) \((r_{\ast}=r_{+}=r_{-})\), which together define a boundary curve in the \((b, g)\) parameter plane that separates the BH domain from the naked singularity region.

\section{Constraining model parameters using EHT observations of M87* and Sgr A*}

Following the EHT observations of supermassive BH M87* and Sgr A* \cite{EventHorizonTelescope:2019dse,EventHorizonTelescope:2022wkp}, the apparent diameter of BH shadow provides a powerful probe to constrain model parameters in various modified gravity. Furthermore, we employ the measured shadow diameters of M87* and Sgr A* to place constraints on the DM parameter $b$. By comparing the theoretical prediction of the shadow diameter $d_{\text{sh}}$ with the EHT observational ranges, we will determine the allowed parameter region consistent with observational uncertainties. 

\subsection{Null geodesic equation for Bardeen BH surrounded by PFDM}
The photon sphere, defined as the unstable circular orbit of photons, delineates the boundary between light rays escaping to infinity and those captured by BH. Consequently, the apparent shadow observed by a distant observer is governed by the geometry of these null trajectories. In the context of the PFDM-Bardeen BH, the interplay between the magnetic charge and the DM parameter significantly influences the radius of the photon sphere and the corresponding shadow size. To explore these effects, we derive the null geodesic equation in this BH and analyze the resulting photon dynamics.

The Lagrangian $\mathcal{L}$ for a point particle in the spacetime \eqref{5} is given by

\be
\label{7}
\mathcal{L}=\frac{1}{2} g_{\mu \nu} \frac{d x^\mu}{d \lambda} \frac{d x^\nu}{d \lambda},
\ee
where $\lambda$ represents the affine parameter. Owing to the spacetime’s spherical symmetry, the motion of the particle can, without loss of generality, be restricted to the equatorial plane ($\theta = \pi/2$). Under this condition, two conserved quantities naturally emerge
\be
\1\{\begin{split}
\label{8}
&E=-g_{t t} \frac{d t}{d \lambda}=f(r) \frac{d t}{d \lambda},\\
&L=g_{\phi \phi} \frac{d \phi}{d \lambda}=r^2 \frac{d \phi}{d \lambda}.
\end{split}\2.
\ee
For \( \mathcal{L} = 0 \), by combining Eqs. \eqref{7} and \eqref{8}, one obtains the governing equations that describe the propagation of light in this BH
\be
\1\{\begin{split}
\label{9}
&\frac{d t}{d \lambda^{\prime}}=\frac{1}{\bar{b} f(r)},\\
&\frac{d \phi}{d \lambda^{\prime}}=\frac{1}{r^2},\\
&\left(\frac{d r}{d \lambda^{\prime}}\right)^2=\frac{1}{\bar{b}^2}-\frac{f(r)}{r^2},
\end{split}\2.
\ee
here, we redefine the affine parameter as $\lambda' = L\lambda$, while $\bar{b} = L/E$ represents the impact parameter associated with the photon trajectory. This parameter holds fundamental importance, as it directly determines the photon’s trajectory and, consequently, the shape and size of BH shadow. Physically, the impact parameter characterizes the perpendicular distance between the photon’s initial direction and the BH's center, thus governing whether the photon will escape to infinity, orbit in the photon sphere, or be captured by the event horizon.

By manipulating the light propagation equation, the radial coordinate \(r\) can be expressed as an ordinary differential equation with respect to the azimuthal angle \(\phi\) within the orbital plane
\be
\label{10}
\left(\frac{d r}{d \phi}\right)^2=r^4\left(\frac{1}{\bar{b}^2}-\frac{f(r)}{r^2}\right) \equiv V_{\rm e f f},
\ee
where $ V_{\rm e f f}$ represents the effective potential governing photon motion, and the left-hand side corresponds to the kinetic energy term. To facilitate the analysis of photon trajectories, we define a new variable \( u = 1/r\), which allows the photon’s equation of motion to be recast in the following form
\be
\label{11}
\left(\frac{d u}{d \phi}\right)^2=\frac{1}{\bar{b}^2}-u^2 f\left(\frac{1}{u}\right) \equiv G(u).
\ee  

\begin{figure}[H]
\centering
\begin{minipage}{0.5\textwidth}
\centering
\includegraphics[scale=1,angle=0]{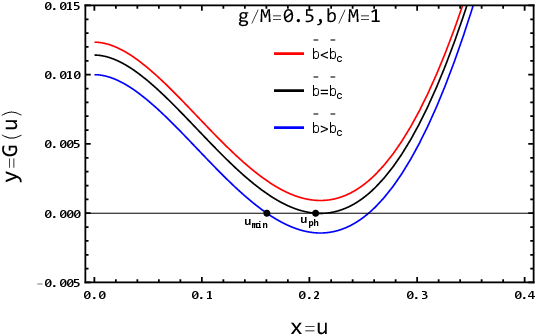}
\end{minipage}
\caption{\label{fig.2}{ The functions $G(u)$ for different values of $\bar{b}$, as a function of $u$.}}
\end{figure}

Figure \ref{fig.2} depicts the dependence of \( G(u) \) on \( u \) for magnetic charge \( g/M = 0.5 \) and DM parameter \( b/M = 1 \). The plot clearly reveals the existence of a critical impact parameter \( \bar{b}_c \) that distinguishes different photon trajectories. For \( \bar{b} > \bar{b}_c \), photons arriving from spatial infinity are gravitationally deflected, reaching a minimum radial distance \( r_{\text{min}} \) before escaping back to infinity (where $r_{\min}$ denotes the closest radial distance of the photon from the BH during its motion). When \( \bar{b} < \bar{b}_c \), photons are inevitably captured by PFDM-Bardeen BH, defining the photon capture cross-section. At the threshold \( \bar{b} = \bar{b}_c \), photons follow unstable circular orbits around PFDM-Bardeen BH, with the corresponding orbital radius identified as the photon sphere radius $r_{\rm ph}$. 

\begin{figure}[H]
\centering
\begin{minipage}{0.5\textwidth}
\centering
\includegraphics[scale=0.9,angle=0]{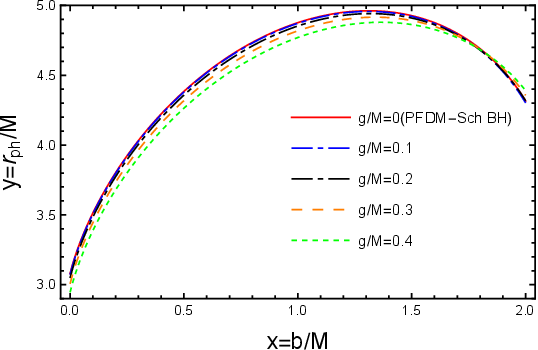}
\end{minipage}%
\begin{minipage}{0.5\textwidth}
\centering
\includegraphics[scale=0.9,angle=0]{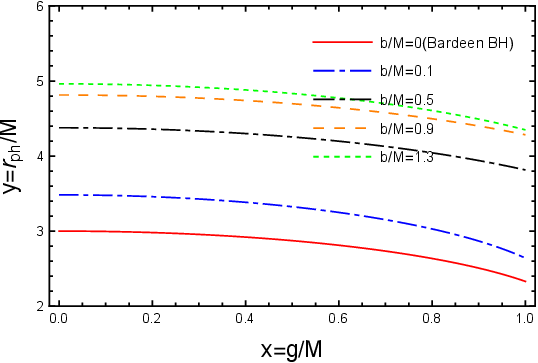}
\end{minipage}%
\caption{\label{fig.3}{The two pictures illustrate the dependence of photon sphere radius $r_{\rm ph}$ on the magnetic charge \( g \) and DM parameter \(b\).
  }}
\end{figure}

Figure \ref{fig.3} illustrates the dependence of the \( r_{\mathrm{ph}}\) on the magnetic charge \(g\) and the DM parameter \( b \). As shown in the left panel, for a fixed magnetic charge \( g \), increasing \( b \) causes \( r_{\mathrm{ph}} \) to increase initially, reach a peak, and then decrease, indicating a non-monotonic relationship. This phenomenon stems from the PFDM‑induced modification of the spacetime geometry via non‑zero energy‑momentum tensor, which introduces a logarithmic term in \(f(r)\) and thereby alters the effective potential \(V_{\mathrm{eff}}(r)\). The non‑monotonic dependence of \(r_{\mathrm{ph}}\) on \(b\) (for fixed \(g\)) arises from the competition between the Bardeen and PFDM contributions to \(V_{\mathrm{eff}}\). For small \(b\), the PFDM term adds a weak attractive correction that shifts the peak of \(V_{\mathrm{eff}}\) outward, increasing \(r_{\mathrm{ph}}\). However, when \(b\) becomes sufficiently large, the logarithmic term in \(f(r)\) modifies the radial gradient of \(V_{\mathrm{eff}}\) in a nonlinear way, eventually causing the peak to move inward. In contrast, for a fixed \( b \), the variation with respect to \( g\) is relatively weak. The right panel further demonstrates that the influence of \( g \) remains modest: within a certain range, an increase in \( b \) at constant \( g\) enlarges \( r_{\mathrm{ph}}\), while the Bardeen BH consistently exhibits a smaller \( r_{\mathrm{ph}} \) than its PFDM–Bardeen counterpart.

To determine the critical impact parameter \( \bar{b}_c \), we first consider the case \( \bar{b} > \bar{b}_c \). In this scenario, when the photon reaches its point of closest approach to the BH \(( r = r_{\text{min}} )\), the relation \( \bar{b} = \frac{r_{\text{min}}}{\sqrt{f(r_{\rm min})}} \) holds. As \( r \rightarrow r_{\rm ph} \), the impact parameter approaches its critical value \( \bar{b}_c \). To identify the radius of the photon sphere, we note that for circular photon orbits the conditions \( \dot{r} = 0 \) (zero radial velocity) and \( \ddot{r} = 0 \) (zero radial acceleration) must be satisfied, which are equivalent to \( V_{\rm eff}|_{r=r_{\rm ph}} = 0 \) and \( dV_{\rm eff}/dr|_{r=r_{\rm ph}} = 0 \), respectively. At this stage, the critical impact parameter \( \bar{b}_c \), corresponding to the photon sphere, can be determined by 
\be
\label{12}
\bar{b}_c = \frac{r_{\text{ph}}}{\sqrt{f(r_{\rm ph})}}.
\ee  

It should be emphasized that the critical impact parameter is fundamentally linked to the radius of BH shadow. The shadow structure can be examined by tracing light rays backward in time from a static observer located at radial coordinate \( r_O \). The shadow boundary is defined by those light rays that asymptotically approach the unstable circular photon orbits at \( r_{\text{ph}} \). For a general static, spherically symmetric, and asymptotically flat spacetime, Refs. \cite{Perlick:2021aok,Perlick:2015vta} demonstrated that when the observer is sufficiently distant from BH (\( r_O \to \infty \)), the apparent shadow radius tends to the critical impact parameter \( \bar{b}_c \). Consequently, this radius is governed by the position of the photon sphere and the form of the metric function \( f(r) \). 

\subsection{Constraining DM parameter via shadow diameter}
Based on the physical implications of Eq. \eqref{12}, this subsection provides stringent constraints on the DM parameter using the EHT observations of supermassive BHs M87* and Sgr A*. For M87*, general relativistic magnetohydrodynamic (GRMHD) simulations indicate that current EHT data cannot effectively distinguish whether the BH is rotating or spherically symmetric based solely on shadow morphology \cite{Mizuno:2018lxz}. The measured shadow size remains consistent with \( 3\sqrt{3}(1 \pm 0.17)M \), independent of whether the model assumes spherical or axisymmetric symmetry \cite{EventHorizonTelescope:2021dqv}. Additionally, for Sgr A*, the spin parameter is sufficiently small such that its influence on the shadow radius can be safely neglected \cite{Vagnozzi:2022moj}. Accordingly, constraints on the PFDM–Bardeen BH can be established using current EHT observations. 

For M87*, the observed shadow exhibits an angular diameter of \(42 \pm 3~\mu\mathrm{as}\), corresponding to a source located approximately 16.8 Mpc away, with a central BH mass estimated at \((6.5 \pm 0.9) \times 10^9 M_\odot\) \cite{EventHorizonTelescope:2019dse}. Meanwhile, for Sgr A*, the EHT collaboration reported angular diameter \(48.7 \pm 7~\mu\mathrm{as}\), with the Galactic Center situated at a distance of \(8277 \pm 33~\mathrm{pc}\) and harboring a supermassive BH of mass \((4.3 \pm 0.013) \times 10^6 M_\odot\) \cite{EventHorizonTelescope:2022wkp}. 

The shadow diameter, expressed in units of mass, can be evaluated from the BH parameters through relation 
\be
\label{13}
d_{\mathrm{sh}}=\frac{D \theta}{M},
\ee  
where \(D\) represents the distance from the observer. Accordingly, the theoretical estimate of the diameter of BH shadow is obtained using $d_{\mathrm{sh}}^{\text {theo }}=2 R_{\mathrm{sh}}=2\bar{b}_{c}$. By applying this relation, we can determine the corresponding shadow diameters for M87* and Sgr A*, as presented by \cite{Bambi:2019tjh,Vagnozzi:2022moj}
\be
\1\{\begin{split}
\label{14}
&d_{\mathrm{sh}}^{\mathrm{M} 87^*}=(11 \pm 1.5) M,\\
& d_{\mathrm{sh}}^{\mathrm{Sgr} . \mathrm{A}^*}=(9.77 \pm 0.67) M .
\end{split}\2.
\ee

\begin{figure*}[h]
\centering
\setlength{\tabcolsep}{0.9cm} 

\begin{tabular}{cc}
\begin{minipage}[t]{0.4\textwidth}
\centering
\begin{overpic}[width=\textwidth]{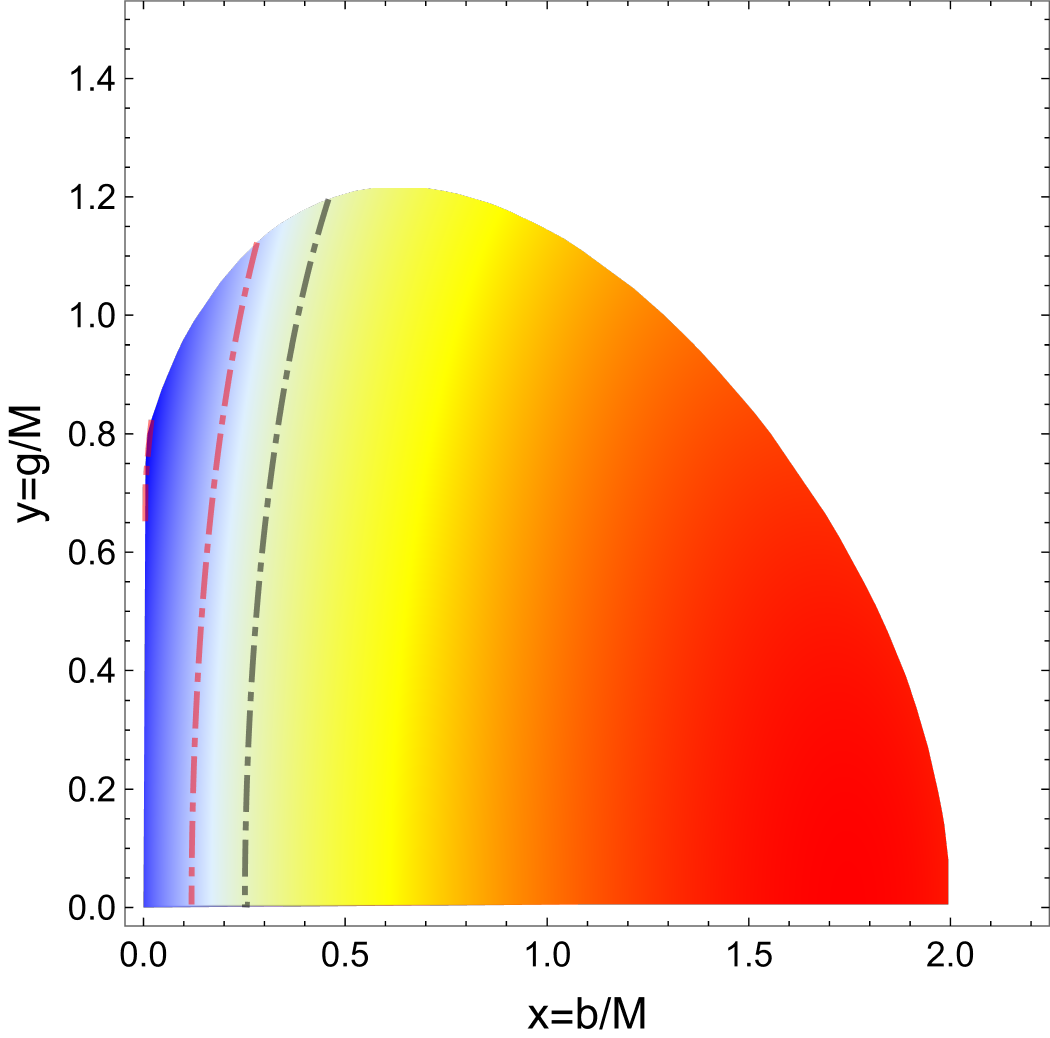}
    \put(100,-1){\includegraphics[width=0.20\textwidth]{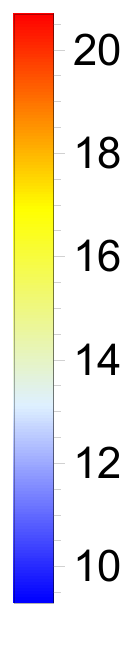}}
    \put(102,105){\color{black}\large $d_{sh}$} 
\end{overpic}
\end{minipage}
&
\begin{minipage}[t]{0.4\textwidth}
\centering
\begin{overpic}[width=\textwidth]{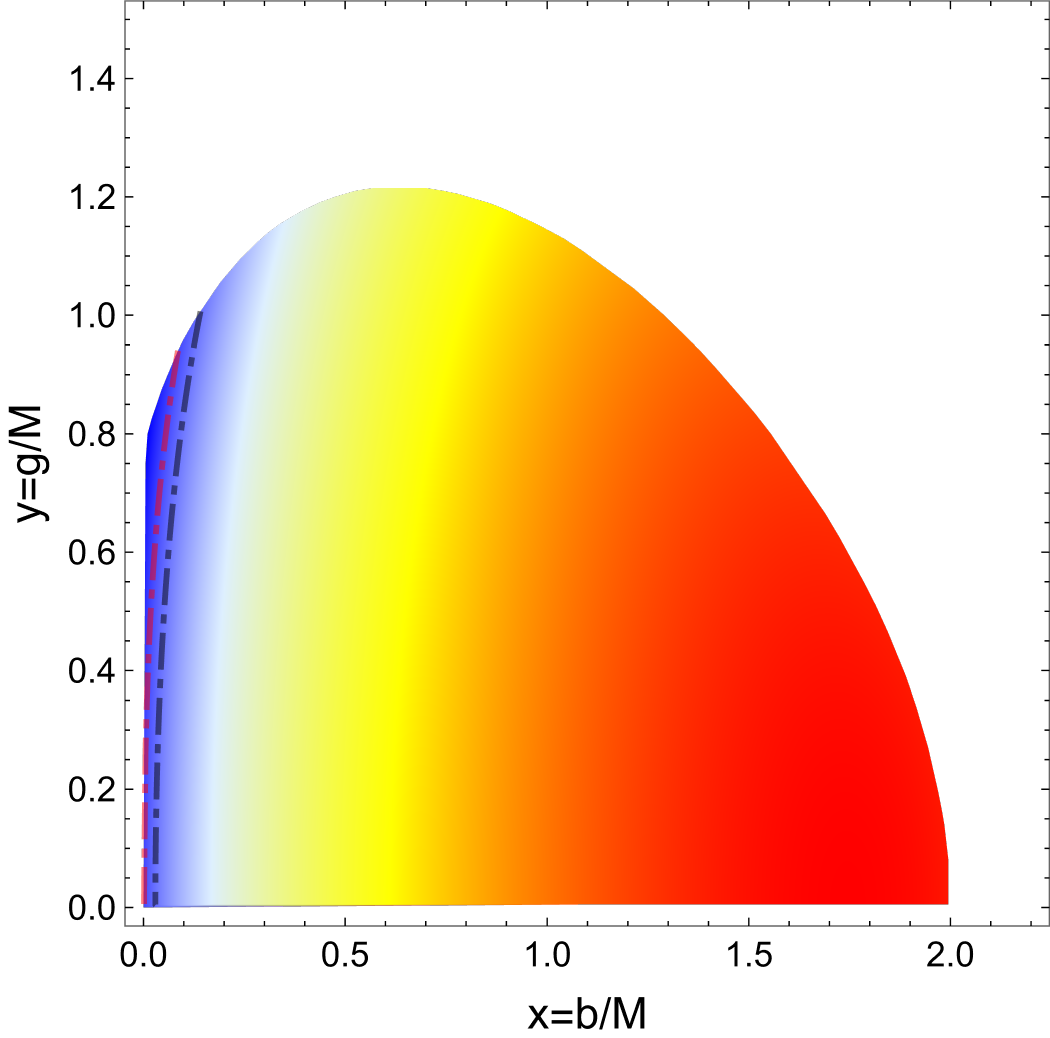}
    \put(100,-1){\includegraphics[width=0.20\textwidth]{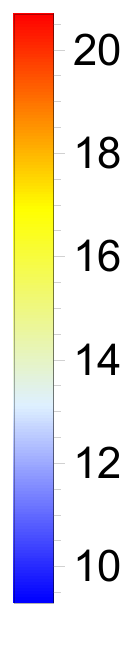}}
    \put(102,105){\color{black}\large $d_{sh}$}
\end{overpic}
\end{minipage}
\end{tabular}

\caption{\label{fig.4}{%
The figure illustrates the contour plots of the shadow diameter within the theoretically allowed parameter space. (Left panel) The red dashed lines correspond to the M87$^*$ shadow diameter within the 1$\sigma$ confidence interval, located at $d_{\rm sh}/M = 9.5$ and $d_{\rm sh}/M = 12.5$. The black dashed line represents the 2$\sigma$ confidence interval at $d_{\rm sh}/M = 14$. (Right panel) For Sgr A$^*$, the red and black dashed lines similarly denote the 1$\sigma$ and 2$\sigma$ confidence regions, corresponding to shadow diameters $d_{\rm sh}/M = 10.44$ and $d_{\rm sh}/M = 11.11$, respectively.}}
\end{figure*}

In Fig. \ref{fig.4}, stringent constraints on the parameter $b$ are derived under the condition that both the inner and outer horizons exist. The left panel illustrates the parameter constraints inferred from the shadow diameter of M87* within the 1$\sigma$ and 2$\sigma$ confidence intervals, where the red dashed curve corresponds to the 1$\sigma$ range and the black dashed curve denotes the 2$\sigma$ range. Likewise, the right panel presents the constraints obtained from the shadow diameter of Sgr A*. From the figure, it can be observed that the allowable range of $g$ is primarily determined by its value at the outer horizon and the lower bound along the constraint curve, indicating that $g$  possesses a definite lower limit. In contrast, the DM parameter $b$  exhibits an upper bound, with its lower bound gradually emerging as $g$  increases. For a clear demonstration of this behavior, Table 1 summarizes the upper and lower bounds of parameter $b$ for both types of BHs at fixed magnetic charge $g$.

Table I reveals four key trends: (i) the upper bound on \(b/M\) increases with the magnetic charge \(g/M\) for both BHs; (ii) a lower bound on \(b\) appears only for M87* at \(g/M\gtrsim0.7\), while no such lower bound exists for Sgr A*; and (iii) the upper bound on \(b\) for Sgr A* is approximately one to two orders of magnitude tighter than that for M87*. (iv) 
Based on these constraints, we infer that if PFDM–Bardeen BH truly exist, the DM parameter \(b/M\) for M87* is constrained to the order of \(\mathcal{O}(10^{-1})\), with part of the parameter space extending down to \(\mathcal{O}(10^{-2})\). In contrast, for Sgr A*, \(b/M\) is much more severely restricted to \(\mathcal{O}(10^{-2})\) or even \(\mathcal{O}(10^{-3})\). Regarding the third point, we conjecture that the tighter upper bound on \(b\) for Sgr A* arises from a combination of two factors: Sgr A* has a smaller observed shadow diameter (\(9.77M\) vs \(11M\)) with much narrower uncertainties, making it more sensitive to any parameter that enlarges the shadow such as \(b\); and it is much closer to Earth with a far more precisely determined mass, resulting in a smaller absolute error in the angular diameter of the shadow and thus imposing stricter constraints on model parameters.

In addition, the constraints in Table I directly translate into an estimate of the DM energy density near the BH via \(\rho_{\mathrm{PFDM}} = b/r^{3}\). Taking the shadow scale \(R_{\rm sh} \sim 5M\), the allowed \(b/M \sim 10^{-3}\)–\(10^{-2}\) for Sgr A* yields \(\rho_{\mathrm{DM}} \sim 0.27\)–\(2.67~\text{g/cm}^{3}\).
Furthermore, if we take 100 pc as the effective range of the DM spike, the estimated density at this scale is on the order of \(10^{-24}\)–\(10^{-25}~\text{g/cm}^{3}\). 
Thus, if PFDM‑Bardeen BH indeed exist, the dramatically higher density inferred near the shadow scale would serve as a distinctive observational signature for testing the presence of such DM.

\begin{table*}[h]
\centering
\caption{\label{Tab.1} The table presents the values of \( b/M \) at the \( 1\sigma \) and \( 2\sigma \) confidence levels for M87* and Sgr A*, with \( g \) held fixed.}
\begin{tabular}{|lllll|lllll|}
\hline
\hline
M87*/b & $1\sigma$ & $1\sigma$ & $2\sigma$ & $2\sigma$ &
Sgr A*/b & $1\sigma$ & $1\sigma$ & $2\sigma$ & $2\sigma$ \\
\hline
g & Upper & Lower & Upper & Lower &
g & Upper & Lower & Upper & Lower \\
\hline
0.1M & 0.12025 & - & 0.25262 & - & 0.1M & 0.00151 & - & 0.03027 & - \\
\hline
0.2M & 0.12327 & - & 0.25602 & - & 0.2M & 0.00324 & - & 0.03270 & - \\
\hline
0.3M & 0.12837 & - & 0.26176 & - & 0.3M & 0.00630 & - & 0.03689 & - \\
\hline
0.4M & 0.13566 & - & 0.26992 & - & 0.4M & 0.01105 & - & 0.04298 & - \\
\hline
0.5M & 0.14532 & - & 0.28066 & - & 0.5M & 0.01788 & - & 0.05122 & - \\
\hline
0.6M & 0.15756 & - & 0.29419 & - & 0.6M & 0.02720 & - & 0.06192 & - \\
\hline
0.7M & 0.17268 & 0.00322 & 0.31078 & - & 0.7M & 0.03949 & - & 0.07547 & - \\
\hline
0.8M & 0.19109 & 0.01593 & 0.33082 & - & 0.8M & 0.05537 & - & 0.09239 & - \\
\hline
\end{tabular}
\end{table*}

\section{Observed appearance, shadow, and photon rings in thin disk accretion flow}
In this section, we explore the optical appearance of the PFDM–Bardeen BH when it is illuminated by a thin  accretion disk \cite{Uniyal:2022vdu,Peng:2020wun,Yang:2022btw}. The observer is assumed to be positioned along the polar axis, viewing the disk face-on in the equatorial plane. Photons emitted from various regions of the disk follow distinct trajectories, leading to different observable features in the BH image. To understand these effects, we first categorize the possible photon trajectories around PFDM–Bardeen BH before examining the corresponding image formation. Hence, within the parameter space constrained by current observational data, our analysis focuses on elucidating how the presence of the DM parameter and magnetic charge modifies photon propagation and the resulting image morphology.

\subsection{Categories of photon trajectories: Direct emission, Lensing rings, and Photon rings}
To establish the foundation for this investigation, we first analyze the trajectories of photons propagating in the vicinity of the BH. Equation \eqref{11} provides the general form of the photon’s trajectory. When the impact parameter satisfies $\bar{b} < \bar{b}_c$, the photon trajectory stays outside the event horizon. In this regime, the total variation of the azimuthal angle $\phi$ along the photon’s path can be obtained by integrating the orbital equation
\be
\label{15}
\phi=\int_0^{u_{\rm H}} \frac{1}{\sqrt{G(u)}} d u, \quad \bar{b}<\bar{b}_c,
\ee
the upper integration limit is $u_{\rm H} = 1 / r_{\rm H} $, where $r_{\rm H} $ represents the radial position of the BH's outer event horizon. When $\bar{b} > \bar{b}_c$, the light ray experiences a turning point determined by the smallest positive real root of $G(u) = 0$ (This root is expressed as $u_{\rm min} = 1/r_{\min}$). Consequently, the total deflection angle of the photon, measured as the accumulated change in the azimuthal coordinate $\phi$ outside the event horizon, can be calculated through the integral form that depends on the specific value of the impact parameter $\bar{b}$ 
\be
\label{16}
\phi=2 \int_0^{u_{\rm min }} \frac{1}{\sqrt{G(u)}} d u, \quad \bar{b}>\bar{b}_c.
\ee

To elucidate the origin of the radiation detected near BHs, photon trajectories are often divided into three distinct classes—direct emission, lensing ring, and photon ring—as initially proposed in Ref. \cite{Gralla:2019xty}. Within this framework, one can characterize the orbital motion of photons by defining the turn number $n(\bar{b}) = \phi / 2\pi$, which quantifies how many complete revolutions a photon performs around the BH prior to reaching the observer. The value of $n(\bar{b})$ depends sensitively on the impact parameter $\bar{b}$, and as demonstrated in Refs. \cite{Peng:2020wun,Yang:2022btw}, it shows discrete transitions corresponding to different photon path families. The function $n(\bar{b})$ can be defined as 
\be
\label{17}
n(\bar{b})=\frac{2 m-1}{4}, \quad m \in \mathbb{Z}^{+} .
\ee

The boundaries separating the different photon trajectory families are determined by the critical impact parameters $\bar{b}_m^{\pm}$. Specifically, $\bar{b}_m^{-}$ (with $\bar{b}_m^{-} < \bar{b}_c$) and \( \bar{b}_m^{+} \) (with \( \bar{b}_m^{+} > \bar{b}_c \)) represent, respectively, the smallest and largest allowed impact parameters associated with the \( m \)-th order photon orbits, where \( \bar{b}_c \) denotes the critical value. These parameters serve as thresholds that distinguish distinct classes of photon paths, which can be systematically categorized as

$\bullet$ direct emission: $\frac{1}{4}<n<\frac{3}{4} \Rightarrow \bar{b} \in\left(0, \bar{b}_2^{-}\right) \cup\left(\bar{b}_2^{+}, \infty\right)$;

$\bullet$ lensing ring: $\frac{3}{4}<n<\frac{5}{4} \Rightarrow \bar{b} \in\left(\bar{b}_2^{-}, \bar{b}_3^{-}\right) \cup\left(\bar{b}_3^{+}, \bar{b}_2^{+}\right)$;

$\bullet$  photon ring: $n>\frac{5}{4} \Rightarrow \bar{b} \in\left(\bar{b}_3^{-}, \bar{b}_3^{+}\right)$.

Direct emission corresponds to photons that travel from the accretion disk to the observer after a single intersection with the disk surface. In contrast, lensing rings arise from photons that undergo two disk crossings due to gravitational deflection before reaching the observer. Finally, photon rings are formed by photons that orbit the BH multiple times—intersecting the disk three or more times—before eventually escaping toward the observer, producing the innermost lensed features in observed image.

\begin{figure}[H]
\centering

\begin{minipage}{0.3\textwidth}
    \centering
    \includegraphics[scale=0.5]{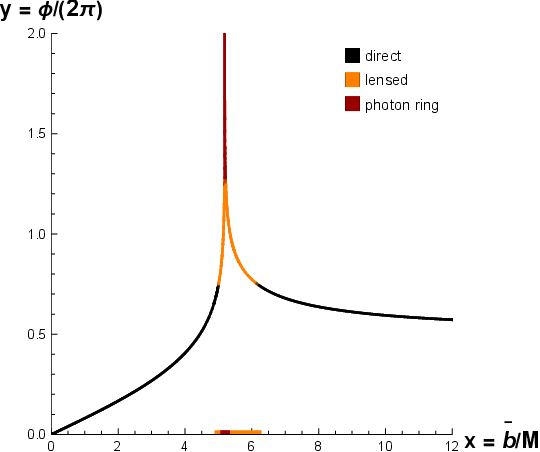}
    \subcaption{Bardeen BH}
\end{minipage}
\hspace{0.1cm}
\begin{minipage}{0.3\textwidth}
    \centering
    \includegraphics[scale=0.5]{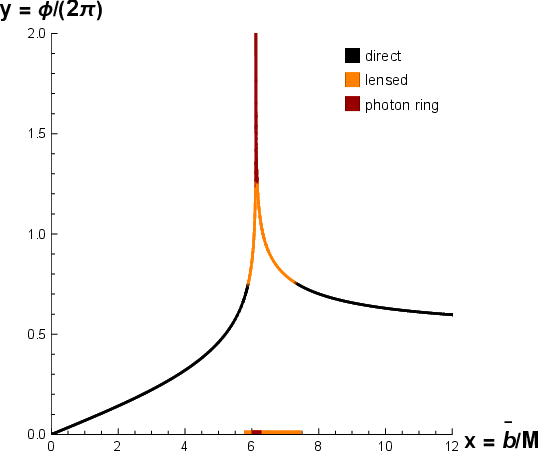}
    \subcaption{PFDM-Sch BH}
\end{minipage}
\hspace{0.1cm}
\begin{minipage}{0.3\textwidth}
    \centering
    \includegraphics[scale=0.5]{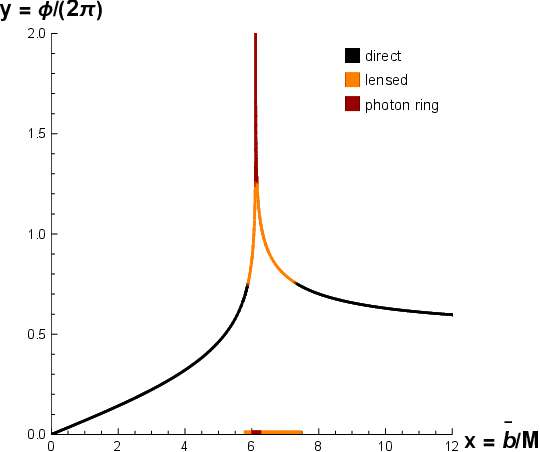}
    \subcaption{PFDM-Bardeen BH}
\end{minipage}

\medskip

\begin{minipage}{0.3\textwidth}
    \centering
    \includegraphics[scale=0.5]{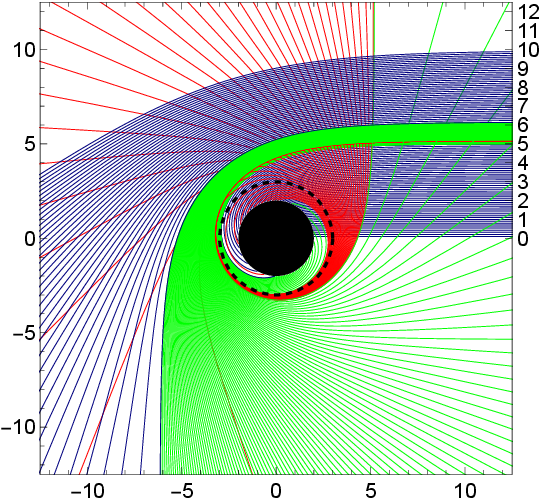}
    \subcaption{Bardeen BH}
\end{minipage}
\hspace{0.1cm}
\begin{minipage}{0.3\textwidth}
    \centering
    \includegraphics[scale=0.5]{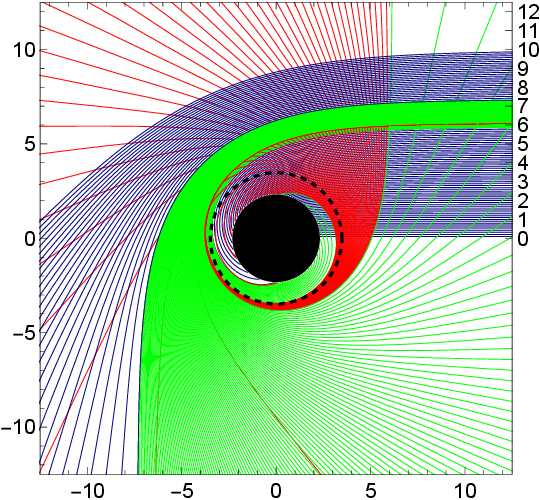}
    \subcaption{PFDM-Sch BH}
\end{minipage}
\hspace{0.1cm}
\begin{minipage}{0.3\textwidth}
    \centering
    \includegraphics[scale=0.5]{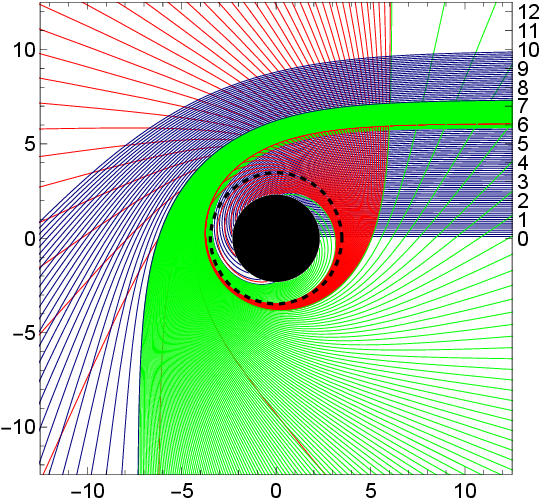}
    \subcaption{PFDM-Bardeen BH}
\end{minipage}

\caption{\label{fig.5}
The photon dynamics for the Bardeen, PFDM-Sch, and PFDM-Bardeen BHs are shown as functions of \( \bar{b} \). The top panels display the winding number \( n=\phi/2\pi \), used to classify trajectories: direct emission (\( n<3/4 \), black), lensed ring (\( 3/4<n<5/4 \), yellow), and photon ring (\( n>5/4 \), red). The bottom panels illustrate representative paths in (\(r,\phi)\) coordinates, with impact parameter 1/10, 1/100, and 1/1000. The BH is shown as a filled disk, and the dashed circle marks the photon orbit. The three cases correspond to \( g/M=0.1 \) (Bardeen BH), \( b/M=0.1 \) (PFDM-Sch BH), and \( b/M=0.1, g/M=0.1 \) (PFDM-Bardeen BH).
}
\end{figure}

Figure \ref{fig.5} illustrates the photon trajectories for three BH spacetimes: the Bardeen BH (\(g/M = 0.1\)), the PFDM-Sch BH (\(b/M = 0.1\)), and the PFDM-Bardeen BH (\(g/M = b/M = 0.1\)). The first row shows how the orbital winding number around the thin accretion disk varies with the impact parameter $\bar{b}$, where red curves denote photon ring trajectories, yellow curves represent lensing ring trajectories, and black curves correspond to direct emission. The second row presents the associated photon paths, with direct emission shown in blue, lensing ring trajectories in green, and photon ring trajectories in red, thereby completing the classification of photon motion in these geometries. In addition, table II reveals clear quantitative distinctions among the Bardeen, PFDM-Sch, and PFDM–Bardeen BHs. The shadow radius \(\bar{b}_c\) shows clear and non-degenerate differences among the three BHs. As indicated in Table II, the Bardeen BH yields the smallest shadow scale, with \(\bar{b}_c/M\approx5.19\). When PFDM is introduced, the shadow becomes significantly larger: the PFDM–Bardeen and PFDM–Sch BHs give \(\bar{b}_c/M\approx6.11\) and \(6.12\), respectively, representing an increase of about 17–18\% relative to the Bardeen case. A similar hierarchy is observed for the other characteristic quantity: both \(r_{\rm H}\) and \(r_{\rm ph}\) consistently exceed the corresponding values of the Bardeen BH.

Table III shows how the DM parameter and the magnetic charge affect the horizon radius, photon sphere radius and the shadow scale. When the magnetic charge is fixed at \(g/M = 0.1\), increasing the DM parameter \(b\) consistently enlarges the key geometrical scales of BH. The outer horizon shifts outward, the photon sphere radius increases, and the corresponding shadow radius becomes larger. This trend indicates that PFDM effectively deepens the gravitational potential at large radii, thereby expanding both the causal boundary and the region accessible to unstable photon orbits. In contrast, when the DM parameter is fixed at \(b/M = 0.1\), variations in the magnetic charge exert only a modest influence. Increasing \(g\) slightly decreases the horizon radius, the photon sphere radius, and the shadow size, and these changes are significantly weaker than those driven by \(b\). As for why the magnetic charge contributes so little to the dynamics, we speculate that several factors may be responsible. In the metric \eqref{5}, both \(g\) and \(r\) appear in the denominator. The influence of the magnetic charge becomes appreciable only when its magnitude is comparable to the radial coordinate. In the large‑scale regions probed by our study—in particular at the photon sphere and shadow scales \(r \sim 5M\)—we typically have \(r \gg g\). Consequently, the Bardeen term \( -2Mr^2/(r^2+g^2)^{3/2} \) expands as \( -2M/r + \mathcal{O}(g^2/r^3) \), so the magnetic charge only enters at order \(g^2/r^3\), a strongly suppressed correction. The role of \(g\) is therefore noticeable only near the central region, where it serves to remove the curvature singularity inherent in the Bardeen BH; however, this extreme inner zone is not directly accessible to EHT imaging. In contrast, the PFDM contributes significantly to the overall spacetime geometry through the logarithmic term \( -(b/r)\ln(r/b) \). 
As shown quantitatively in Table III, varying \(b/M\) from 0 to 0.08 increases the photon sphere radius by about 14$\%$, while varying \(g/M\) from 0 to 0.3 decreases it by only 1.7$\%$. Thus, on macroscopic scales, the gravitational influence of the PFDM dominates over that of the magnetic charge, effectively masking the subtle effects of \(g\) in the dynamics. This explains why all our observational signatures—shadow size, photon ring brightness, and redshift distribution—are primarily controlled by \(b\), with \(g\) playing only a negligible role.
Overall, \(b\) serves as the primary parameter governing the global BH scale and shadow diameter, while the magnetic charge introduces only a minor, secondary correction that mildly contracts these characteristic radii.

\begin{table*}[htbp]
\centering
\caption{\label{Tab.2} Table summarizes the dimensionless physical quantities for the three BHs.}
\setlength{\tabcolsep}{2pt} 
\renewcommand{\arraystretch}{1.05} 
\begin{tabular}{|l|cccccccc|}
\hline
BHs & $r_{\rm H}/M$ & $r_{\rm ph}/M$& $\bar{b}_{c}/M$ & $\bar{b}^{-}_{1}/M$ &
$\bar{b}^{-}_{2}/M$  & $\bar{b}^{+}_{2}/M$ & $\bar{b}^{-}_{3}/M$& $\bar{b}^{+}_{3}/M$  \\
\hline
Bardeen ($g/M=0.1$,$b/M=0.0$) 
& 1.99247 & 2.99164 & 5.18747 & 2.83799 & 5.00485 & 6.16226 & 5.17897 & 5.21962 \\
\hline
PFDM-Sch ($g/M=0.0$,$b/M=0.1$) 
&2.31416 &3.48255 &6.12047 & 3.30518 &5.88831 &7.36368 &6.10897 &6.16345\\
\hline
PFDM-Bardeen ($g/M=b/M=0.1$) 
&2.30829 &3.47608 &6.11374 &3.29758 &5.88022 &7.35976 & 6.10210 &6.15704 \\
\hline
\end{tabular}
\end{table*}

\begin{table*}[htbp]
\centering
\caption{\label{Tab.3}Table examines how the model parameters affect various dimensionless physical quantities under two controlled scenarios: \(g/M=0.1\) or \(b/M=0.1\).
 }
\setlength{\tabcolsep}{4.3pt} 
\renewcommand{\arraystretch}{1.05} 
\begin{tabular}{|l|cccccccc|}
\hline
{\bf Magnetic charge  g/M=0.1}& $r_{\rm H}/M$ & $r_{\rm ph}/M$& $\bar{b}_{c}/M$ & $\bar{b}^{-}_{1}/M$ &
$\bar{b}^{-}_{2}/M$  & $\bar{b}^{+}_{2}/M$ & $\bar{b}^{-}_{3}/M$& $\bar{b}^{+}_{3}/M$  \\
\hline
b/M=0.00 (Bardeen)
& 1.99247 & 2.99164 & 5.18747 & 2.83799 & 5.00485 & 6.16226 & 5.17897 & 5.21962 \\
\hline
b/M=0.04
&2.15297 &3.23645 &5.64572 &3.07044 &5.44002&6.74339&5.63586 &5.68274\\
\hline
b/M=0.08
&2.26149 &3.40360 &5.97039 &3.22894 &5.74579 &7.16907 & 5.95933 &6.01163 \\
\hline
{\bf DM parameter b/M=0.1}& $r_{\rm H}/M$ & $r_{\rm ph}/M$& $\bar{b}_{c}/M$ & $\bar{b}^{-}_{1}/M$ &
$\bar{b}^{-}_{2}/M$  & $\bar{b}^{+}_{2}/M$ & $\bar{b}^{-}_{3}/M$& $\bar{b}^{+}_{3}/M$  \\
\hline
g/M=0.00 (PFDM-Sch)
&2.31416 &3.48255 &6.12047 & 3.30518 &5.88831 &7.36368 &6.10897 &6.16345\\
\hline
g/M=0.15
&2.30091 &3.46794 &6.10530 &3.28803 &5.87004 &7.35485 &6.09348 &6.14901\\
\hline
g/M=0.30
&2.26008 &3.42313 &6.05894 &3.23515 &5.81376 &7.32820 & 6.04608 &6.10501 \\
\hline

\end{tabular}
\end{table*}

\subsection{Observed specific intensities inferred through transfer functions}
To quantify the radiation received by a distant observer, we employ transfer functions, which encode the mapping between emission processes in the accretion flow and their corresponding appearance on the observer’s screen. These functions incorporate the full relativistic effects experienced by photons—such as gravitational redshift, light bending, and multiple-orbit lensing—and therefore provide a direct bridge between the local emission of thin disk and observed specific intensity. By integrating the photon trajectories weighted by the appropriate transfer functions, one can systematically construct the observed brightness distribution, enabling a consistent comparison of BHs.

Following the setup of the previous subsection, we model the emission as originating from a static accretion disk lying in the equatorial plane of PFDM-Bardeen BH, with isotropic local radiation. Photons are traced along null geodesics to a distant polar observer, who measures the radiation at frequency \( \nu_o \). The corresponding specific intensity is obtained through the relativistic transfer relation of Ref. \cite{Uniyal:2022vdu,Peng:2020wun,Yang:2022btw,Wang:2023vcv,Zeng:2020vsj,Gralla:2019xty}, which follows
\be
\label{18}
I_o\left(r, v_o\right)=f(r)^{\frac{3}{2}} I_e\left(r, v_e\right),
\ee
where \(I_{\rm e}(r,\nu_{e})\) and \( \nu_{\rm e}\) denote the specific intensity and emission frequency at the disk surface, whereas \(I_{o}(r,\nu_{o})\) refers to the corresponding quantity measured by an observer at the north pole. The relation follows from Liouville’s theorem, which guarantees that the quantity \(I_{\rm e}(r,\nu_{e})/\nu_{\rm e}^{3}\) remains conserved along photon geodesics. The total observed intensity \(I_{\rm obs}(r)\) at an image radius is subsequently obtained by integrating the observer frame specific intensity \(I_{o}(r,\nu_{o})\) over all frequencies.
\be
\label{19}
I_{\rm obs}(r)=\int I_o\left(r, v_o\right) d v_o=f(r)^2 I_{\rm e m}(r),
\ee
here the total emitted intensity defined as
\(I_{\rm em}(r) = \int I_{\rm e}(r,\nu_{\rm e}) d\nu_{\rm e}\) ,
which represents the frequency integrated radiation output from each point on the disk. It should be stressed that the observed signal accounts only for radiation emitted by the disk, without including absorption, scattering, or reflection effects. Each time a photon trajectory intersects the equatorial plane, it acquires an additional contribution to the luminosity received by a distant observer. As outlined earlier, the degree of orbital winding determines the number of such intersections: weakly deflected rays cross the disk once, moderately wound trajectories intersect it twice after bending behind the BH, and highly wound paths encounter the disk three or more times as they loop around the compact object. The total observed intensity is therefore obtained by summing the emission gathered at each intersection along the photon path, yielding 
\be
\label{20}
I_{\rm o b s}(\bar{b})=\left.\sum_m f(r)^2 I_{\rm e m}(r)\right|_{r=r_m(\bar{b})},
\ee
where transfer function \(r_m(\bar{b})\) provides a mapping between the photon impact parameter and the radial location at which the photon intersects the thin emitting disk. Its derivative $d r/d \bar{b}$, often referred to as the demagnification factor, quantifies how strongly the transfer function compresses or stretches this mapping. In essence, the slope characterizes the degree of demagnification encoded in the transfer function itself \cite{Gralla:2019xty}. A steeper slope of the transfer function restricts the range of allowed impact parameters, thereby suppressing its contribution to the BH’s total observed intensity. Although photons near the photon ring may, in principle, intersect the accretion disk multiple times, the extreme steepness of their transfer functions renders their intensity contribution effectively negligible. Consequently, our analysis focuses solely on the first three intersections (\(m = 1, 2, 3\)). The corresponding transfer functions can be written as
\be
\1\{\begin{split}
\label{21}
& r_1(\bar{b})=\frac{1}{u\left(\frac{\pi}{2}, \bar{b}\right)}, \quad \bar{b} \in\left(\bar{b}_1^{-},+\infty\right), \\
& r_2(\bar{b})=\frac{1}{u\left(\frac{3 \pi}{2}, \bar{b}\right)}, \quad \bar{b} \in\left(\bar{b}_2^{-}, \bar{b}_2^{+}\right),\\
&r_3(\bar{b})=\frac{1}{u\left(\frac{5 \pi}{2}, \bar{b}\right)}, \quad \bar{b} \in\left(\bar{b}_3^{-}, \bar{b}_3^{+}\right).
\end{split}\2.
\ee

\begin{figure}[H]
\centering
\begin{minipage}{0.3\textwidth}
    \centering
    \includegraphics[scale=0.5]{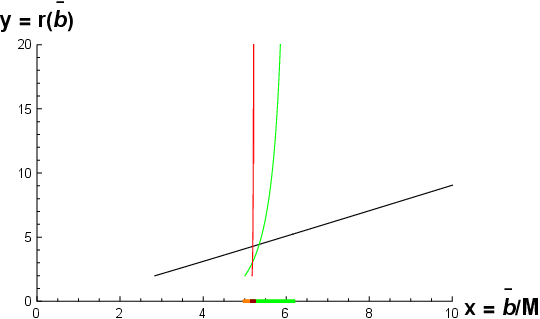}
    \subcaption{$g/M=0.1,b/M=0$;\\
    $\tan{\th_{b}}=0.986608$;\\
    $\tan{\th_{g}}=75.067$;\\
    $\tan{\th_{r}}=6698.88$.}
\end{minipage}
\hspace{0.1cm}
\begin{minipage}{0.3\textwidth}
    \centering
    \includegraphics[scale=0.5]{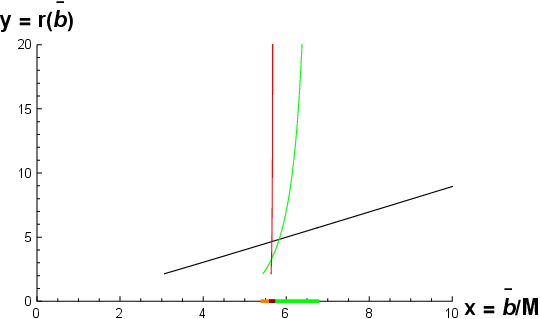}
    \subcaption{$g/M=0.1,b/M=0.04$;\\ $\tan{\th_{b}}=0.982185$;\\
    $\tan{\th_{g}}=100.331$;\\
    $\tan{\th_{r}}=3062.18$.}
\end{minipage}
\hspace{0.1cm}
\begin{minipage}{0.3\textwidth}
    \centering
    \includegraphics[scale=0.5]{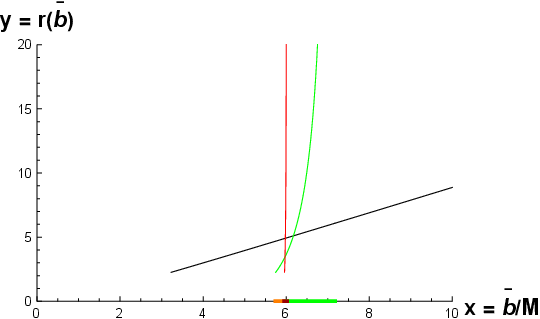}
    \subcaption{$g/M=0.1,b/M=0.08$;\\ $\tan{\th_{b}}=0.978153$;\\
    $\tan{\th_{g}}=101.946$;\\
    $\tan{\th_{r}}=1322.79$.}
\end{minipage}
\medskip
\begin{minipage}{0.3\textwidth}
    \centering
    \includegraphics[scale=0.5]{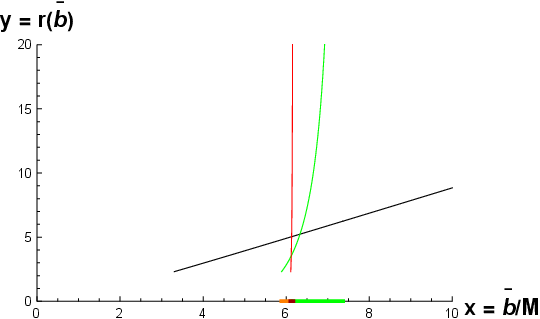}
    \subcaption{$g/M=0,b/M=0.1$;\\ $\tan{\th_{b}}=0.976615$;\\
    $\tan{\th_{g}}=79.564$;\\
    $\tan{\th_{r}}=2465.63$.}
\end{minipage}
\hspace{0.1cm}
\begin{minipage}{0.3\textwidth}
    \centering
    \includegraphics[scale=0.5]{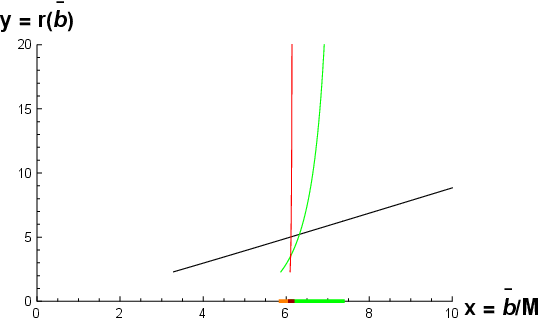}
    \subcaption{$g/M=0.15,b/M=0.1$;\\ $\tan{\th_{b}}=0.976195$;\\
    $\tan{\th_{g}}=68.629$;\\
    $\tan{\th_{r}}=1174.87$.}
\end{minipage}
\hspace{0.1cm}
\begin{minipage}{0.3\textwidth}
    \centering
    \includegraphics[scale=0.5]{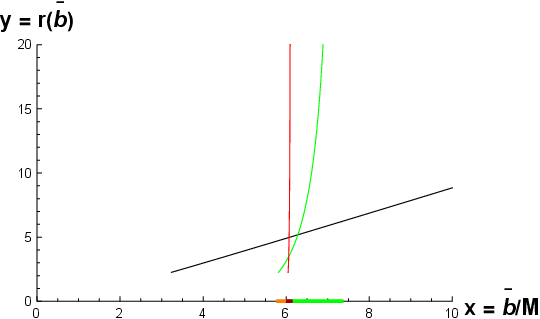}
    \subcaption{$g/M=0.3,b/M=0.1$;\\ $\tan{\th_{b}}=0.974893$;\\
    $\tan{\th_{g}}=77.761$;\\
    $\tan{\th_{r}}=2433.77$.}
\end{minipage}
\caption{\label{fig.6} The first three transfer functions for BHs with different parameter choices are displayed. The top row shows the cases \((g/M = 0.1, b/M = 0,0.04, 0.08)\) from left to right, while the bottom row corresponds to \((g/M = 0,0.15,0.3, b/M = 0.1)\).} 
\end{figure}

Subsequently, we display these three classes of transfer functions in Fig. \ref{fig.6}. The black curve denotes the first transfer function, whose nearly constant and relatively small slope produces the direct image of the emitting disk. The Green curve corresponds to the second transfer function; its steep slope yields a strongly demagnified image originating from the far side of the disk, associated with the lensing ring. The red curve represents the third transfer function, characterized by an almost divergent slope, which generates an even more severely demagnified image from the front side of the disk, identified as the photon ring. 

In addition, we list the slopes \(\tan{\theta_b}\), \(\tan{\theta_r}\), and \(\tan{\theta_g}\) corresponding to the black, red, and green curves in each panel. When focusing solely on the direct emission branch (black curve), the influence of the DM parameter \(b\) remains modest: Its increase produces only a slight reduction in the slope, which may in turn yield a modest enhancement in the oberserved brightness. In contrast, variations in the magnetic charge \(g\) generate changes that are even smaller—confined to the third decimal place and therefore negligible for practical purposes. Consequently, distinguishing the luminosity distributions of three BH configurations, as well as identifying how each parameter shapes the observed brightness, requires a more detailed analysis, which we carry out in the next subsection.

\subsection{Optical appearance of PFDM-Bardeen BH with thin accretion disk}
To investigate the optical appearance of the PFDM–Bardeen BH, we construct its observable image by projecting photon trajectories onto a two-dimensional screen placed at the observer’s location. Under various emission models, the visible structure of the spacetime is determined by how photons propagate in the strong-gravity region. In particular, the shadow boundary is defined by those rays that asymptotically spiral toward the unstable photon sphere. By tracing these critical geodesics from the vicinity of the BH and mapping them onto the observer’s image plane, we obtain the shadow silhouette expressed in the celestial coordinates. This procedure allows a direct visualization of how the DM parameter and magnetic charge modify both the shadow geometry and the surrounding brightness distribution.

Accordingly, once the emissivity profile of thin disk is specified, the received intensity can be evaluated through the transfer function formalism and then mapped onto the observer’s image plane in celestial coordinates, yielding the full optical appearance of the BH. In our analysis, we explore three idealized, static emission models for the accretion disk. In the first emission model, the disk luminosity rises steeply from large radii and reaches the innermost stable circular orbit $(r_{\rm isco})$, after which the profile transitions into a smooth quadratic falloff. In other words, the emission is sharply enhanced near the inner edge of the disk and then progressively diminishes as one moves outward.
\be
\label{22}
I_{\rm em}\left(r\right)= \begin{cases} I_0\left(\frac{1}{r-\left(r_{\mathrm{isco}}-1\right)}\right)^2, & r>r_{\mathrm{isco}} \\ 0, & r<r_{\mathrm{isco}}\end{cases}
\ee
where $I_0$ denotes the maximum value of the emitted intensity. In the second model, the emissivity is prescribed to peak precisely at the $r_{\rm ph}$, after which it decreases following a cubic falloff. Thus, the brightest emission originates from the photon sphere region, with the intensity rapidly diminishing as the radius moves away from this location.
\be
\label{23}
I_{\rm e m}(r)= \begin{cases}I_0\left(\frac{1}{r-\left(r_{\rm p h}-1\right)}\right)^3, & r>r_{\rm p h} \\ 0, & r \leq r_{\rm p h}\end{cases}
\ee

In the third model, the disk brightness decreases smoothly from the event horizon outward to the $(r_{\rm isco})$. In this prescription, the emission gradually weakens across this radial range, producing a continuous, monotonic decline without a pronounced peak.
\be
\label{24}
I_{\rm e m}(r)= \begin{cases}I_0 \frac{\frac{\pi}{2}-\arctan \left(r-\left(r_{\rm i s c o}-1\right)\right)}{\frac{\pi}{2}-\arctan \left(r_{\rm H}-\left(r_{\rm i s c o}-1\right)\right)}, & r>r_{\rm H} \\ 0. & r \leq r_{\rm H}\end{cases}
\ee

Figures \ref{7} and \ref{8} display how the optical appearance of PFDM-Bardeen BH evolves as the DM parameter and magnetic charge. For each emission prescription, the first two columns illustrate the radial profiles of the intrinsic disk emissivity and the corresponding intensity measured by a distant observer. The final column translates these results into the observer’s sky by mapping the radiation field onto the two-dimensional celestial coordinate plane, thereby showing the resulting images of the BH under each emission model. In Fig. \ref{7}, the black curve represents the Bardeen BH, whereas the green, and red curves correspond to PFDM-Bardeen BH with \(g/M= 0.1\) and DM parameter \(b/M = 0.04\) and \(0.08\), respectively. 
The key trends emerging from these comparisons are summarized below.


Starting with Model 1 (top row), the emitted intensity \(I_{\rm em}\) reaches its maximum at the $r_{\rm isco}$ and then decays rapidly with increasing radius, with the peak shifting outward as the DM parameter \(b\) increases. This outward shift reflects a key physical effect: PFDM deepens the gravitational potential at intermediate radii, effectively pushing stable circular orbits outward and relocating the region of peak dissipation. The corresponding observed intensity \(I_{\rm obs}(\bar{b})\) (middle column) reveals a striking qualitative difference. For the pure Bardeen BH (\(b=0\), black curve), three distinct peaks appear: a dominant direct‑emission peak, a weaker lensing‑ring peak, and an even fainter photon‑ring peak. In contrast, for PFDM‑Bardeen BHs (\(b>0\), green and red curves), only two peaks remain, since the photon ring contribution is almost entirely suppressed. The physical origin of this suppression may lie in the modified effective potential: the PFDM term broadens the centrifugal barrier and increases the critical impact parameter \(\bar{b}_c\), thereby reducing the trapping efficiency of null geodesics near the $r_{\rm ph}$. Photons that would otherwise execute multiple orbits are instead captured or deflected with fewer windings, extinguishing the high‑order ring. The suppression of the photon ring provides a robust discriminator between the Bardeen and PFDM‑Bardeen BHs. When PFDM is present, the photon ring peak disappears in Model 1, whereas the presence of a photon ring peak would strongly suggest the absence of PFDM and point instead to a pure Bardeen BH.

In Model 2 (middle row), the emission profile again shifts outward as the DM parameter \(b\) increases. Yet the corresponding observed intensity displays only two peaks—arising from the direct emission and the lensing ring—for both the Bardeen and PFDM–Bardeen BHs. The Bardeen case remains the brightest, while the presence of DM consistently reduces the overall luminosity. Furthermore, in Model 2 the photon ring is absent even without PFDM, so the number of peaks cannot be used to distinguish between the two types of BHs—a consequence that likely arises from the emissivity profile itself. Therefore, for Model 2, if an observed system exhibits only two peaks, one cannot immediately conclude that PFDM is present.
Moreover, the model 3 (bottom row) yields an analogous behavior: the observed flux is dominated entirely by the direct emission, with contributions from the lensing and photon rings effectively absent. Overall, these results demonstrate that the DM parameter \(b\) exerts a clear suppressive effect on the observed brightness. 


Figure \ref{fig.8} examines the influence of the magnetic charge \(g\) by fixing \(b/M = 0.1\) and considering \(g/M = 0,\ 0.15,\ 0.3\). Overall, the magnetic charge induces only minor modifications in both the emitted and observed intensity, as well as in the overall optical appearance, indicating that the gravitational effect of \(g\) is subdominant compared to the DM parameter on scales \(r \sim 5M\). In Model 1 and Model 2, both the PFDM‑Schwarzschild ( \(g=0\) ) and PFDM‑Bardeen ( \(g>0\) ) BHs exhibit only two peaks, with the PFDM‑Sch case being slightly brighter but qualitatively identical. This may be because in these two models the emissivity peaks at either the $r_{\rm isco}$(Model 1) or the $r_{\rm ph}$ (Model 2), where the metric modification due to \(g\) is of order \(g^2/r^3\) and thus too small to alter the peak structure. Model 3, however, reveals a notable distinction: the PFDM‑Sch BH displays three small but discernible peaks, whereas the PFDM‑Bardeen BH completely lack these higher‑order features. We cannot definitively determine why Model 3 exhibits such a behavior; it may be attributed to numerical errors in our calculations or point to deeper physical mechanisms yet to be uncovered. Nevertheless, it is clear that on the scale of the bright rings, the influence of the NLED is subdominant compared to that of PFDM.

\begin{figure}[H]
	\centering
	
	\begin{subfigure}[b]{0.31\textwidth}
		\centering
		\begin{overpic}[width=\linewidth]{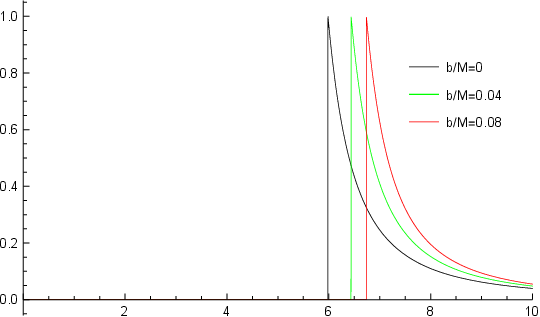}
			\put(99,2){\color{black} $r/M$}
			\put(0,60){\color{black} $I_{\text{em}}/I_{0}$}
		\end{overpic}
	\end{subfigure}
	\hfill
	\begin{subfigure}[b]{0.31\textwidth}
		\centering
		\begin{overpic}[width=\linewidth]{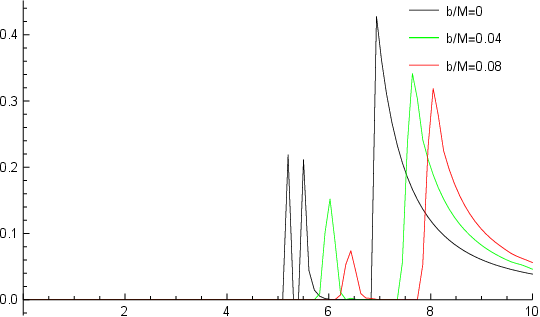}
			\put(103,2){\color{black} $\bar{b}/M$}
			\put(0,60){\color{black} $I_{\text{obs}}/I_{0}$}
		\end{overpic}
	\end{subfigure}
	\hfill
	\begin{subfigure}[b]{0.31\textwidth}
		\centering
		\includegraphics[width=0.7\linewidth]{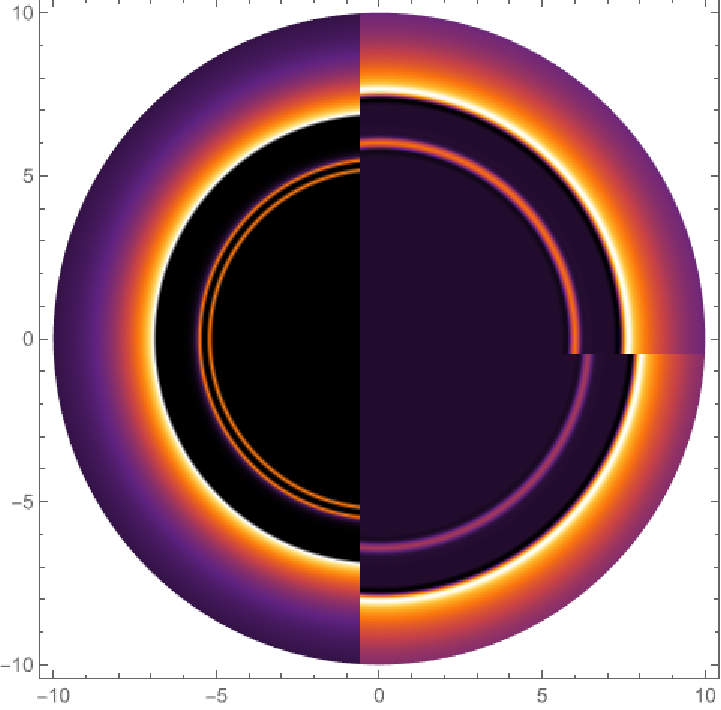}
		\hspace{0.1cm}
		\includegraphics[width=0.11\linewidth]{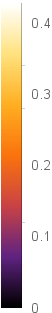}
	\end{subfigure}
	
	\vspace{0.05cm} 
	\begin{subfigure}[b]{0.31\textwidth}
		\centering
		\begin{overpic}[width=\linewidth]{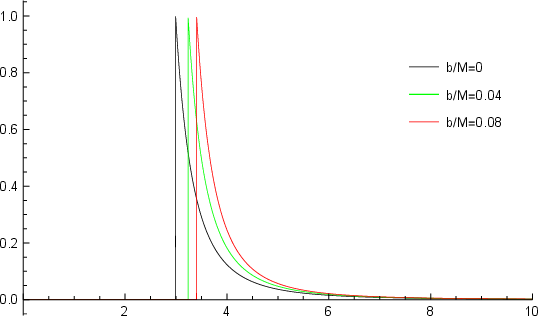}
			\put(99,2){\color{black} $r/M$}
			\put(0,60){\color{black} $I_{\text{em}}/I_{0}$}
		\end{overpic}
	\end{subfigure}
	\hfill
	\begin{subfigure}[b]{0.31\textwidth}
		\centering
		\begin{overpic}[width=\linewidth]{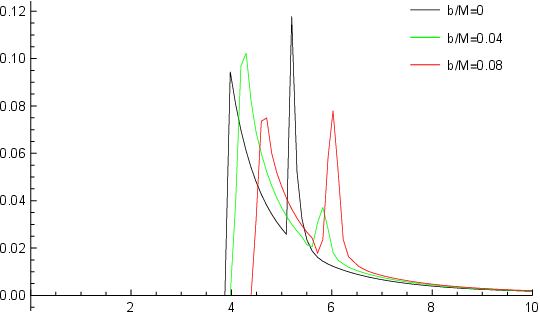}
			\put(103,2){\color{black} $\bar{b}/M$}
			\put(0,63){\color{black} $I_{\text{obs}}/I_{0}$}
		\end{overpic}
	\end{subfigure}
	\hfill
	\begin{subfigure}[b]{0.31\textwidth}
		\centering
		\includegraphics[width=0.7\linewidth]{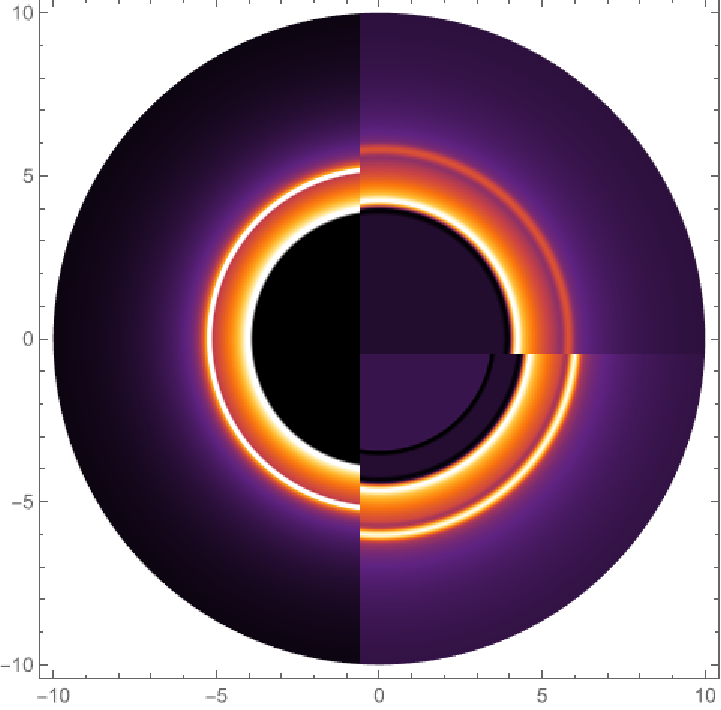}
		\hspace{0.1cm}
		\includegraphics[width=0.12\linewidth]{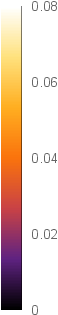}
	\end{subfigure}
	
	\vspace{0.05cm}
	
	\begin{subfigure}[b]{0.31\textwidth}
		\centering
		\begin{overpic}[width=\linewidth]{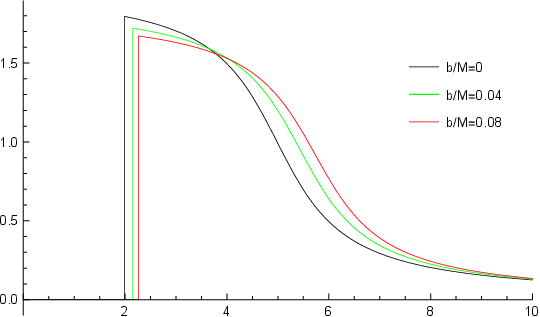}
			\put(99,2){\color{black} $r/M$}
			\put(0,60){\color{black} $I_{\text{em}}/I_{0}$}
		\end{overpic}
	\end{subfigure}
	\hfill
	\begin{subfigure}[b]{0.31\textwidth}
		\centering
		\begin{overpic}[width=\linewidth]{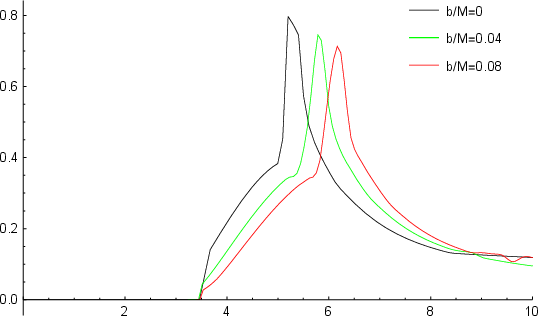}
			\put(103,2){\color{black} $\bar{b}/M$}
			\put(0,64){\color{black} $I_{\text{obs}}/I_{0}$}
		\end{overpic}
	\end{subfigure}
	\hfill
	\begin{subfigure}[b]{0.31\textwidth}
		\centering
		\includegraphics[width=0.7\linewidth]{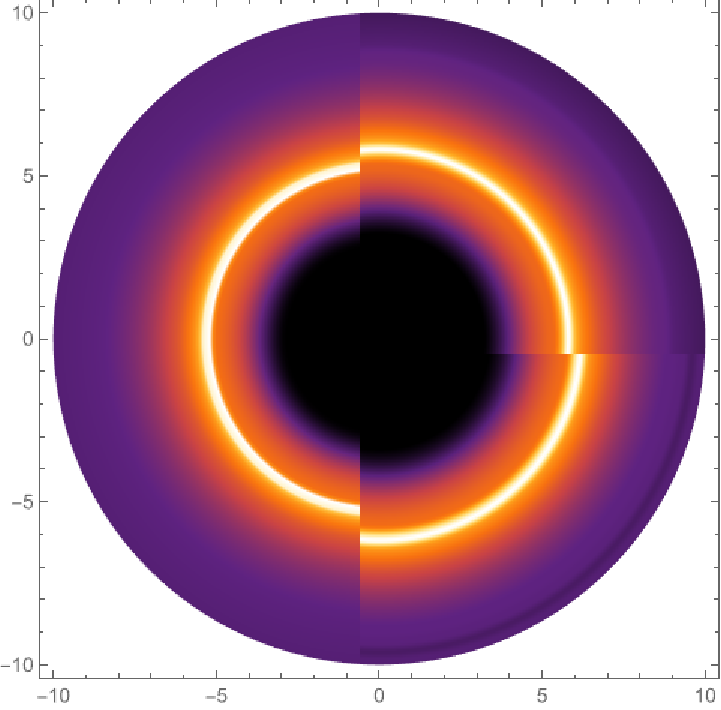}
		\hspace{0.1cm}
		\includegraphics[width=0.11\linewidth]{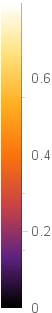}
	\end{subfigure}
    \caption{\label{fig.7} The face-on optical appearance of the thin disk is displayed for three distinct emission prescriptions. The top, middle, and bottom rows correspond to Models 1, 2, and 3, respectively. In each row, the black, green, and red images represent configurations with fixed magnetic charge \(g/M = 0.1\) and DM parameters \(b/M = 0\), \(0.04\) and \(0.08\). For all panels, both the emitted intensity 
    \(I_{\mathrm{em}}\) and the observed intensity
    \(I_{\mathrm{obs}}\) are rescaled by the maximum emission \(I_{0}\) outside the event horizon, ensuring a uniform normalization across models and parameter choices.}
\end{figure}

\begin{figure}[H]
	\centering
	
	\begin{subfigure}[b]{0.31\textwidth}
		\centering
		\begin{overpic}[width=\linewidth]{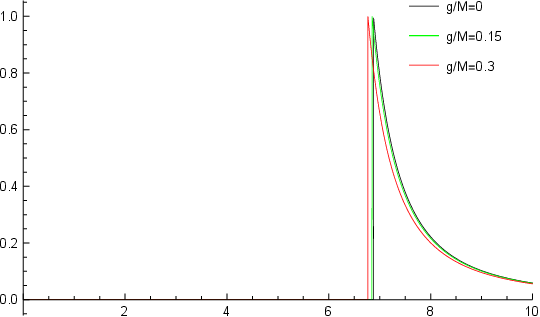}
			\put(99,2){\color{black} $r/M$}
			\put(0,60){\color{black} $I_{\text{em}}/I_{0}$}
		\end{overpic}
	\end{subfigure}
	\hfill
	\begin{subfigure}[b]{0.31\textwidth}
		\centering
		\begin{overpic}[width=\linewidth]{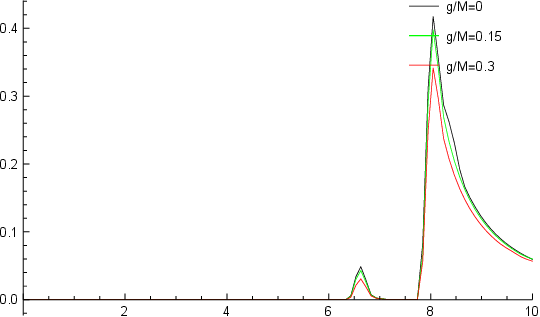}
			\put(103,2){\color{black} $\bar{b}/M$}
			\put(0,60){\color{black} $I_{\text{obs}}/I_{0}$}
		\end{overpic}
	\end{subfigure}
	\hfill
	\begin{subfigure}[b]{0.31\textwidth}
		\centering
		\includegraphics[width=0.7\linewidth]{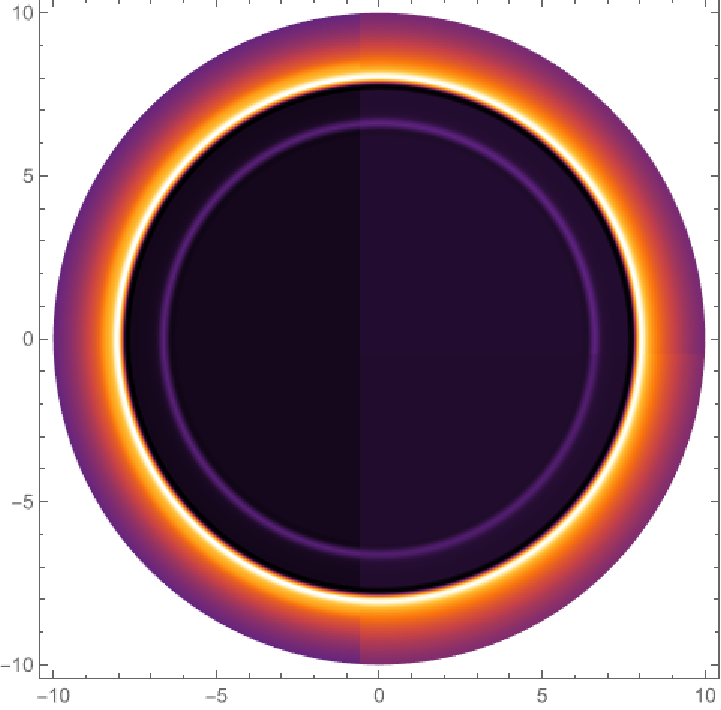}
		\hspace{0.1cm}
		\includegraphics[width=0.11\linewidth]{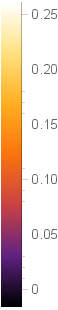}
	\end{subfigure}
	
	\vspace{0.05cm} 
	\begin{subfigure}[b]{0.31\textwidth}
		\centering
		\begin{overpic}[width=\linewidth]{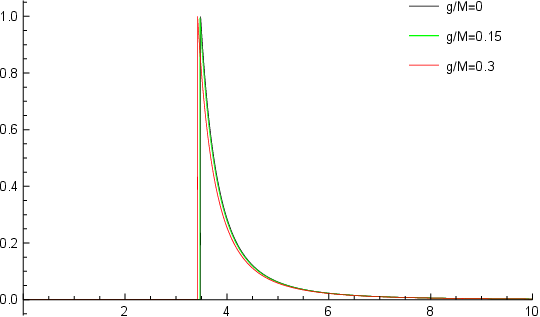}
			\put(99,2){\color{black} $r/M$}
			\put(0,60){\color{black} $I_{\text{em}}/I_{0}$}
		\end{overpic}
	\end{subfigure}
	\hfill
	\begin{subfigure}[b]{0.31\textwidth}
		\centering
		\begin{overpic}[width=\linewidth]{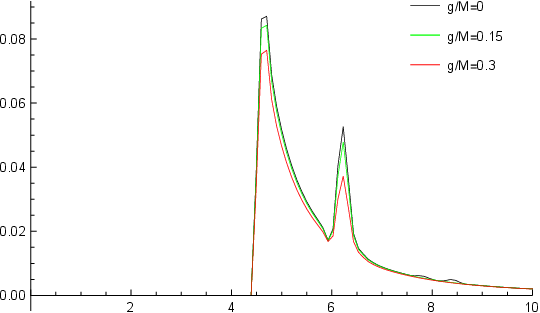}
			\put(103,2){\color{black} $\bar{b}/M$}
			\put(0,63){\color{black} $I_{\text{obs}}/I_{0}$}
		\end{overpic}
	\end{subfigure}
	\hfill
	\begin{subfigure}[b]{0.31\textwidth}
		\centering
		\includegraphics[width=0.7\linewidth]{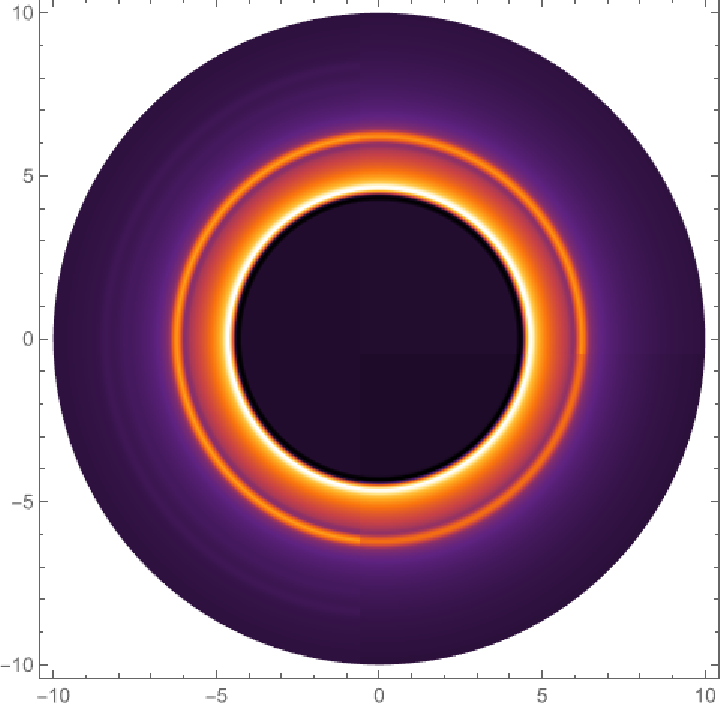}
		\hspace{0.1cm}
		\includegraphics[width=0.12\linewidth]{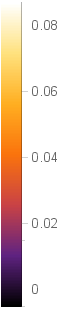}
	\end{subfigure}
	
	\vspace{0.05cm}
	
	\begin{subfigure}[b]{0.31\textwidth}
		\centering
		\begin{overpic}[width=\linewidth]{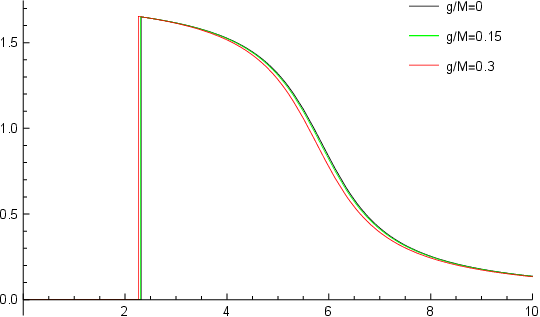}
			\put(99,2){\color{black} $r/M$}
			\put(0,60){\color{black} $I_{\text{em}}/I_{0}$}
		\end{overpic}
	\end{subfigure}
	\hfill
	\begin{subfigure}[b]{0.31\textwidth}
		\centering
		\begin{overpic}[width=\linewidth]{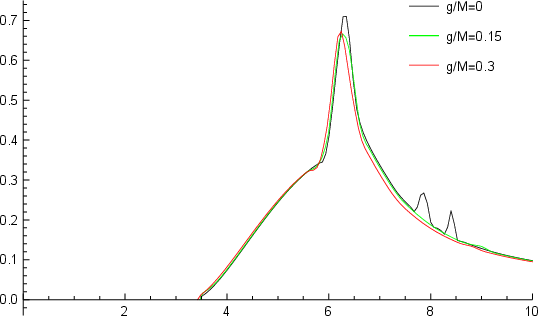}
			\put(103,2){\color{black} $\bar{b}/M$}
			\put(0,64){\color{black} $I_{\text{obs}}/I_{0}$}
		\end{overpic}
	\end{subfigure}
	\hfill
	\begin{subfigure}[b]{0.31\textwidth}
		\centering
		\includegraphics[width=0.7\linewidth]{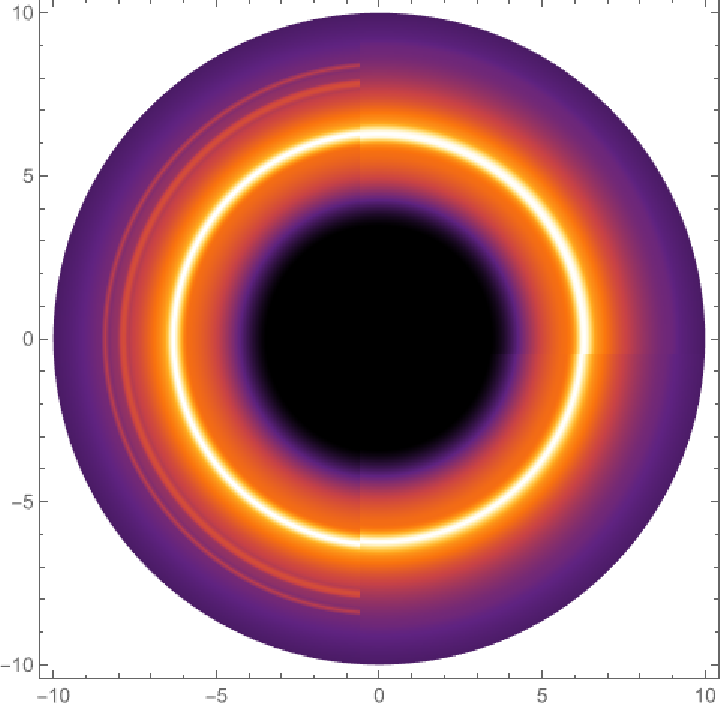}
		\hspace{0.1cm}
		\includegraphics[width=0.11\linewidth]{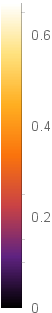}
	\end{subfigure}
    \caption{\label{fig.8} In each row, the black, green, and red panels correspond to configurations with a fixed DM parameter \(b/M = 0.1\) and magnetic charges \(g/M = 0\), \(0.15\), and \(0.30\), respectively. For the intensity calculations, we adopt the corresponding \(r_{\rm isco}/M = 6.87751\), \(6.85051\), and \(6.76815\). The optical appearances obtained for these three parameter choices are then assembled into a composite image, shown in the rightmost column.}
\end{figure}

\section{Primary and secondary imaging features of thin accretion disks}

In this chapter, we investigate the imaging properties of thin accretion disks by systematically analyzing both the first-order (primary) and second-order (secondary) emission. These two levels of imaging encode distinct aspects of photon trajectories in curved spacetime: the primary image is dominated by direct photons reaching the observer with minimal deflection, while the secondary image arises from strongly lensed photons that execute additional orbits around BH. By comparing these structures across different parameter choices—such as the magnetic charge $g$, DM parameter $b$, we aim to clarify how the underlying BH shapes the optical appearance of the disk. 

\subsection{Observation coordinate system}
\begin{figure}[H]
\centering
\begin{minipage}{1.0\textwidth}
\centering
\includegraphics[scale=0.50,angle=0]{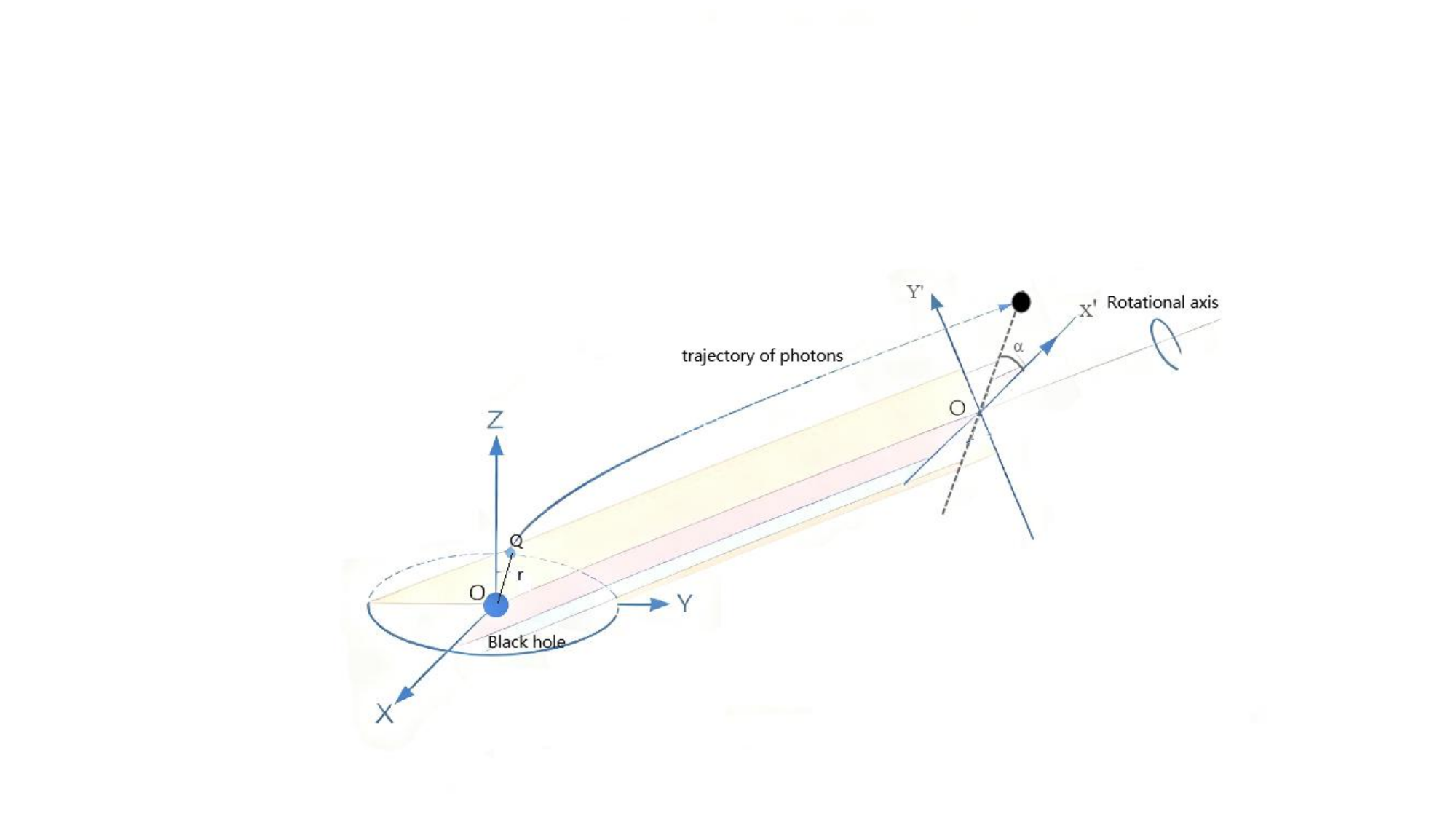}
\end{minipage}
\caption{\label{fig.9}{The \(XYZ\) coordinate system is embedded in the BH. A photon emitted from a radius \(r\) propagates toward the observer’s coordinate system \(X'OY'\), forming an angle \(\alpha\) in the observer’s frame. The black point corresponds to the coordinate \(q(\bar{b}, \alpha)\).}}
\end{figure}

The imaging properties of a geometrically thin accretion disk can be examined by introducing an observer-adapted coordinate system, as shown in Fig. \ref{fig.9}. In the BH's spherical coordinates \((r, \theta, \phi)\), the observer is positioned at \((\infty, \theta, 0)\), with the origin located at the BH center \((r = 0)\).
Within this observational frame \(X'OY'\), consider an image point \(q(\bar{b}, \alpha)\), characterized by the photon’s impact parameter and its apparent angle \(\alpha\) on the observer’s sky. Extending the associated light ray backward along its geodesic leads to an intersection with the accretion disk at point \(Q(r, \pi/2, \phi)\). By invoking the reversibility of null geodesics, any photon emitted from the disk at \(Q(r, \pi/2, \phi)\) will traverse the same trajectory forward in time and arrive at the observer as the image point \(q(\bar{b}, \alpha)\). This one-to-one correspondence between disk emission points and image coordinates provides the foundation for analyzing both primary and higher-order images.

Fixing the radial coordinate  defines a family of photon trajectories that map onto a closed curve of constant radius on the observer’s image plane. As illustrated on the Fig. \ref{9}, an orbit of fixed \(r\) in the equatorial plane intersects any plane spanned by angles \(\alpha\) and \(\alpha+\pi\) at two distinct points, separated by an azimuthal shift of \(\pi\). For convenience, the directions \(\alpha = 0\) and \(\phi = 0\) are aligned with the \(X'\)-axis of the observer and the \(X\)-axis of the BH coordinate system, respectively.

From simple spatial geometry, one can determine the angle \(\varphi\) between the rotation axis and the line segment (OQ), which connects the observer to the emission point on the orbit. Geometrically, the angle $\varphi$ can be expressed as \cite{You:2024uql,Cai:2025pan,Cai:2025rst}
\be
\label{25}
\varphi=\frac{\pi}{2}+\arctan (\tan \theta \sin \alpha).
\ee
As the parameter \(\bar{b}\) approaches the critical value \(\bar{b}_{c}\), photons experience increasingly pronounced gravitational deflection, allowing a single emission point \(Q\) on the disk to generate multiple distinct image locations in the observer’s sky. These image locations are indexed by ordering their azimuthal deflections, so that \(q^{n}\) \((n \in \mathbb{N})\) designates the image produced after the \(n\)-th accumulated winding, with \(n\) representing the image order.

In the \(X'OY'\) plane, all even-order images are projected onto the same side as the emission direction \(\alpha\), whereas odd-order images are reflected to the opposite side, appearing around \(\alpha+\pi\). The cumulative azimuthal shift that gives rise to the \(n\)-th image is denoted by \(\varphi^{n}\), defined as 
\be
\label{26}
\varphi^n= \begin{cases}n \pi+(-1)^n\left[\frac{\pi}{2}+\arctan (\tan \theta \sin \alpha)\right] &, n \in  \text {even} \\ (n+1)\pi+(-1)^n\left[\frac{\pi}{2}+\arctan (\tan \theta \sin \alpha)\right] &, n \in \text { odd }\end{cases}
\ee

When \(n = 0\), the image corresponds to the observer’s direct view of the disk—its primary appearance without any additional winding of the photon trajectory. Increasing \(n\) to 1, 2, 3, and beyond designates progressively higher-order images, each arising from photons that have undergone additional azimuthal winding before reaching the observer, thereby producing secondary, tertiary, and subsequent visual repetitions of the emitting region. In the following subsection, we provide a detailed examination of the primary and secondary images produced by the PFDM–Bardeen BH, emphasizing how variations in the model parameters and the observer’s viewing angle modulate the resulting imaging features.

\subsection{Investigation of direct and secondary images }
Photons arriving from large distances follow trajectories determined by their impact parameter. As each photon approaches PFDM-Bardeen BH, its path is progressively bent by the gravitational field, and the total angular deviation accumulated by the time it reaches its point of closest approach $r_{\rm min}$ is encoded in the function \(\varphi_{1}(\bar{b})\). This curve serves as a reference: trajectories that accumulate less angular advance than \(\varphi_{1}(\bar{b})\) form the family we denote by \(\varphi_{2}(\bar{b})\), whereas those undergoing greater angular advance constitute the branch \(\varphi_{3}(\bar{b})\).
Accordingly, the quantities may be defined as 

\begin{equation}
\label{27}
\left\{
\begin{aligned}
    \varphi_1(\bar{b}) &= \int_0^{u_{\rm min }} \frac{1}{\sqrt{G(u)}} d u, \\
    \varphi_2(\bar{b}) &=  \int_0^{u_r} \frac{1}{\sqrt{G(u)}} d u, \\
    \varphi_3(\bar{b}) &= 2 \int_0^{u_{\rm min }} \frac{1}{\sqrt{G(u)}} d u-\int_0^{u_r} \frac{1}{\sqrt{G(u)}} d u.
\end{aligned}
\right.
\end{equation}

\begin{figure}[htbp]
\centering
\begin{minipage}{0.5\textwidth}
\centering
\includegraphics[scale=1,angle=0]{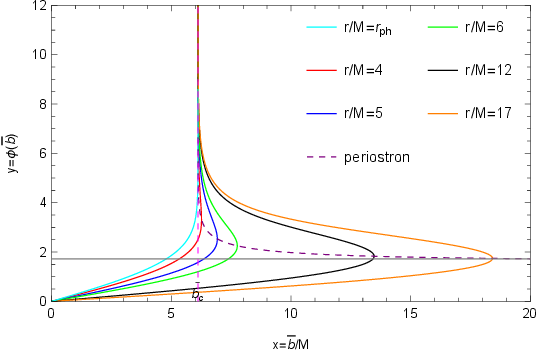}
\end{minipage}
\caption{\label{fig.10}{The deflection \(\varphi(b)\) is shown for various orbital radii \(r\), illustrating how the bending angle associated with each intersection point evolves as a function of the impact parameter \(\bar{b}\).}}
\end{figure}

\begin{figure}[htbp]
\centering

\begin{minipage}{0.3\textwidth}
    \centering
    \includegraphics[scale=0.5]{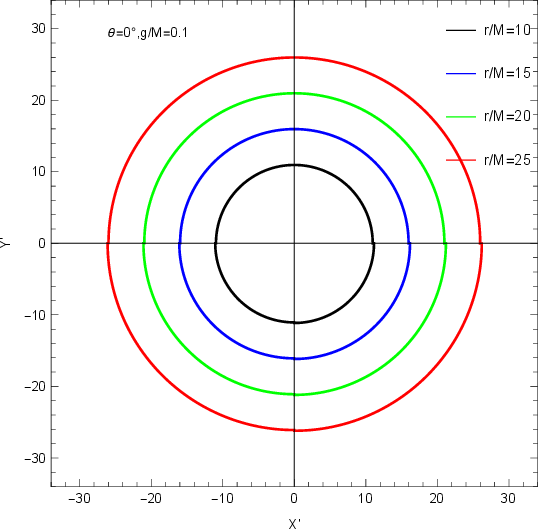}
    \end{minipage}
\hspace{0.1cm}
\begin{minipage}{0.3\textwidth}
    \centering
    \includegraphics[scale=0.5]{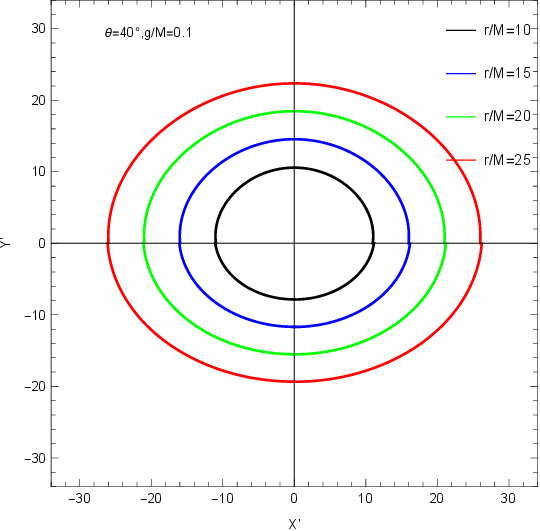}
   \end{minipage}
\hspace{0.1cm}
\begin{minipage}{0.3\textwidth}
    \centering
    \includegraphics[scale=0.5]{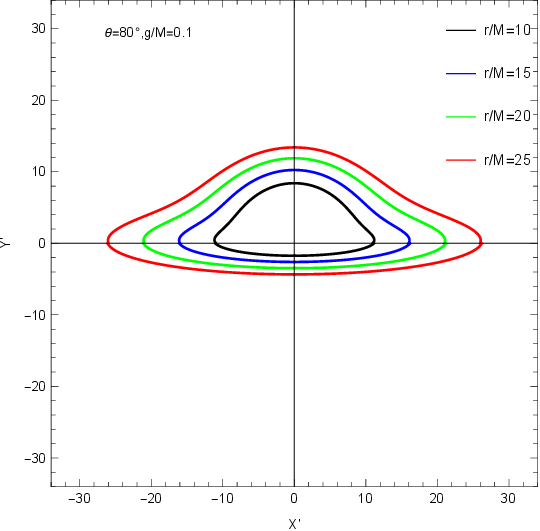}
   \end{minipage}

\medskip

\begin{minipage}{0.3\textwidth}
    \centering
    \includegraphics[scale=0.5]{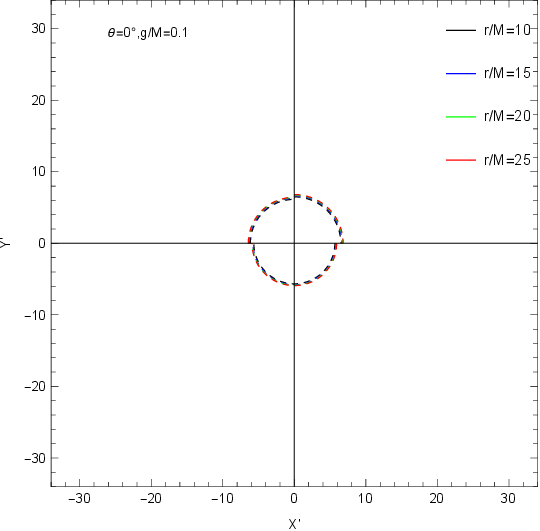}
  \end{minipage}
\hspace{0.1cm}
\begin{minipage}{0.3\textwidth}
    \centering
    \includegraphics[scale=0.5]{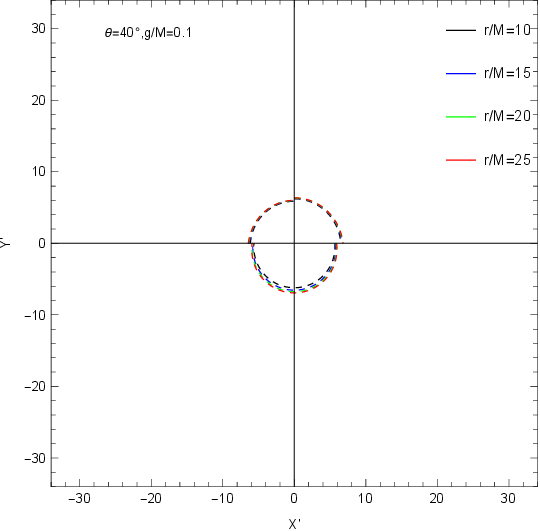}
  \end{minipage}
\hspace{0.1cm}
\begin{minipage}{0.3\textwidth}
    \centering
    \includegraphics[scale=0.5]{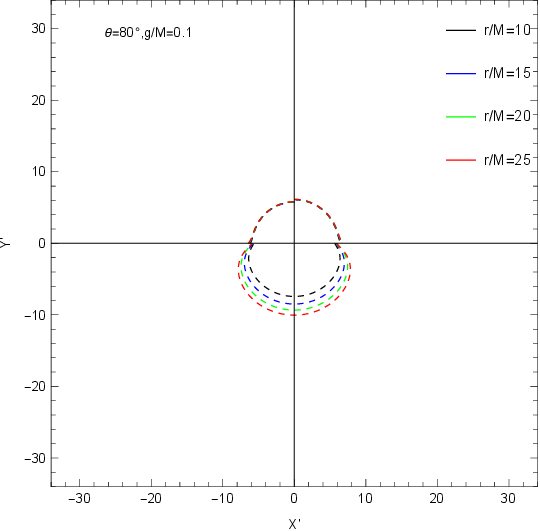}
   \end{minipage}

\medskip

\begin{minipage}{0.3\textwidth}
    \centering
    \includegraphics[scale=0.5]{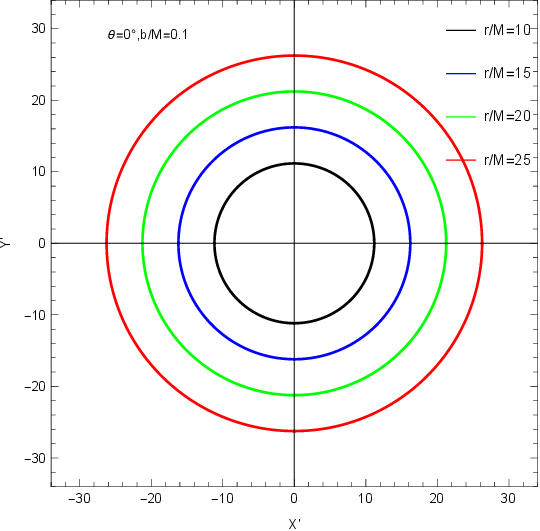}
  \end{minipage}
\hspace{0.1cm}
\begin{minipage}{0.3\textwidth}
    \centering
    \includegraphics[scale=0.5]{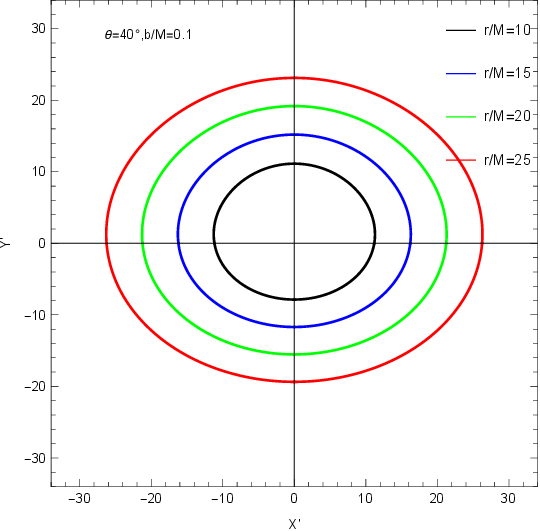}
  \end{minipage}
\hspace{0.1cm}
\begin{minipage}{0.3\textwidth}
    \centering
    \includegraphics[scale=0.5]{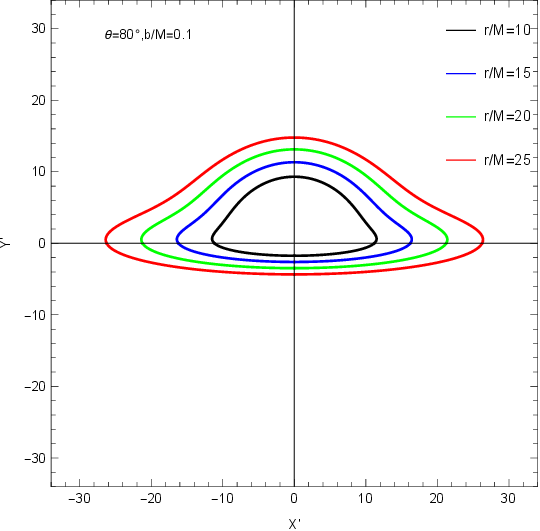}
   \end{minipage}

\medskip

\begin{minipage}{0.3\textwidth}
    \centering
    \includegraphics[scale=0.5]{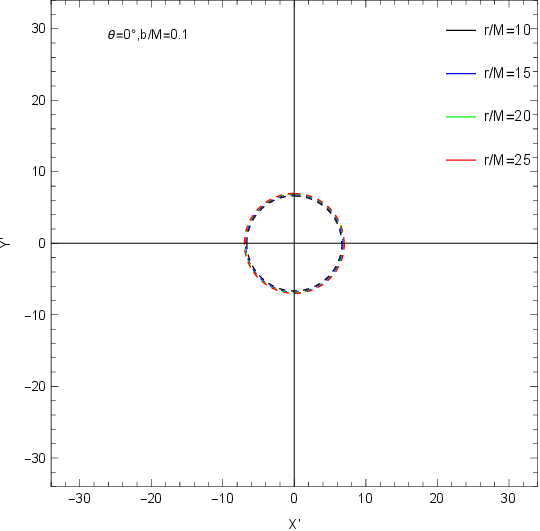}
  \end{minipage}
\hspace{0.1cm}
\begin{minipage}{0.3\textwidth}
    \centering
    \includegraphics[scale=0.5]{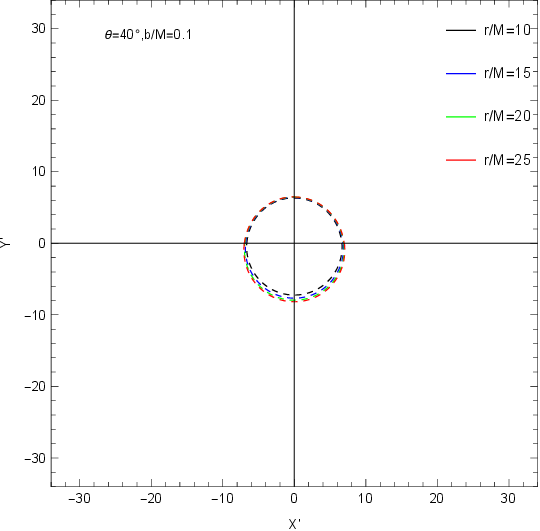}
  \end{minipage}
\hspace{0.1cm}
\begin{minipage}{0.3\textwidth}
    \centering
    \includegraphics[scale=0.5]{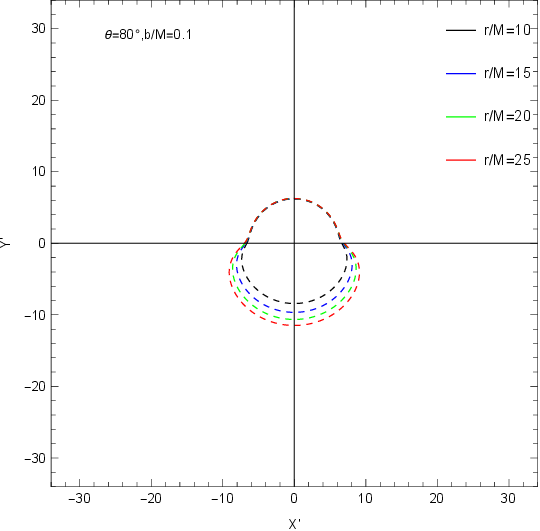}
   \end{minipage}
   \caption{\label{fig.11}{The top two rows display the primary and secondary images for fixed \(g/M = 0.1\) at viewing angles \(\theta = 0^\circ,40^\circ,80^\circ\). In each panel, the cases \(b/M = 0,0.04,0.08\) are combined into a single composite image for direct comparison. The bottom two rows show the corresponding primary and secondary images obtained by fixing \(b/M = 0.1\) and varying the viewing angle over the same set \((0^\circ, 40^\circ, 80^\circ)\). Here, the configurations with \(g/M = 0, 0.15, 0.3\) are merged into each panel, enabling a unified assessment of how the magnetic charge influences the imaging structure.}}
\end{figure}

Using Eq. \eqref{27}, we numerically integrate the photon trajectories and obtain the dependence of \(\varphi(\bar{b})\) on the impact parameter \(\bar{b}\) for the PFDM–Bardeen BH, evaluated at fixed radius \(r\) with parameters \(g/M = b/M = 0.1\). As illustrated in Fig. \ref{fig.10}, the pink dashed curve represents the constant-\(r\) contour associated with the critical impact parameter \(\bar{b}_{c}\), while the purple dashed curve shows the dependence of \(\varphi\) on \(\bar{b}\) evaluated at the photon’s point of closest approach to the BH. The contour curves become increasingly convex as the radius \(r\) grows, and their intersections with the purple curve identify the corresponding values of \(\varphi\) at these turning points. Using this family of contours curves, we reconstruct the primary and secondary images of the accretion disk, as presented in Fig. \ref{fig.11}.

The first two rows of Fig. \ref{fig.11} display the primary and secondary images of the accretion disk for a fixed magnetic charge \(g/M = 0.1\), viewed at inclination angles \(\theta = 0^\circ\), \(40^\circ\), and \(80^\circ\). The primary images are significantly sharper and more distinct than the secondary ones, and as the viewing angle increases, the contours gradually evolve into the characteristic appearance of thin accretion disk. In each panel, the cases \(b/M = 0\), \(0.04\), and \(0.08\) are juxtaposed, revealing that larger values of \(b\) systematically enlarge the resulting image contours. This dependence on the DM parameter becomes even more pronounced in the secondary images, where its influence is markedly stronger.

The third and fourth rows present the evolution of the image contours as the viewing inclination varies, with the DM parameter fixed at \(b/M = 0.01\). Each panel combines the three magnetic charge configurations \(g/M = 0\), \(0.15\), and \(0.3\). The contour morphology changes appreciably with the observer’s angle; however, the influence of the magnetic charge is negligible—its impact on both the primary and secondary images is practically indistinguishable. This is reasonable because, on the distance scales relevant to the appearance of the BH accretion disk, the contribution of the magnetic charge is negligibly weak. This also implies that any observational discrimination of a PFDM–Bardeen BH must primarily rely on the value of the DM parameter \(b\). The presence of DM produces a clear enlargement of the accretion disk image.

\subsection{Radiative intensity profile of accretion Disk}
In this subsection, we analyze the radial distribution of the observed radiative flux emitted by the accretion disk within the Novikov–Thorne framework. The model is based on three essential assumptions (1): The accretion flow is a geometrically thin, optically thick disk in a steady state with a constant mass accretion rate. The gas follows nearly Keplerian circular orbits between the \( r_{\text{isco}} \) and outer radius \( r_{\text{out}} \), with the disk lying on the equatorial plane and aligned with the BH spin axis. The self-gravity of the disk is negligible. (2): The dissipated energy is efficiently converted into radiation, allowing the disk to emit locally as a blackbody in thermal equilibrium. The emergent flux is perpendicular to the disk surface, and the continuum spectrum is dominated by thermal emission.
(3): The underlying geometry is stationary, axisymmetric, asymptotically flat, and symmetric with respect to the equatorial plane, ensuring that the particle motion and emitted radiation are governed solely by the background metric.

The radiative energy flux $F(r)$ can be defined as
\be
\label{28}
F(r)=-\frac{\dot{M}}{4 \pi \sqrt{-g}} \frac{\Omega_{, r}}{(\tilde{E}-\Omega \tilde{L})^2} \int_{r_{\rm isco}}^r(\tilde{E}-\Omega \tilde{L}) \tilde{L}_{, r} \dif r.
\ee
This expression is commonly adopted in the literature and is applicable specifically within cylindrical coordinates. When rewritten in spherical coordinates and with physical dimensions reinstated, it takes the following form \cite{Collodel:2021gxu}
\be
\label{29}
F(r)=-\frac{c^2\dot{M}}{4 \pi \sqrt{-g/g_{\th\th}}} \frac{\Omega_{, r}}{(\tilde{E}-\Omega \tilde{L})^2} \int_{r_{\rm isco}}^r(\tilde{E}-\Omega \tilde{L}) \tilde{L}_{, r} \dif r,
\ee
where \(c\) denotes the speed of light. Subsequently, we consider accretion onto BH of mass \(M = 2 \times 10^{6}~M_{\odot}\), assuming the accretion rate \(\dot{M} = 2 \times 10^{-6}~M_{\odot}\mathrm{yr^{-1}}\). In addition, \(\tilde{E}\), \(\tilde{L}\), and \(\Omega\) denote the specific energy, specific angular momentum, and angular velocity of test particle, respectively. Their explicit definitions can be found in our previous work \cite{Feng:2022bst,Feng:2024iqj,Wu:2024sng,Cai:2025rst,Cai:2025pan}.

Owing to the distinct gravitational environments at the disk and the observer, together with their relative motion, the emitted photons undergo a corresponding frequency shift. Consequently, the radiation flux measured by the observer is given by \cite{Luminet:1979nyg,Huang:2023ilm}
\be
\label{30}
F_{\rm o b s}=\frac{F(r)}{(1+z)^4},
\ee
where $z$ denotes the redshift accumulated by photons as they propagate from the disk to the distant observer. The corresponding factor \(1+z\) incorporates the gravitational redshift produced by the spacetime metric. Its explicit form is \cite{You:2024uql}
\be
\label{31}
1+z=\frac{1+\Omega \bar{b} \sin \theta \cos \alpha}{\sqrt{-g_{t t}-\Omega^2 g_{\phi \phi}}}.
\ee

\begin{figure}[H]
\centering
\begin{minipage}{0.53\textwidth}
\centering
\includegraphics[scale=0.97,angle=0]{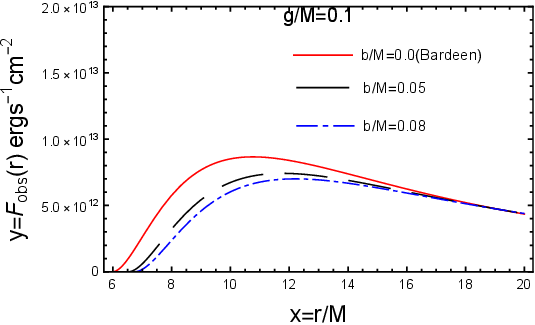}
\end{minipage}%
\begin{minipage}{0.53\textwidth}
\centering
\includegraphics[scale=0.9,angle=0]{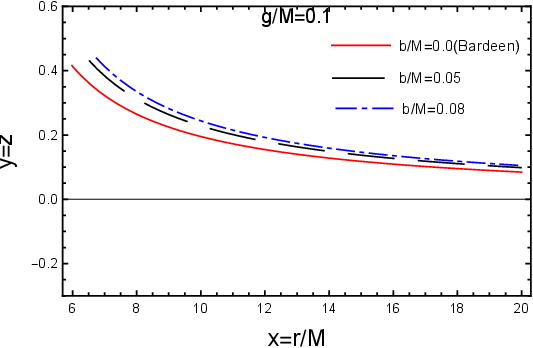}
\end{minipage}%
\caption{\label{fig.12}{The figure presents the observed flux \(F_{\mathrm{obs}}\) and the redshift \(z\) as functions of \(r\), with the observation angle fixed at \(\theta = 0^{\circ}\).}} 
\end{figure}

Figure \ref{fig.12} quantitatively presents the observed flux \(F_{\mathrm{obs}}\) and the redshift \(z\) as functions of \(r\).  Moreover, we provide the peak flux and the corresponding radius: for \(b/M = 0\), \(F_{\mathrm{max,obs}} = 8.66481 \times 10^{12}\) \(\rm erg  s^{-1} cm^{-2}\) at a radius of \(r/M = 10.7633\); for \(b/M = 0.05\), \(F_{\mathrm{max,obs}} = 7.42073 \times 10^{12}\)\(\rm erg  s^{-1} cm^{-2}\) at \(r/M = 11.7795\); and for \(b/M = 0.08\), \(F_{\mathrm{max,obs}} = 7.00601 \times 10^{12}\)\(\rm erg  s^{-1} cm^{-2}\)  at \(r/M = 12.2032\). The consistent conclusion is that increasing the DM parameter \(b\) indeed reduces the observed value of \(F_{\mathrm{obs}}\). This phenomenon may be explained by the factor. As $b$ increases, the velocity and density of the accreting matter change, thereby altering the energy required for photons to escape and consequently affecting the flux distribution received by a distant observer.

\begin{figure}[H]
	\centering
	\begin{tabular}{ccc}
		\begin{minipage}[t]{0.3\textwidth}
			\centering
			\begin{overpic}[width=0.75\textwidth]{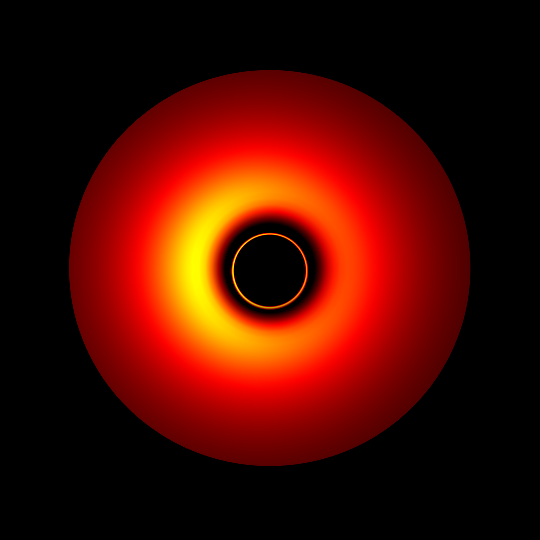} 
				\put(1,103){\color{black}\large $b/M=0, \theta=10^{\circ}$} 
				\put(-12,48){\color{black} Y'}
				\put(48,-10){\color{black} X'}
			\end{overpic}
		\end{minipage}
		&
		\begin{minipage}[t]{0.3\textwidth}
			\centering
			\begin{overpic}[width=0.75\textwidth]{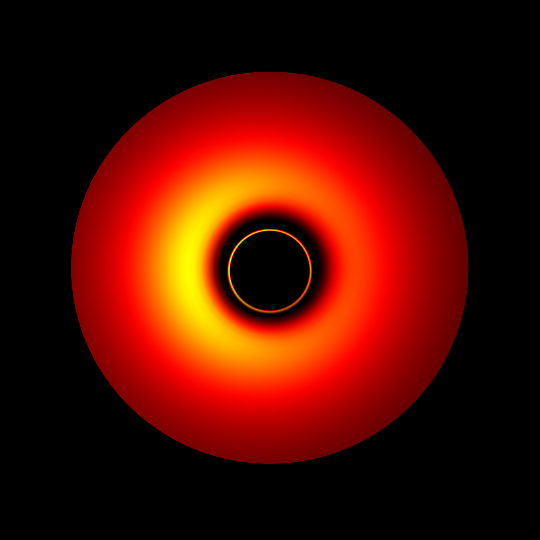} 
				\put(1,103){\color{black}\large $b/M=0.05, \theta=10^{\circ}$} 
				\put(-12,48){\color{black} Y'}
				\put(48,-10){\color{black} X'}
			\end{overpic}
		\end{minipage}
		&
		\begin{minipage}[t]{0.3\textwidth}
			\centering
			\begin{overpic}[width=0.75\textwidth]{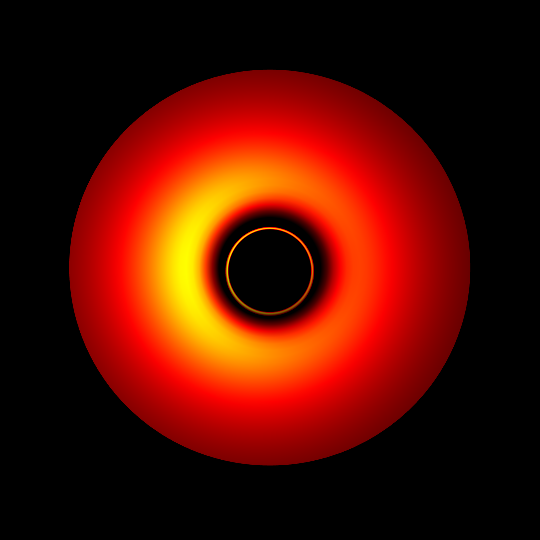}
				\put(1,103){\color{black}\large $b/M=0.08, \theta=10^{\circ}$} 
				\put(-12,48){\color{black} Y'}
				\put(48,-10){\color{black} X'}
			\end{overpic}
		\end{minipage}
		\vspace{40pt} 
		\\ 
		\begin{minipage}[t]{0.3\textwidth}
			\centering
			\begin{overpic}[width=0.75\textwidth]{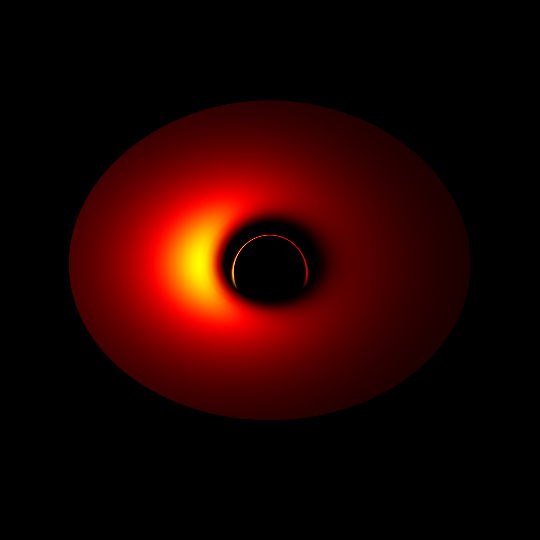} 
				\put(1,103){\color{black}\large $b/M=0, \theta=40^{\circ}$} 
				\put(-12,48){\color{black} Y'}
				\put(48,-10){\color{black} X'}
			\end{overpic}
		\end{minipage}
		&
		\begin{minipage}[t]{0.3\textwidth}
			\centering
			\begin{overpic}[width=0.75\textwidth]{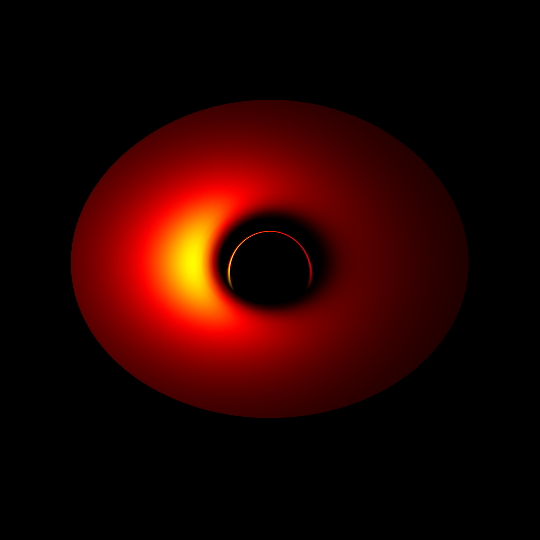} 
				\put(1,103){\color{black}\large $b/M=0.05, \theta=40^{\circ}$} 
				\put(-12,48){\color{black} Y'}
				\put(48,-10){\color{black} X'}
			\end{overpic}
		\end{minipage}
		&
		\begin{minipage}[t]{0.3\textwidth}
			\centering
			\begin{overpic}[width=0.75\textwidth]{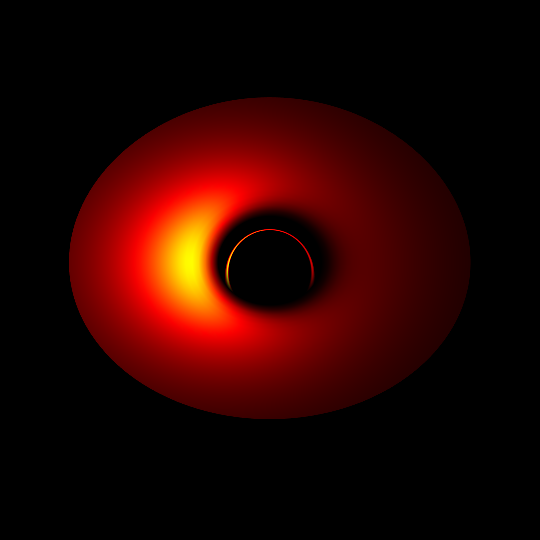} 
				\put(1,103){\color{black}\large $b/M=0.08, \theta=40^{\circ}$} 
				\put(-12,48){\color{black} Y'}
				\put(48,-10){\color{black} X'}
			\end{overpic}
		\end{minipage}
		\vspace{40pt} 
		\\
		\begin{minipage}[t]{0.3\textwidth}
			\centering
			\begin{overpic}[width=0.75\textwidth]{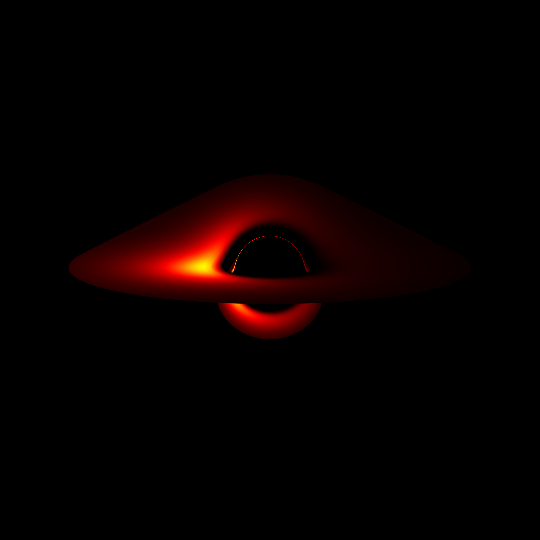}
				\put(1,103){\color{black}\large $b/M=0,\theta=80^{\circ}$}
				\put(-12,48){\color{black} Y'}
				\put(48,-10){\color{black} X'}
			\end{overpic}
		\end{minipage}
		&
		\begin{minipage}[t]{0.3\textwidth}
			\centering
			\begin{overpic}[width=0.75\textwidth]{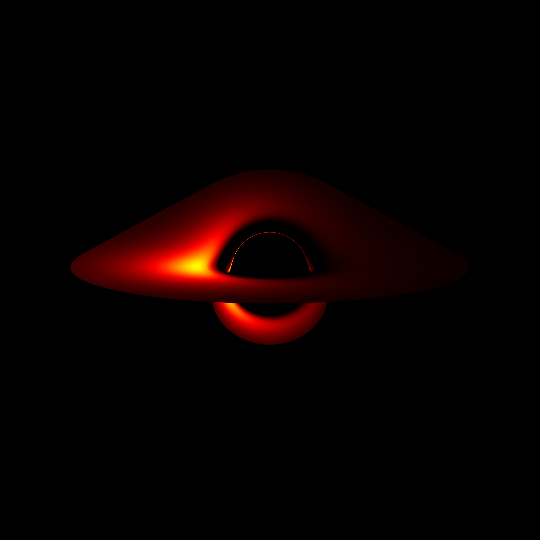}
				\put(1,103){\color{black}\large $b/M=0.05, \theta=80^{\circ}$} 
				\put(-12,48){\color{black} Y'}
				\put(48,-10){\color{black} X'}
			\end{overpic}
		\end{minipage}
		&
		\begin{minipage}[t]{0.3\textwidth}
			\centering
			\begin{overpic}[width=0.75\textwidth]{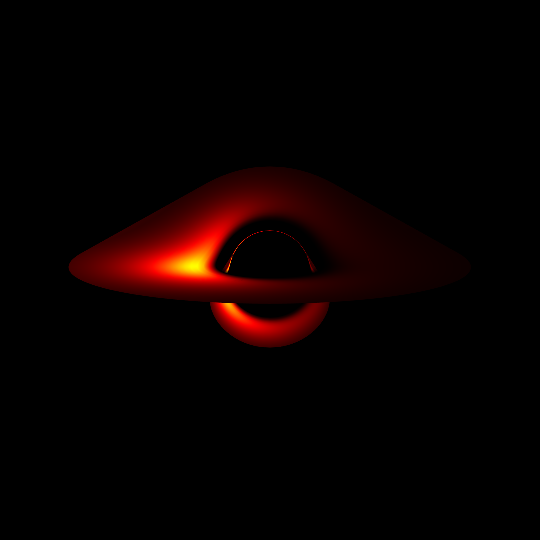}
				\put(1,103){\color{black}\large $b/M=0.08, \theta=80^{\circ}$}
				\put(-12,48){\color{black} Y'}
				\put(48,-10){\color{black} X'}
			\end{overpic}
		\end{minipage}
     \end{tabular}
	\caption{\label{fig.13} This figure presents the distribution of \(F_{\text{obs}}\), with the primary and secondary images combined. It illustrates the cases for fixed magnetic charge \(g/M = 0.1\), considering viewing angles of \(10^\circ\), \(40^\circ\), and \(80^\circ\), and comparing the DM parameters \(b/M = 0\), \(0.05\), and \(0.08\). }
	\end{figure}

\begin{figure}[H]
	\centering
	\begin{tabular}{ccc}
		\begin{minipage}[t]{0.3\textwidth}
			\centering
			\begin{overpic}[width=0.75\textwidth]{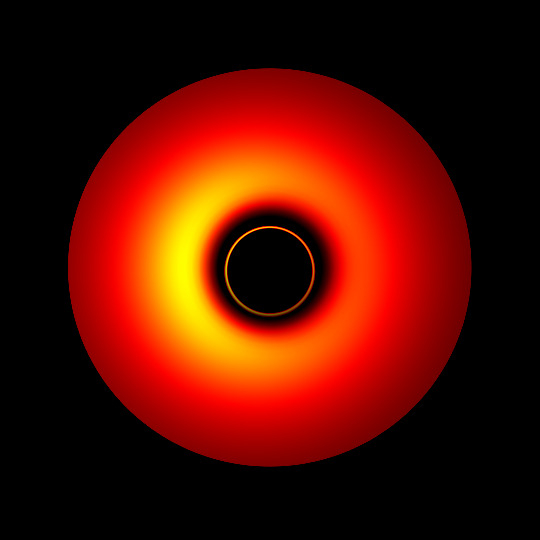} 
				\put(1,103){\color{black}\large $g/M=0, \theta=10^{\circ}$} 
				\put(-12,48){\color{black} Y'}
				\put(48,-10){\color{black} X'}
			\end{overpic}
		\end{minipage}
		&
		\begin{minipage}[t]{0.3\textwidth}
			\centering
			\begin{overpic}[width=0.75\textwidth]{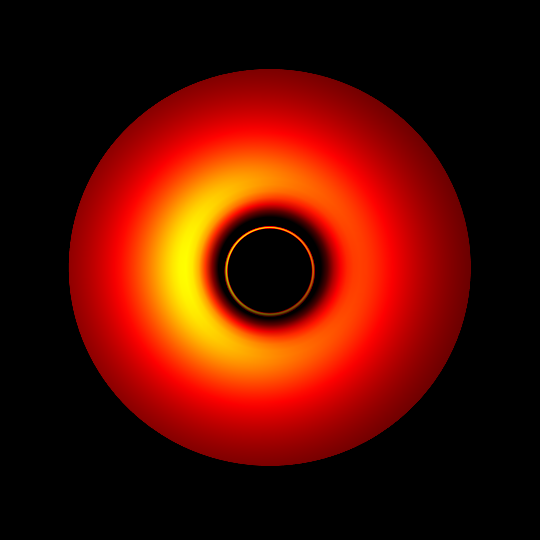} 
				\put(1,103){\color{black}\large $g/M=0.3, \theta=10^{\circ}$} 
				\put(-12,48){\color{black} Y'}
				\put(48,-10){\color{black} X'}
			\end{overpic}
		\end{minipage}
		&
		\begin{minipage}[t]{0.3\textwidth}
			\centering
			\begin{overpic}[width=0.75\textwidth]{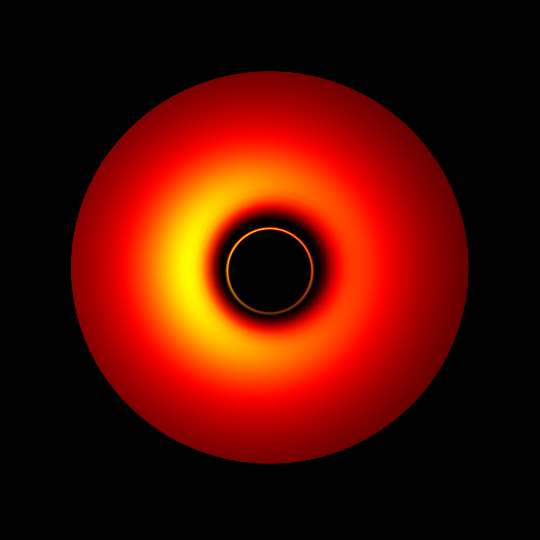}
				\put(1,103){\color{black}\large $g/M=0.6, \theta=10^{\circ}$} 
				\put(-12,48){\color{black} Y'}
				\put(48,-10){\color{black} X'}
			\end{overpic}
		\end{minipage}
		\vspace{40pt} 
		\\ 
		\begin{minipage}[t]{0.3\textwidth}
			\centering
			\begin{overpic}[width=0.75\textwidth]{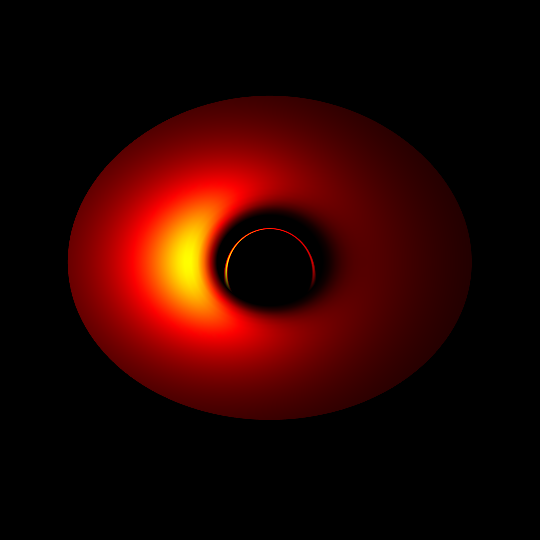} 
				\put(1,103){\color{black}\large $g/M=0, \theta=40^{\circ}$} 
				\put(-12,48){\color{black} Y'}
				\put(48,-10){\color{black} X'}
			\end{overpic}
		\end{minipage}
		&
		\begin{minipage}[t]{0.3\textwidth}
			\centering
			\begin{overpic}[width=0.75\textwidth]{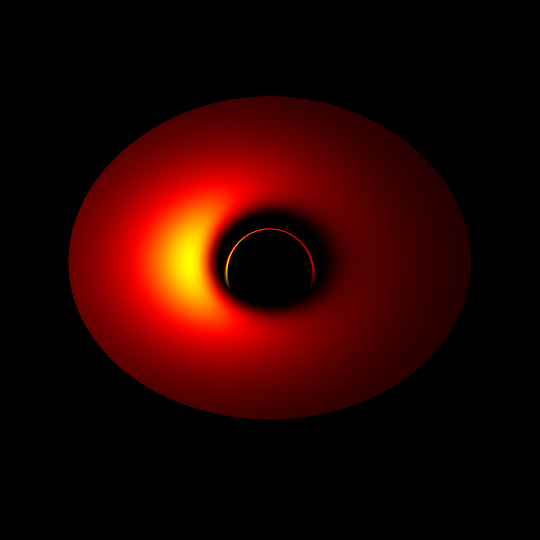} 
				\put(1,103){\color{black}\large $g/M=0.3, \theta=40^{\circ}$} 
				\put(-12,48){\color{black} Y'}
				\put(48,-10){\color{black} X'}
			\end{overpic}
		\end{minipage}
		&
		\begin{minipage}[t]{0.3\textwidth}
			\centering
			\begin{overpic}[width=0.75\textwidth]{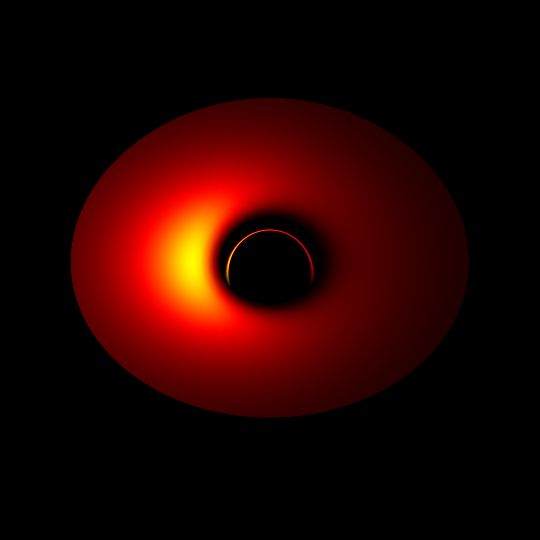} 
				\put(1,103){\color{black}\large $g/M=0.6, \theta=40^{\circ}$} 
				\put(-12,48){\color{black} Y'}
				\put(48,-10){\color{black} X'}
			\end{overpic}
		\end{minipage}
		\vspace{40pt} 
		\\
		\begin{minipage}[t]{0.3\textwidth}
			\centering
			\begin{overpic}[width=0.75\textwidth]{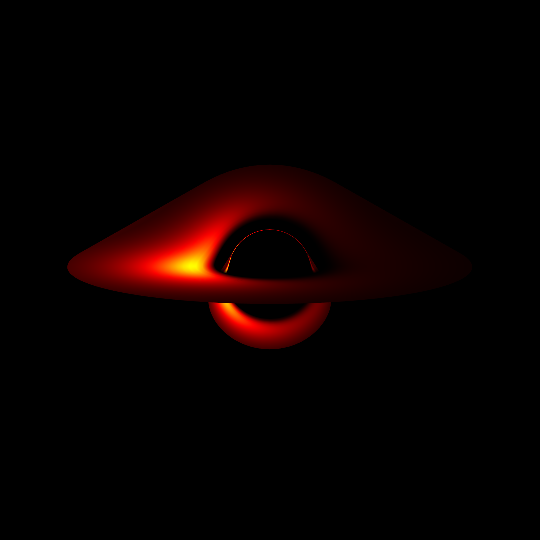}
				\put(1,103){\color{black}\large $g/M=0,\theta=80^{\circ}$}
				\put(-12,48){\color{black} Y'}
				\put(48,-10){\color{black} X'}
			\end{overpic}
		\end{minipage}
		&
		\begin{minipage}[t]{0.3\textwidth}
			\centering
			\begin{overpic}[width=0.75\textwidth]{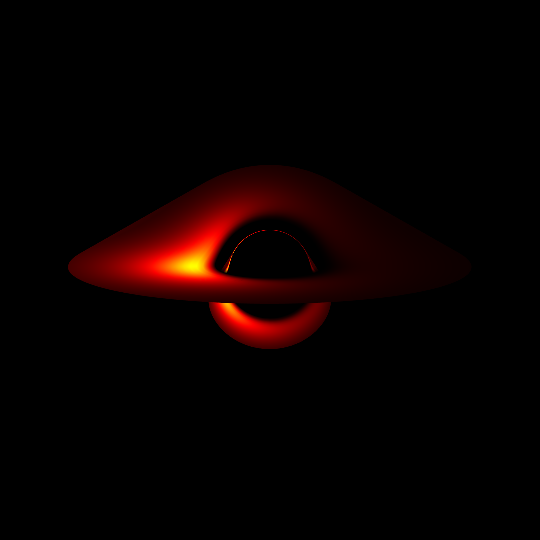}
				\put(1,103){\color{black}\large $g/M=0.3, \theta=80^{\circ}$} 
				\put(-12,48){\color{black} Y'}
				\put(48,-10){\color{black} X'}
			\end{overpic}
		\end{minipage}
		&
		\begin{minipage}[t]{0.3\textwidth}
			\centering
			\begin{overpic}[width=0.75\textwidth]{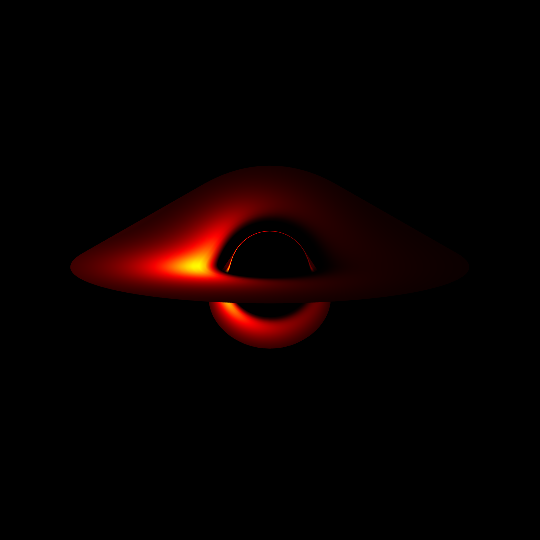}
				\put(1,103){\color{black}\large $g/M=0.6, \theta=80^{\circ}$}
				\put(-12,48){\color{black} Y'}
				\put(48,-10){\color{black} X'}
			\end{overpic}
		\end{minipage}
     \end{tabular}
	\caption{\label{fig.14} The figure shows the cases with fixed DM parameter \(b/M = 0.1\), for viewing angles of \(10^\circ\), \(40^\circ\), and \(80^\circ\), considering magnetic charges \(g/M = 0\), \(0.3\), and \(0.6\).}
	\end{figure}

\begin{figure}[H]
	\centering
	\begin{tabular}{ccc}
		\begin{minipage}[t]{0.3\textwidth}
			\centering
			\begin{overpic}[width=0.75\textwidth]{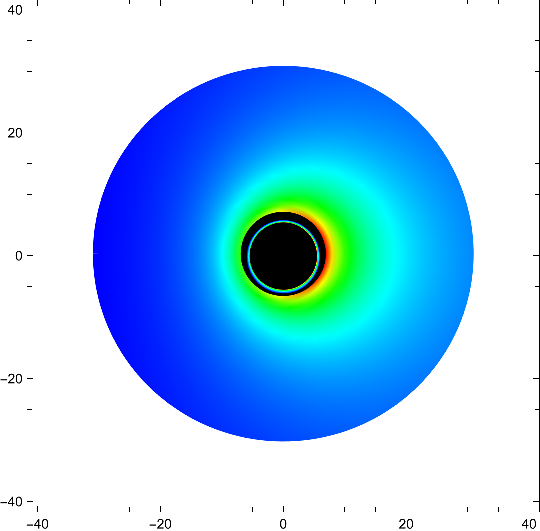}
				\put(1,100){\color{black}\large $b/M=0, \theta=10^{\circ}$}
				\put(-8,48){\color{black} Y'}
				\put(48,-10){\color{black} X'}
			\end{overpic}
			\raisebox{0.5\height}{ 
				\begin{overpic}[width=0.07\textwidth]{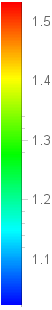}
				\end{overpic}
			}
		\end{minipage}
		&
		\begin{minipage}[t]{0.3\textwidth}
			\centering
			\begin{overpic}[width=0.75\textwidth]{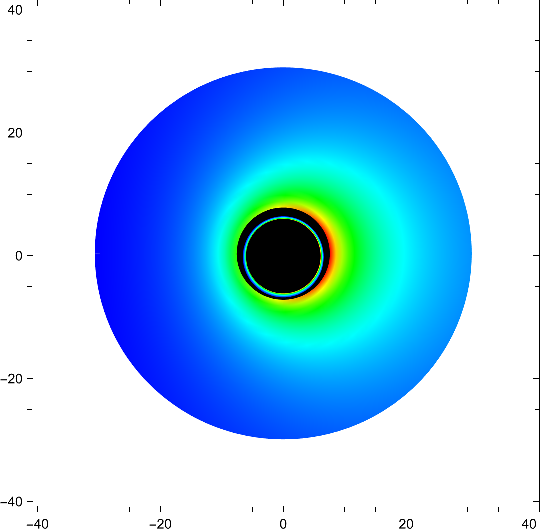}
				\put(1,100){\color{black}\large $b/M=0.05, \theta=10^{\circ}$} 
				\put(-8,48){\color{black} Y'}
				\put(48,-10){\color{black} X'}
			\end{overpic}
			\raisebox{0.5\height}{ 
				\begin{overpic}[width=0.07\textwidth]{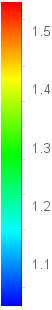} 
				\end{overpic}
			}			
		\end{minipage}
		&
		\begin{minipage}[t]{0.3\textwidth}
			\centering
			\begin{overpic}[width=0.75\textwidth]{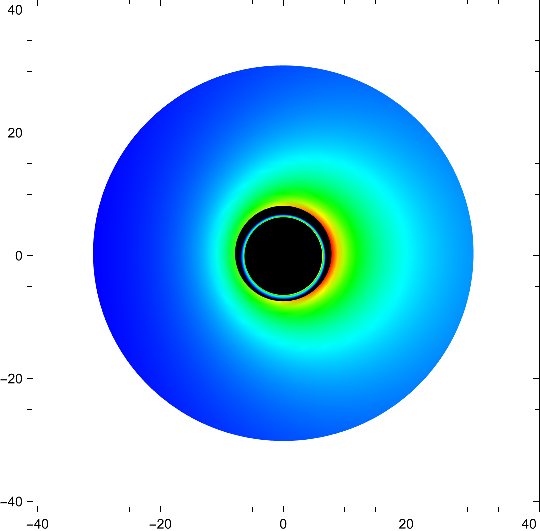}
				\put(1,100){\color{black}\large $b/M=0.08, \theta=10^{\circ}$} 
				\put(-8,48){\color{black} Y'}
				\put(48,-10){\color{black} X'}
			\end{overpic}
			\raisebox{0.5\height}{
				\begin{overpic}[width=0.07\textwidth]{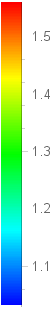}
				\end{overpic}
			}
		\end{minipage}
		\vspace{40pt} 
		\\ 
		\begin{minipage}[t]{0.3\textwidth}
			\centering
			\begin{overpic}[width=0.75\textwidth]{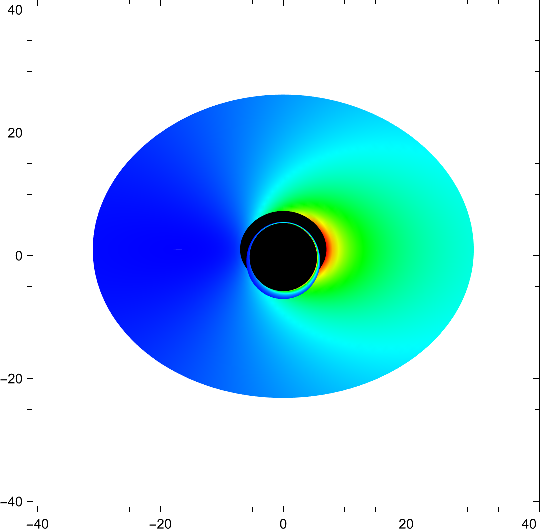}
				\put(1,100){\color{black}\large $b/M=0, \theta=40^{\circ}$} 
				\put(-8,48){\color{black} Y'}
				\put(48,-10){\color{black} X'}
			\end{overpic}
			\raisebox{0.5\height}{
				\begin{overpic}[width=0.07\textwidth]{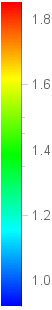} 
				\end{overpic}
			}
		\end{minipage}
		&
		\begin{minipage}[t]{0.3\textwidth}
			\centering
			\begin{overpic}[width=0.75\textwidth]{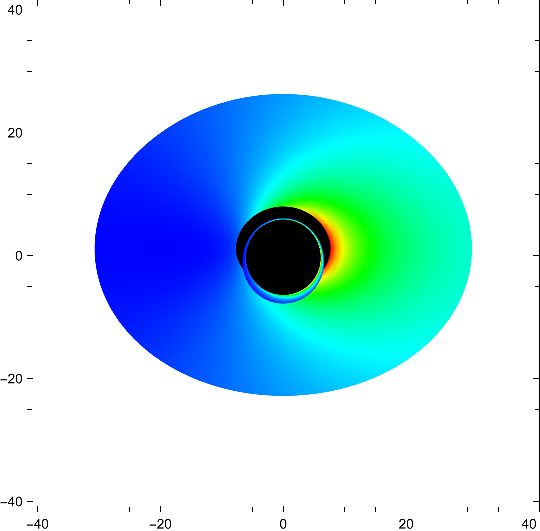} 
				\put(1,100){\color{black}\large $b/M=0.05, \theta=40^{\circ}$} 
				\put(-8,48){\color{black} Y'}
				\put(48,-10){\color{black} X'}
			\end{overpic}
			\raisebox{0.5\height}{ 
				\begin{overpic}[width=0.07\textwidth]{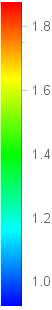} 
				\end{overpic}
			}
		\end{minipage}
		&
		\begin{minipage}[t]{0.3\textwidth}
			\centering
			\begin{overpic}[width=0.75\textwidth]{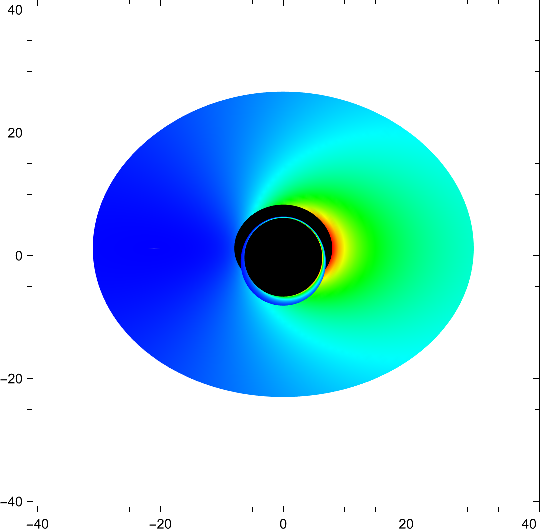 } 
				\put(1,100){\color{black}\large $b/M=0.08, \theta=40^{\circ}$} 
				\put(-8,48){\color{black} Y'}
				\put(48,-10){\color{black} X'}
			\end{overpic}
			\raisebox{0.5\height}{
				\begin{overpic}[width=0.07\textwidth]{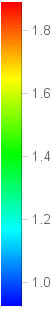}
				\end{overpic}
			}
		\end{minipage}
		\vspace{40pt} 
		\\ 
		\begin{minipage}[t]{0.3\textwidth}
			\centering
			\begin{overpic}[width=0.75\textwidth]{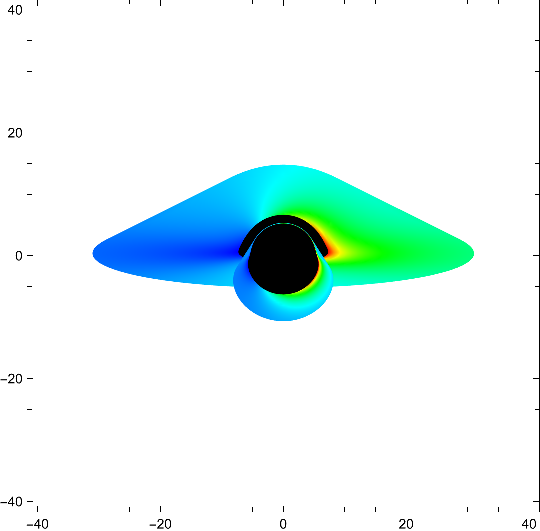} 
				\put(1,100){\color{black}\large $b/M=0, \theta=80^{\circ}$}
				\put(-8,48){\color{black} Y'}
				\put(48,-10){\color{black} X'}
			\end{overpic}
			\raisebox{0.5\height}{ 
				\begin{overpic}[width=0.07\textwidth]{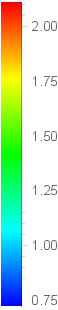}
				\end{overpic}
			}
		\end{minipage}
		&
		\begin{minipage}[t]{0.3\textwidth}
			\centering
			\begin{overpic}[width=0.75\textwidth]{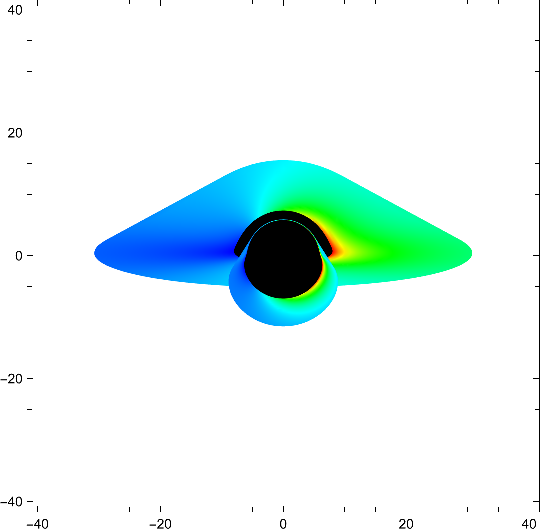} 
				\put(1,100){\color{black}\large $b/M=0.05, \theta=80^{\circ}$} 
				\put(-8,48){\color{black} Y'}
				\put(48,-10){\color{black} X'}
			\end{overpic}
			\raisebox{0.5\height}{ 
				\begin{overpic}[width=0.07\textwidth]{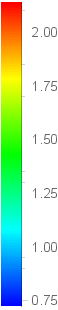} 
				\end{overpic}
			}
		\end{minipage}
		&
		\begin{minipage}[t]{0.3\textwidth}
			\centering
			\begin{overpic}[width=0.75\textwidth]{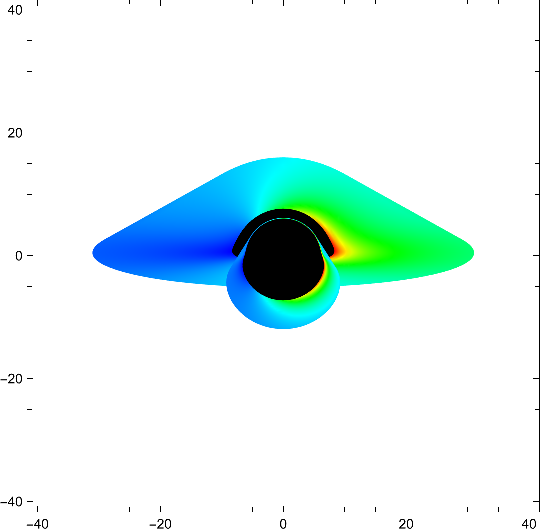}
				\put(1,100){\color{black}\large $b/M=0.08, \theta=80^{\circ}$}
				\put(-8,48){\color{black} Y'}
				\put(48,-10){\color{black} X'}
			\end{overpic}
			\raisebox{0.5\height}{ 
				\begin{overpic}[width=0.07\textwidth]{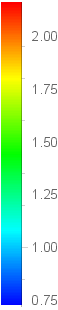} 
				\end{overpic}}
		\end{minipage}
	\end{tabular}
	\caption{\label{fig.15} This figure illustrates the distribution of the redshift \(z\) across the accretion disk for different viewing angles. The magnitude of the redshift is indicated by the color bar on the right. As before, we fix \(g/M = 0.1\) and examine how variations in the DM parameter \(b\) modify the disk appearance under different observational inclinations.}
	\end{figure}

\begin{figure}[H]
	\centering
	\begin{tabular}{ccc}
		\begin{minipage}[t]{0.3\textwidth}
			\centering
			\begin{overpic}[width=0.75\textwidth]{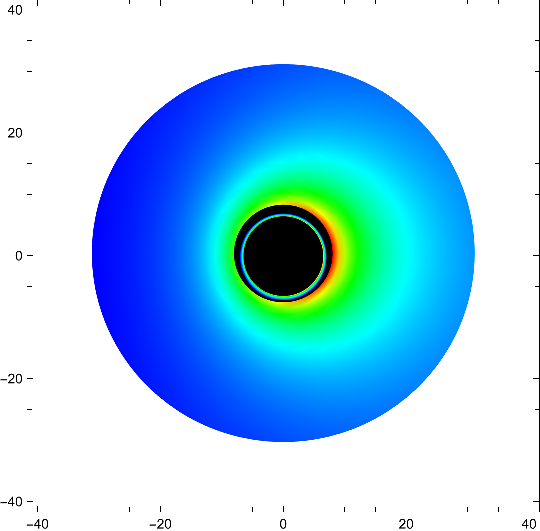}
				\put(1,100){\color{black}\large $g/M=0, \theta=10^{\circ}$}
				\put(-8,48){\color{black} Y'}
				\put(48,-10){\color{black} X'}
			\end{overpic}
			\raisebox{0.5\height}{ 
				\begin{overpic}[width=0.07\textwidth]{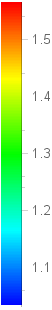}
				\end{overpic}
			}
		\end{minipage}
		&
		\begin{minipage}[t]{0.3\textwidth}
			\centering
			\begin{overpic}[width=0.75\textwidth]{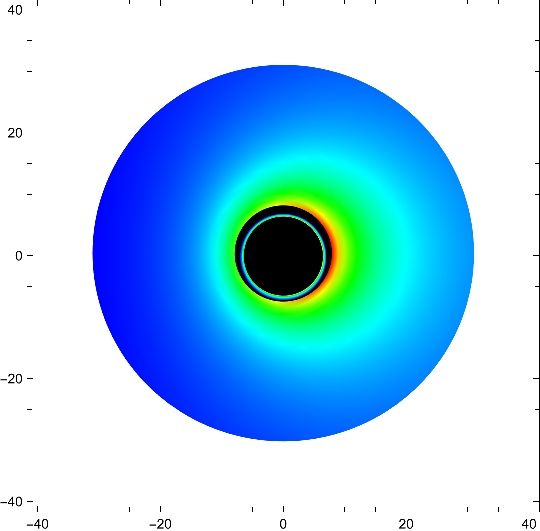}
				\put(1,100){\color{black}\large $g/M=0.3, \theta=10^{\circ}$} 
				\put(-8,48){\color{black} Y'}
				\put(48,-10){\color{black} X'}
			\end{overpic}
			\raisebox{0.5\height}{ 
				\begin{overpic}[width=0.07\textwidth]{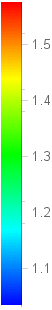} 
				\end{overpic}
			}			
		\end{minipage}
		&
		\begin{minipage}[t]{0.3\textwidth}
			\centering
			\begin{overpic}[width=0.75\textwidth]{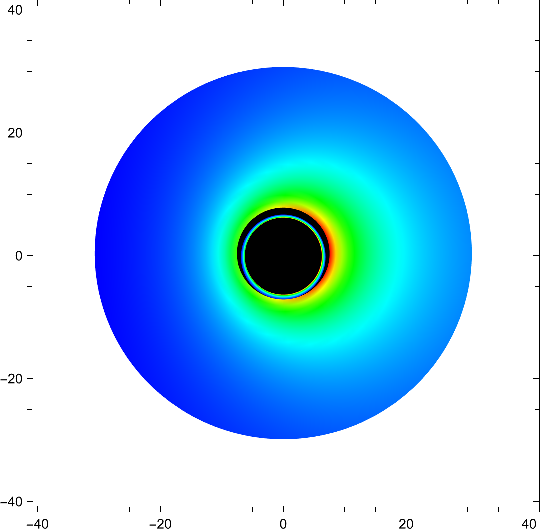}
				\put(1,100){\color{black}\large $g/M=0.6,\theta=10^{\circ}$} 
				\put(-8,48){\color{black} Y'}
				\put(48,-10){\color{black} X'}
			\end{overpic}
			\raisebox{0.5\height}{
				\begin{overpic}[width=0.07\textwidth]{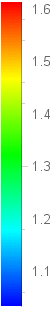}
				\end{overpic}
			}
		\end{minipage}
		\vspace{40pt} 
		\\ 
		\begin{minipage}[t]{0.3\textwidth}
			\centering
			\begin{overpic}[width=0.75\textwidth]{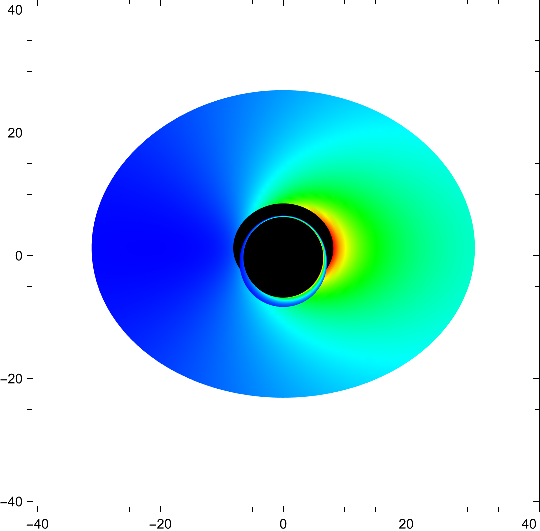}
				\put(1,100){\color{black}\large $g/M=0, \theta=40^{\circ}$} 
				\put(-8,48){\color{black} Y'}
				\put(48,-10){\color{black} X'}
			\end{overpic}
			\raisebox{0.5\height}{
				\begin{overpic}[width=0.07\textwidth]{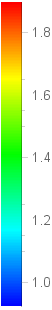     } 
				\end{overpic}
			}
		\end{minipage}
		&
		\begin{minipage}[t]{0.3\textwidth}
			\centering
			\begin{overpic}[width=0.75\textwidth]{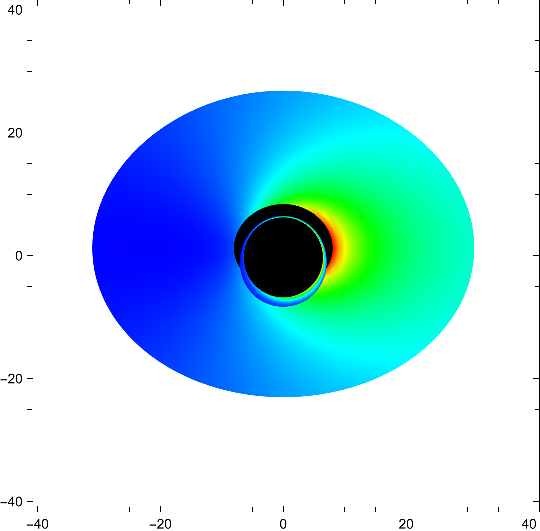} 
				\put(1,100){\color{black}\large $g/M=0.3, \theta=40^{\circ}$} 
				\put(-8,48){\color{black} Y'}
				\put(48,-10){\color{black} X'}
			\end{overpic}
			\raisebox{0.5\height}{ 
				\begin{overpic}[width=0.07\textwidth]{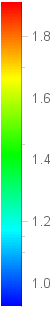} 
				\end{overpic}
			}
		\end{minipage}
		&
		\begin{minipage}[t]{0.3\textwidth}
			\centering
			\begin{overpic}[width=0.75\textwidth]{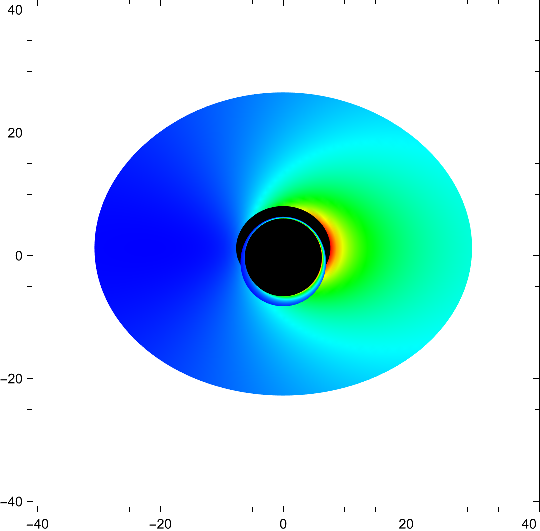} 
				\put(1,100){\color{black}\large $g/M=0.6, \theta=40^{\circ}$} 
				\put(-8,48){\color{black} Y'}
				\put(48,-10){\color{black} X'}
			\end{overpic}
			\raisebox{0.5\height}{
				\begin{overpic}[width=0.07\textwidth]{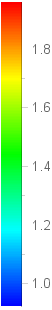}
				\end{overpic}
			}
		\end{minipage}
		\vspace{40pt} 
		\\ 
		\begin{minipage}[t]{0.3\textwidth}
			\centering
			\begin{overpic}[width=0.75\textwidth]{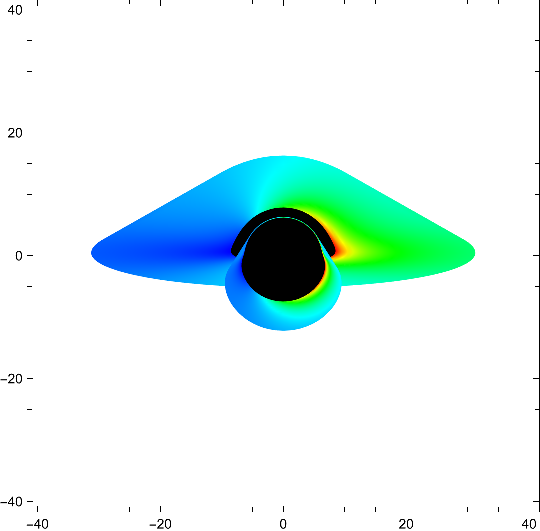} 
				\put(1,100){\color{black}\large $g/M=0, \theta=80^{\circ}$}
				\put(-8,48){\color{black} Y'}
				\put(48,-10){\color{black} X'}
			\end{overpic}
			\raisebox{0.5\height}{ 
				\begin{overpic}[width=0.07\textwidth]{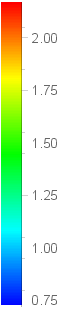}
				\end{overpic}
			}
		\end{minipage}
		&
		\begin{minipage}[t]{0.3\textwidth}
			\centering
			\begin{overpic}[width=0.75\textwidth]{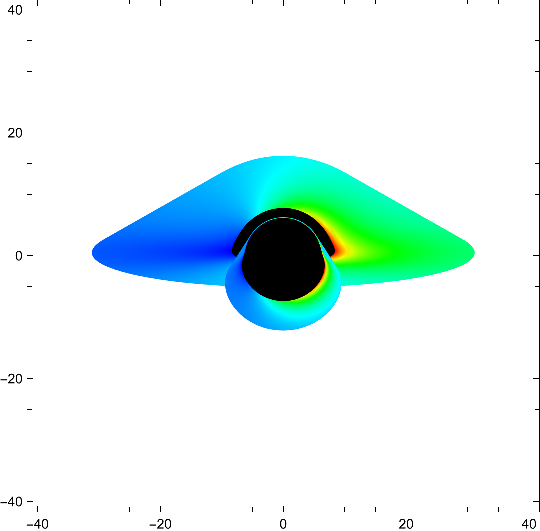} 
				\put(1,100){\color{black}\large $g/M=0.3, \theta=80^{\circ}$} 
				\put(-8,48){\color{black} Y'}
				\put(48,-10){\color{black} X'}
			\end{overpic}
			\raisebox{0.5\height}{ 
				\begin{overpic}[width=0.07\textwidth]{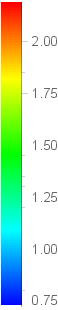} 
				\end{overpic}
			}
		\end{minipage}
		&
		\begin{minipage}[t]{0.3\textwidth}
			\centering
			\begin{overpic}[width=0.75\textwidth]{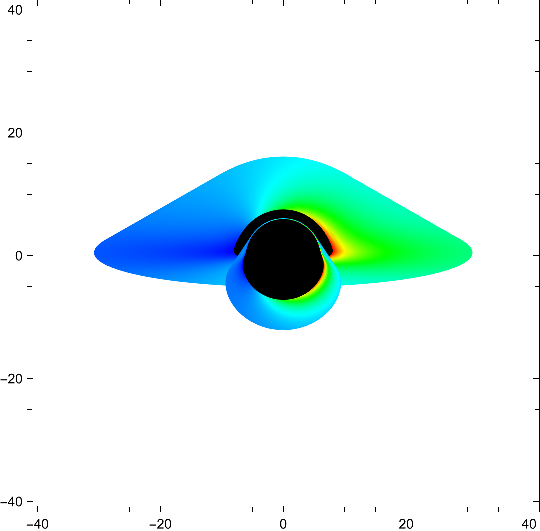}
				\put(1,100){\color{black}\large $g/M=0.6, \theta=80^{\circ}$}
				\put(-8,48){\color{black} Y'}
				\put(48,-10){\color{black} X'}
			\end{overpic}
			\raisebox{0.5\height}{ 
				\begin{overpic}[width=0.07\textwidth]{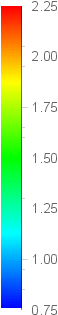} 
				\end{overpic}}
		\end{minipage}
	\end{tabular}
	\caption{\label{fig.16} Three representative viewing angles are fixed together with \(b/M = 0.1\), allowing us to examine how variations in the magnetic charge \(g/M\) influence the morphology of both the primary and secondary images.}
	\end{figure}

Figures \ref{fig.13} and \ref{fig.14} present the primary and secondary images associated with the optical appearance of accretion disk surrounding the PFDM–Bardeen BH. In these panels, the observed radiation flux \(F_{\mathrm{obs}}\) is color-mapped, where brighter, yellowish regions indicate higher flux, while the black background corresponds to vanishing emission. A well defined disk outline emerges when the viewing angle approaches \(90^\circ\), whereas the morphology becomes nearly circular as the inclination tends toward \(0^\circ\). Moreover, because the accretion disk rotates and the observation angle is non‑zero, light on the left side moves toward the observer while that on the right side moves away, causing an asymmetric flux enhancement on the left. To further examine these effects, we compute the spatial distribution of the redshift factor across the accretion disk images. As shown in Figures \ref{fig.15},\ref{fig.16} and Table IV, the primary and secondary images are displayed together, with larger values of redshift factor corresponding to regions rendered in deeper red tones.  Furthermore, we further examine the redshift distribution as a function of the observation angle and find an intriguing phenomenon: for the PFDM–Bardeen BH, blueshift appears only in the images at sufficiently high inclination, while the secondary image shows redshifted emission dominated, and even the primary image remains globally redshift dominated.

The absence of blueshift at low inclination angles arises from two factors. First, the static, spherically symmetric PFDM-Bardeen BH lacks Kerr‑type spin-dragging, so no strong Doppler boost occurs on the disk side facing the observer. Consequently, at low inclination angles, the projected component of the orbital velocity of the accretion disk matter along the line of sight is very small, making the Doppler effect negligible, and the gravitational redshift due to photons escaping from the strong gravitational field dominates, resulting in a net redshift throughout the image with no blueshift. As the inclination angle increases, the matter moving toward the observer acquires a significant line-of-sight velocity component, producing a clear Doppler blueshift, which forms a blueshifted region in the primary image. For the secondary image, the light rays typically originate from the far side of the accretion disk (the side moving away from the observer) and undergo multiple strong gravitational bendings near the BH; each crossing of the disk involves varying velocity directions and cumulative gravitational redshifts, so the net redshift always dominates, making blueshift difficult to observe even at high inclination angles. Second, the PFDM term shifts the ISCO and the emission region outward, deepening the gravitational potential and enhancing gravitational redshift. This provides a falsifiable prediction: if future observations detect any significant blueshifted emission at low inclination angles from the accretion disk of Sgr A* or M87*, the PFDM–Bardeen scenario may be ruled out.

\begin{table*}[htbp]
\centering
\caption{\label{Tab.4} The table presents the redshift values at \(r/M = 15\) and \(\alpha = \pi/4\) for different observation angles and BH parameters.}
\setlength{\tabcolsep}{4.3pt} 
\renewcommand{\arraystretch}{1.05} 
\begin{tabular}{|l|ccccc|}
\hline
{\bf $z(g/M=0.1,r/M=15)$}& $\theta=0^{\circ}$ &  $\theta=20^{\circ}$&  $\theta=40^{\circ}$&  $\theta=60^{\circ}$ &
 $\theta=80^{\circ}$  \\
\hline
b/M=0.00 (Bardeen)
& 0.118018 & 0.192739 & 0.252562 & 0.276795 & 0.260952  \\
\hline
b/M=0.05
&0.137267 &0.201935 &0.25088 &0.34339 &0.287228\\
\hline
b/M=0.08
&0.146461 &0.213659 &0.264209 &0.360144 &0.302925 \\
\hline
{\bf $z(b/M=0.1,r/M=15)$}&$\theta=0^{\circ}$ &  $\theta=20^{\circ}$&  $\theta=40^{\circ}$&  $\theta=60^{\circ}$ &
 $\theta=80^{\circ}$\\
\hline
g/M=0.00 (PFDM-Sch)
&0.152224 &0.220988 &0.27252 &0.370568 &0.312744\\
\hline
g/M=0.30
&0.152072 &0.220748 &0.272211 &0.37021 &0.312279\\
\hline
g/M=0.60
&0.151615 &0.220029 &0.271483 &0.369141 &0.310886 \\
\hline

\end{tabular}
\end{table*}

\subsection{Comparison of Black Hole in Different Dark Matter halos}
The preceding analysis was carried out for PFDM characterized by the density profile $\rho_{\rm PFDM}=b/r^3$ and the metric function $f(r)=1-2Mr^2/(r^2+g^2)^{3/2}-(b/r)\ln(r/|b|)$. Moreover, as shown in Sections IV and V, the influence of the magnetic charge $g$ in the PFDM-Bardeen BH is negligible compared to the DM parameter $b$. Therefore, in this subsection we focus on the observational effects of different DM models on the redshift of the accretion disk around Schwarzschild (SCh) BH, and compare them with the PFDM case to explore such differences. To this end, we replace PFDM with three qualitatively different DM models: 

\subsubsection{The CDM halo with NFW profile}

The CDM model, first introduced by Peebles \cite{Peebles1982LargescaleBT}, occupies a particularly important place. The term "cold" indicates that the constituent particles move with velocities much lower than the speed of light, a property that endows CDM with a crucial role in shaping the large‑scale structure of the universe. Indeed, current observations of the cosmic large‑scale structure are broadly consistent with the predictions of the CDM paradigm. Therefore, investigating the influence of CDM on BH environments can provide valuable insights into the fundamental nature of DM \cite{Xu:2018wow}. For this class of CDM, our principal findings concerning the BH DM halo interaction are  twofold: (i) In the relativistic regime, the density profile inevitably exhibits a “cusp” feature on small scales for both CDM models considered; (ii) the BH singularity remains unaffected by DM. Consequently, the NFW profile proves to be the effective description for the density distribution of such DM. The spacetime geometry for a pure NFW halo (without BH) has already been derived and extensively discussed in the literature \cite{Matos:2003nb,Fay:2004vw,Matos:2004je}. In those works, the pure DM spacetime is assumed to be “almost flat” because of the low DM density and the absence of relativistic motions. Based on these results, various dynamical processes in a pure dm halo-such as tidal disruption events (TDEs) and stellar motions-can be investigated via geometric methods.

The density profile of CDM is the NFW profile obtained from numerical simulations based on the CDM paradigm \cite{Dubinski:1991bm,Navarro:1996gj}. The expression for this density distribution is given by
\be
\label{32}
\rho_{\mathrm{NFW}}(r) = \frac{\rho_c}{\left(\frac{r}{R_s}\right)
\left(1+\frac{r}{R_s}\right)^2},
\ee
where \(\rho_c\) is density of the CDM halo collapse and \(R_s\) is the scale radius. For a BH of mass \(M\), the enclosed DM mass within a given radius \(r\) can then be written as
\be
\label{33}
M_{\mathrm{DM}}(r) = 4\pi \int_0^r \rho_{\mathrm{NFW}}(r')\, r'^2 dr'.
\ee

Furthermore, Ref. \cite{Xu:2018wow} presents an analytical solution for BH embedded in an NFW profile derived from the Einstein field equations, which is expressed as
\be
\label{34}
d s^2=-\left(\left(1+\frac{r}{R_s}\right)^{-\frac{8 \pi \rho_c R_s^3}{ r}}-\frac{2M}{r }\right) d t^2+\left(\left(1+\frac{r}{R_s}\right)^{-\frac{8 \pi \rho_c R_s^3}{r}}-\frac{2M}{r}\right)^{-1} d r^2+r^2 d\Omega^2 .
\ee

Based on the observational data of Sgr A*, we take \(\rho_c = 1.936 \times 10^{7}\, M_\odot/\mathrm{kpc}^{3}\) and \(R_s = 17.46\,\mathrm{kpc}\) as the DM halo parameters, where \(\rho_c\) is assumed to be the central density of the Sgr A* halo and \(R_s\) the core radius. These parameters are used in our analysis. We quantitatively analyze the NFW profile and find that, the DM density at the shadow scale \(R_{\mathrm{sh}} \approx 5M\) is approximately \(2.22 \times 10^{-14}\,\mathrm{g/cm^3}\), which is orders of magnitude weaker than that of the PFDM DM. At \(r \approx 100\,\mathrm{pc}\), the density is about \(2.26 \times 10^{-22}\,\mathrm{g/cm^3}\). The redshift map for this DM‑modified BH is presented in the first row of Fig. \ref{fig.17}, with all dimensionless parameters set to 0.1. As the inclination angle increases, the blueshift becomes more significant, which is qualitatively consistent with the PFDM case.

\subsubsection{The CDM halo with Dehnen-type density profile}
    
In addition to the NFW profile, another widely used model to describe the density distribution of DM halos is the Dehnen profile, originally proposed to model the inner structure of spherical galaxies and DM halos \cite{Dehnen:1993uh}. The Dehnen profile is characterized by a power-law cusp in the innermost region, followed by a steeper fall-off at larger radii. Its general form is given by
\be   
\label{35}
\rho_{\rm Dehnen-type}=\rho_s\left(\frac{r}{r_s}\right)^{-\gamma}\left[\left(\frac{r}{r_s}\right)^\alpha+1\right]^{\frac{\gamma-\beta}{\alpha}},
\ee
where the three parameters \(\alpha\), \(\beta\), and \(\gamma\) are model constants. The Dehnen family encompasses several variants, each determined by the inner slope \(\gamma\). In the generalized double‑power law parameterization with \((\alpha, \beta, \gamma) = (1, 4, \gamma)\), \(\gamma\) is restricted to the interval \((0,3)\). In the present work, we restrict ourselves to the \((1,4,0)\) case. This gives
\be
\label{36}
\rho_{\rm Dehnen-type}=\frac{\rho_s}{\left(\frac{r}{r_s}+1\right)^4},
\ee
here $\rho_s$ and $r_s$ denote the central halo density and the halo core radius respectively. Furthermore, assuming the central BH to be Sch BH, Ref. \cite{Gohain:2024eer} derived the solution for Sch BH embedded in a Dehnen‑type
\((1,4,0)\) DM halo. The metric takes the form

\be
\label{37}
d s^2=-\left(1-\frac{2 M}{r}-\frac{4 \pi r_s^3\left(r_s+2 r\right) \rho_s}{3\left(r_s+r\right)^2}\right)d t^2+\left(1-\frac{2 M}{r}-\frac{4 \pi r_s^3\left(r_s+2 r\right) \rho_s}{3\left(r_s+r\right)^2}\right)^{-1} d r^2+r^2 d\Omega^2.
\ee

Following the same procedure as in the preceding subsection, we estimate that the Dehnen-type DM density at \(r \approx 5M\) is about \(1.31 \times 10^{-24}\,\mathrm{g/cm^3}\), and at \(r = 100\,\mathrm{pc}\) it is approximately \(1.28 \times 10^{-24}\,\mathrm{g/cm^3}\). Notably, the Dehnen-type density profile exhibits little variation from the shadow scale out to 100 pc, a characteristic that clearly distinguishes it from the PFDM DM profiles. The second row of Fig. \ref{fig.17} shows the redshift maps for such DM BH. The inclination angle has a considerably stronger effect than the variation of model parameters, and the blueshift again grows with increasing inclination, which is qualitatively consistent with the PFDM case.

\subsubsection{The CDM halo with Moore profile}
Since the seminal discovery of the universal density profile by NFW, the structure of DM halos emerging from dissipationless hierarchical clustering under cosmological initial conditions has attracted considerable research interest. Nevertheless, subsequent high‑resolution N‑body simulations have suggested potential deviations from the NFW form. Specifically, Fukushige and Makino \cite{Fukushige:1996nr} carried out a simulation employing 768000 particles and found that a galaxy‑sized halo possesses a cusp steeper than \(\rho \propto r^{-1}\). Later, Moore and Ghigna  performed simulations with up to four million particles, revealing that the density profile scales as \(\rho \propto r^{-1.5}\) in both galaxy‑sized and cluster‑sized halos \cite{Moore:1997sg,Ghigna:1999sn}. On the basis of these results, Moore et al. put forward a modified universal profile, often termed the Moore profile, which features an inner cusp \(\rho \sim r^{-3/2}\). As such a steep inner cusp can significantly alter the gravitational field in the vicinity of a BH horizon and thereby influence the redshift distribution, it is worthwhile to compare the observational signatures predicted by the Moore halo with those arising from the NFW, Dehnen, and PFDM models. 

The modified universal Moore profile can be expressed as

\be
\label{38}
\rho_{\text {Moore }}(r)=\frac{\rho_{\text {Moore }}^0}{\left(\frac{r}{R_h}\right)^{\frac{3}{2}}\left(1+\left(\frac{r}{R_h}\right)^{\frac{3}{2}}\right)},
\ee
here \(\rho_0^{\mathrm{Moore}}\)also denotes the central DM halo density, and \(R_h\) is the scale radius. Since the density function does not converge as \(r \to 0\), a "cusp" phenomenon arises. Consequently, this profile belongs to the class of central cusp models. In addition, solving the Einstein field equations yields the BH solution for the Moore profile, which takes the form \cite{Wu:2024hxr}

\be
\label{38}
d s^2=- \left( f_{\text {Moore }}(r)-\frac{2M}{r}    \right) d t^2+\left( f_{\text {Moore }}(r)-\frac{2M}{r}    \right)^{-1} d r^2+r^2 d\Omega^2,
\ee
where $f_{\text {Moore }}(r)$ can be written as
\be
\label{39}
\begin{aligned}
& f_{\text {Moore }}(r)= \left[1+\frac { 1 6 \pi R _ { h } ^ { 3 } \rho _ { \text { Moore } } ^ { 0 } } { 3 } \left(\frac{\log \left(\sqrt{\frac{r}{R_h}}+1\right)}{R_h}-\frac{\log \left(\left(\frac{r}{R_h}\right)^{3 / 2}+1\right)}{r}\right.\right. \\
&-\frac{\log \left(-R_h \sqrt{\frac{r}{R_h}}+R_h+r\right)}{2 R_h} 
 \left.\left.+\frac{\sqrt{3} \tan ^{-1}\left(\frac{2 \sqrt{\frac{r}{R_h}}-1}{\sqrt{3}}\right)}{R_h}\right)\right].
\end{aligned}
\ee

\begin{figure}[H]
	\centering
	\begin{tabular}{@{}ccc@{}}

		\begin{minipage}[t]{0.3\textwidth}
			\centering
			\begin{overpic}[width=0.75\linewidth]{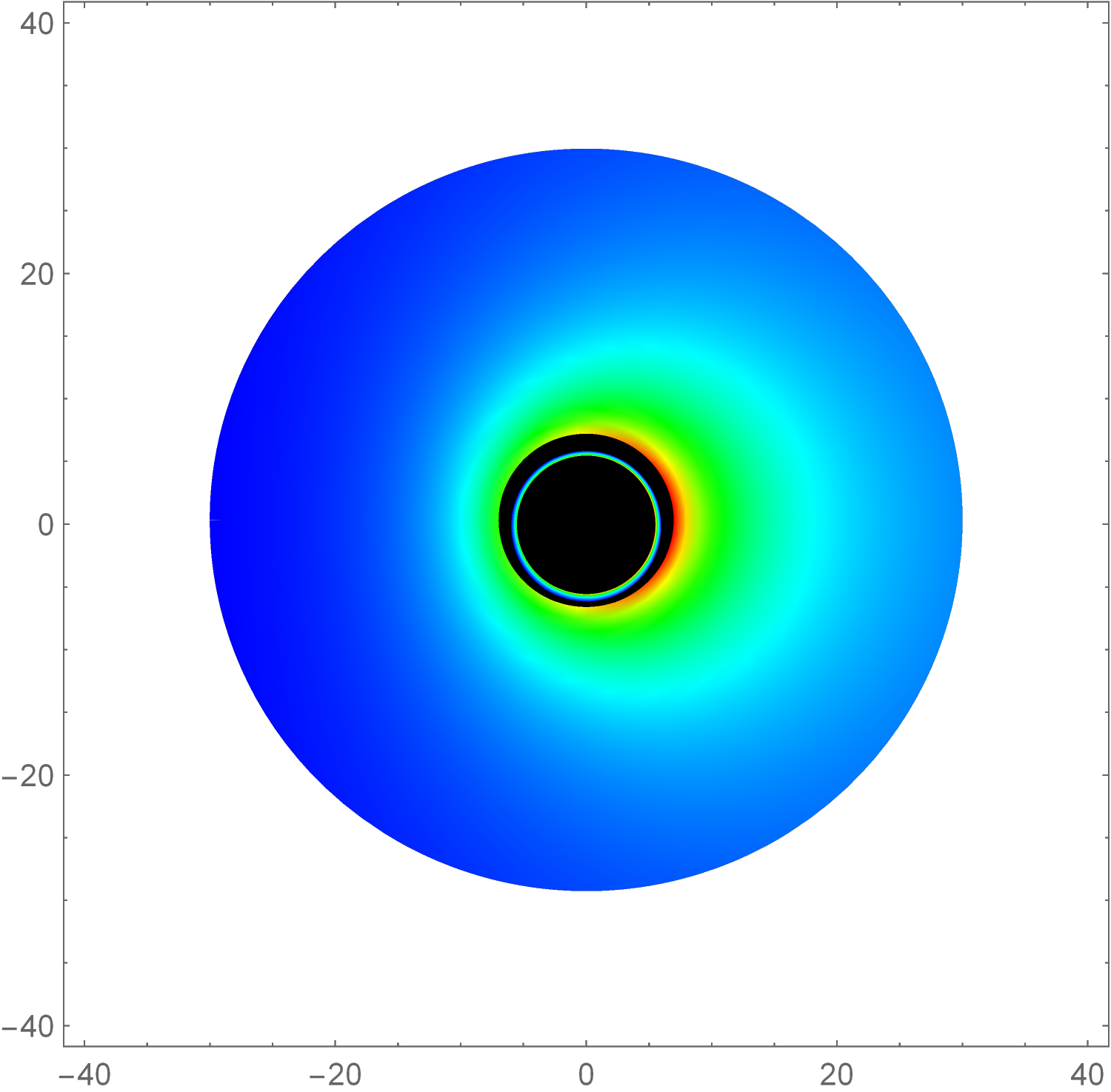}
				\put(1,100){\color{black}\large
					$R_s/M=0.1,\theta=10^{\circ}$}
				\put(-8,48){\color{black}Y'}
				\put(48,-10){\color{black}X'}
			\end{overpic}
			\raisebox{0.1\height}{
				\begin{overpic}[width=0.07\linewidth]{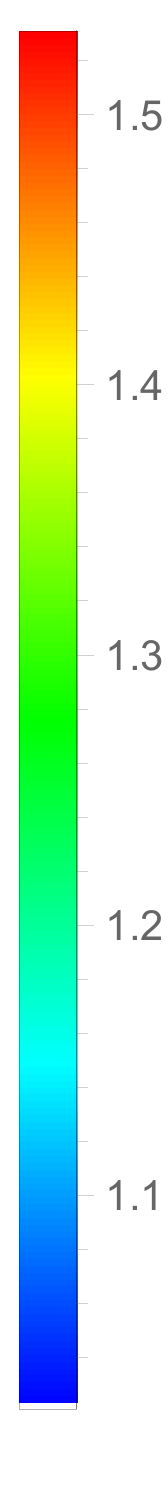}
				\end{overpic}
			}
		\end{minipage}
		&
		\begin{minipage}[t]{0.3\textwidth}
			\centering
			\begin{overpic}[width=0.75\linewidth]{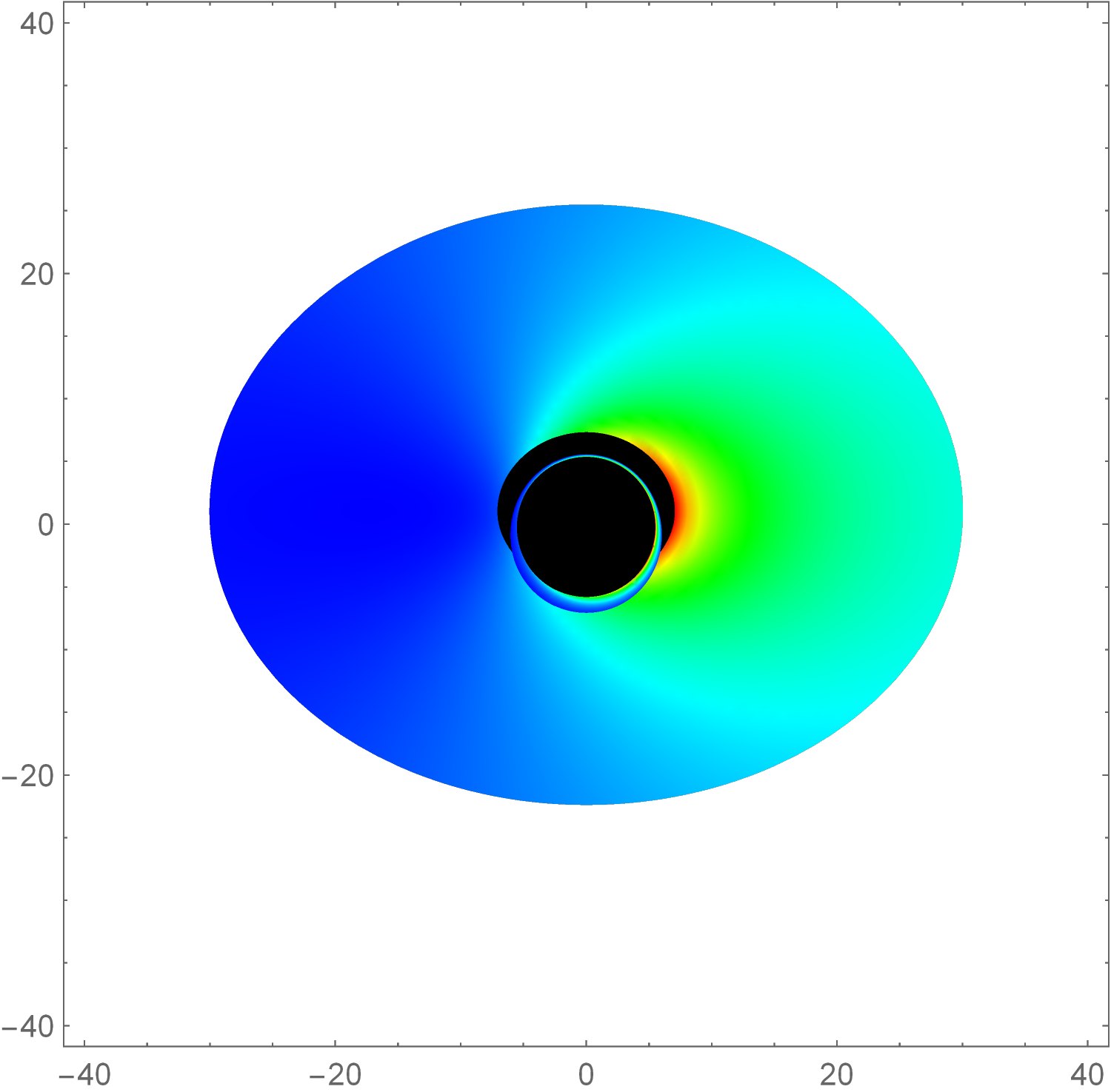}
				\put(1,100){\color{black}\large
					$R_s/M=0.1,\theta=40^{\circ}$}
				\put(-8,48){\color{black}Y'}
				\put(48,-10){\color{black}X'}
			\end{overpic}
			\raisebox{0.1\height}{
				\begin{overpic}[width=0.07\linewidth]{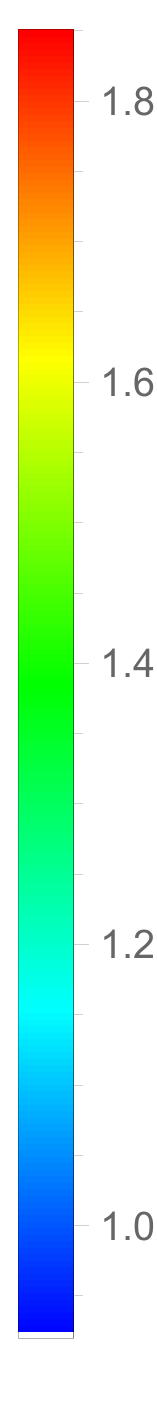}
				\end{overpic}
			}
		\end{minipage}
		&
		\begin{minipage}[t]{0.3\textwidth}
			\centering
			\begin{overpic}[width=0.75\linewidth]{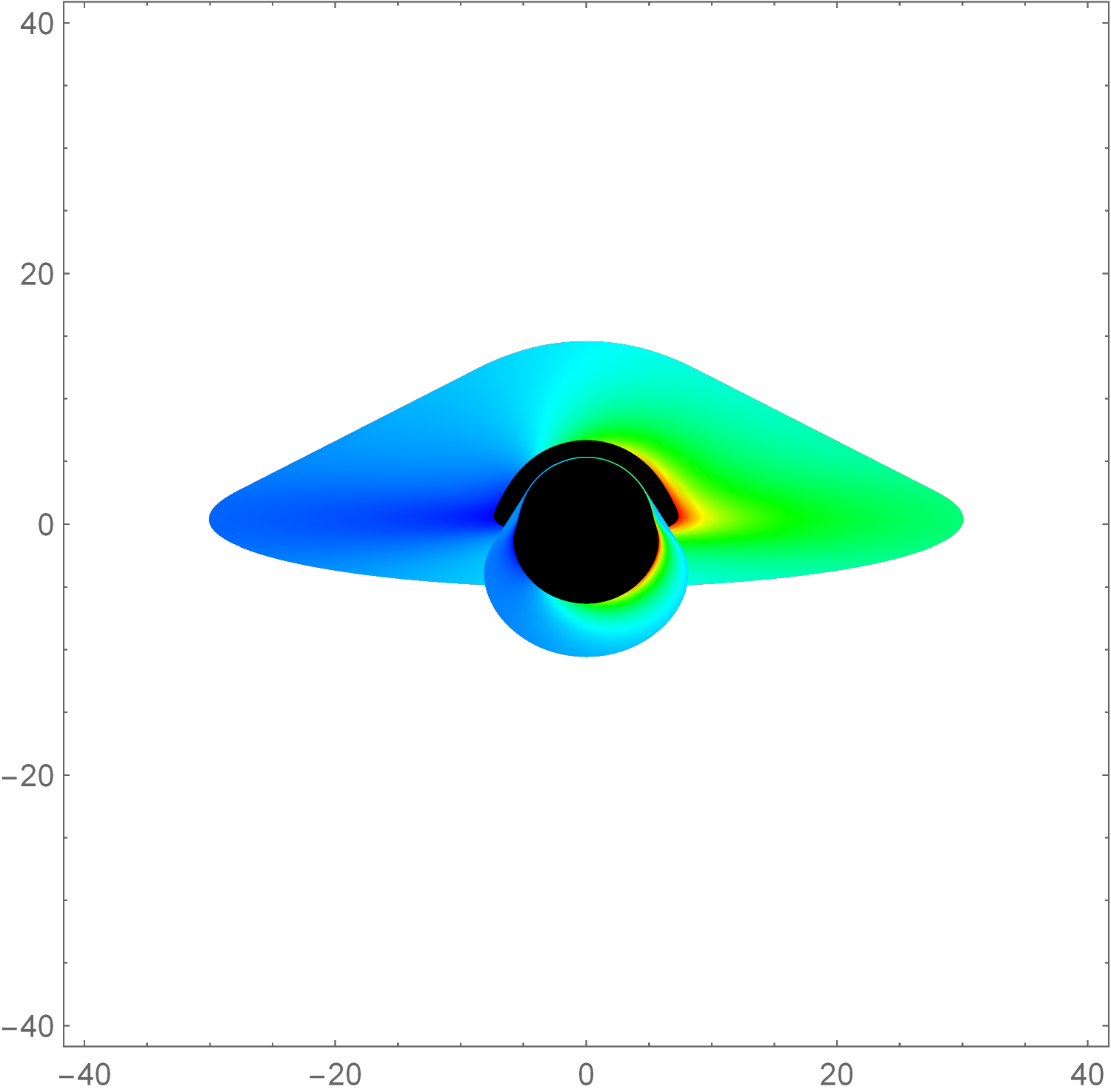}
				\put(1,100){\color{black}\large
					$R_s/M=0.1,\theta=80^{\circ}$}
				\put(-8,48){\color{black}Y'}
				\put(48,-10){\color{black}X'}
			\end{overpic}
			\raisebox{0.2\height}{
				\begin{overpic}[width=0.07\linewidth]{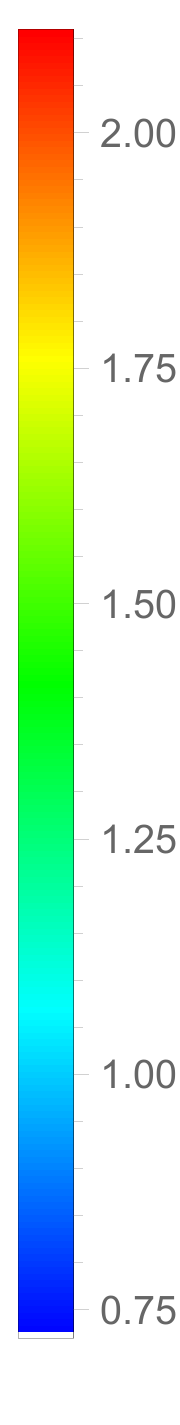}
				\end{overpic}
			}
		\end{minipage}

		\\[40pt]

		\begin{minipage}[t]{0.3\textwidth}
			\centering
			\begin{overpic}[width=0.75\linewidth]{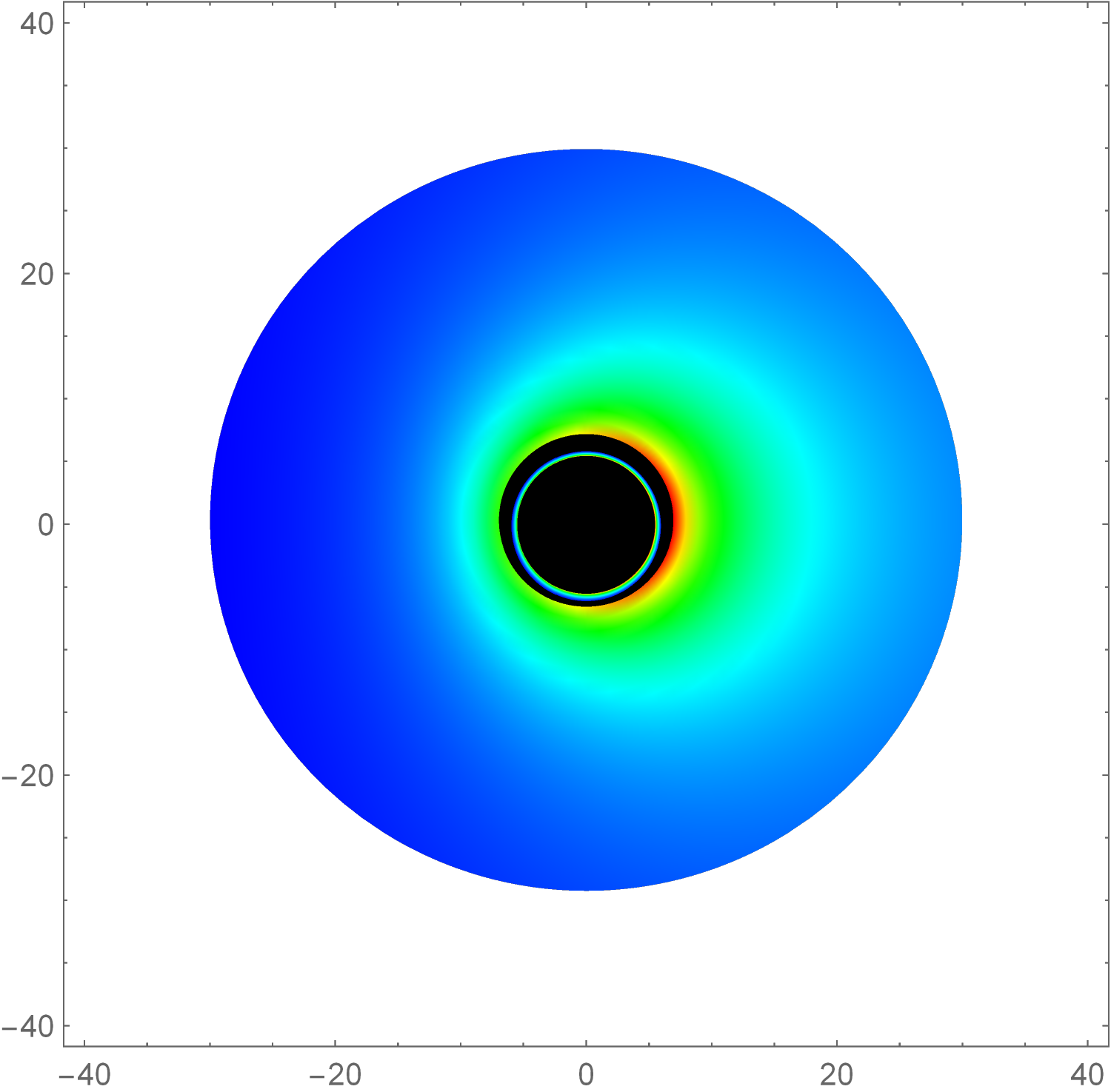}
				\put(1,100){\color{black}\large
					$r_s/M=0.1,\theta=10^{\circ}$}
				\put(-8,48){\color{black}Y'}
				\put(48,-10){\color{black}X'}
			\end{overpic}
			\raisebox{0.1\height}{
				\begin{overpic}[width=0.07\linewidth]{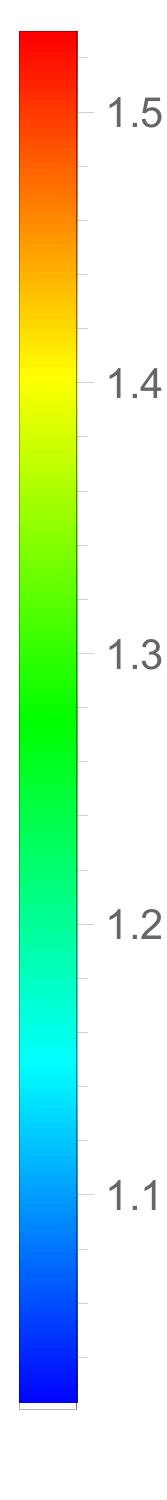}
				\end{overpic}
			}
		\end{minipage}
		&
		\begin{minipage}[t]{0.3\textwidth}
			\centering
			\begin{overpic}[width=0.75\linewidth]{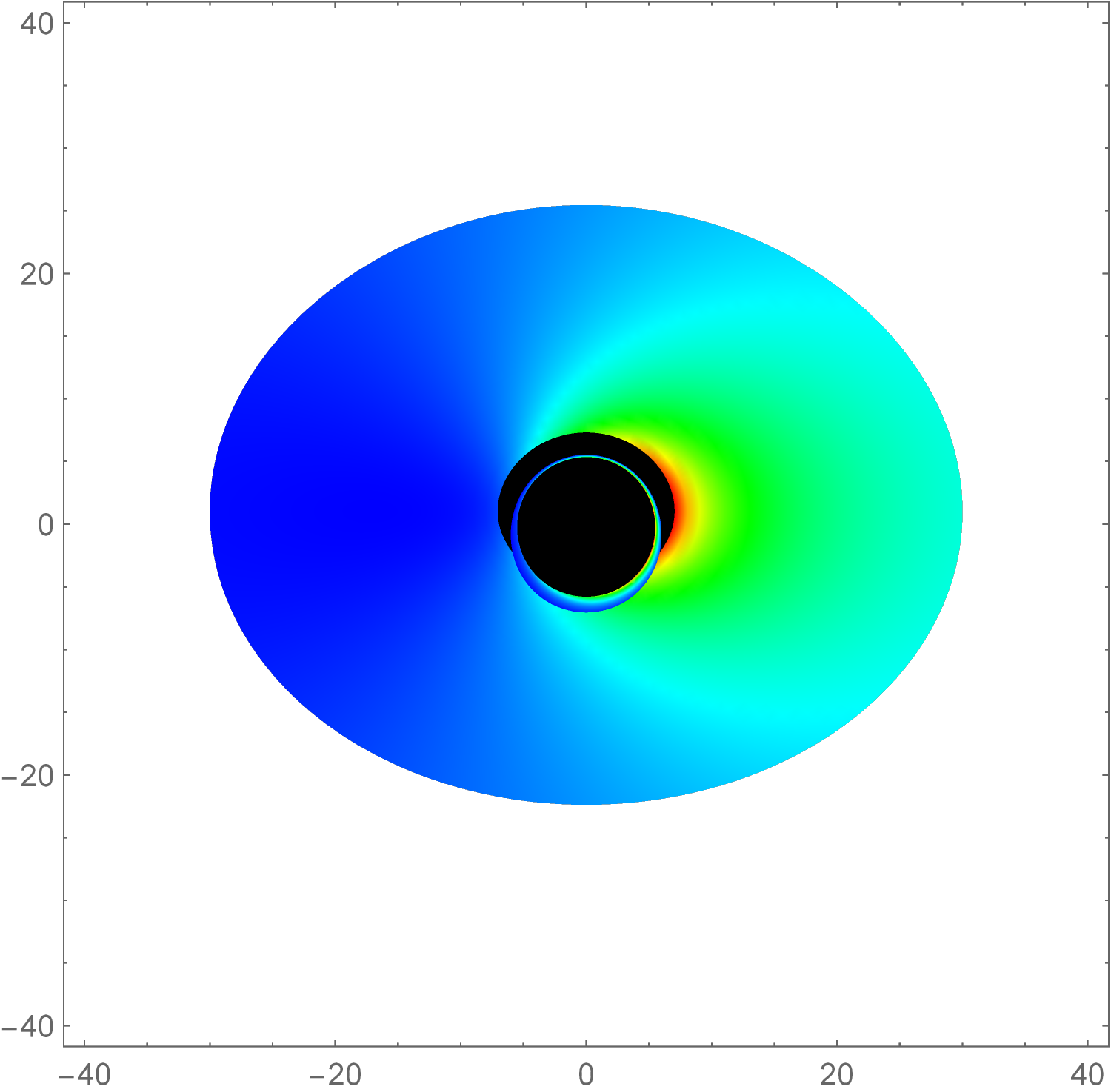}
				\put(1,100){\color{black}\large
					$r_s/M=0.1,\theta=40^{\circ}$}
				\put(-8,48){\color{black}Y'}
				\put(48,-10){\color{black}X'}
			\end{overpic}
			\raisebox{0.1\height}{
				\begin{overpic}[width=0.07\linewidth]{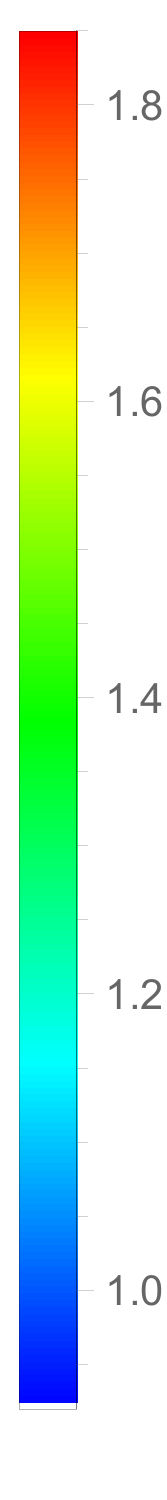}
				\end{overpic}
			}
		\end{minipage}
		&
		\begin{minipage}[t]{0.3\textwidth}
			\centering
			\begin{overpic}[width=0.75\linewidth]{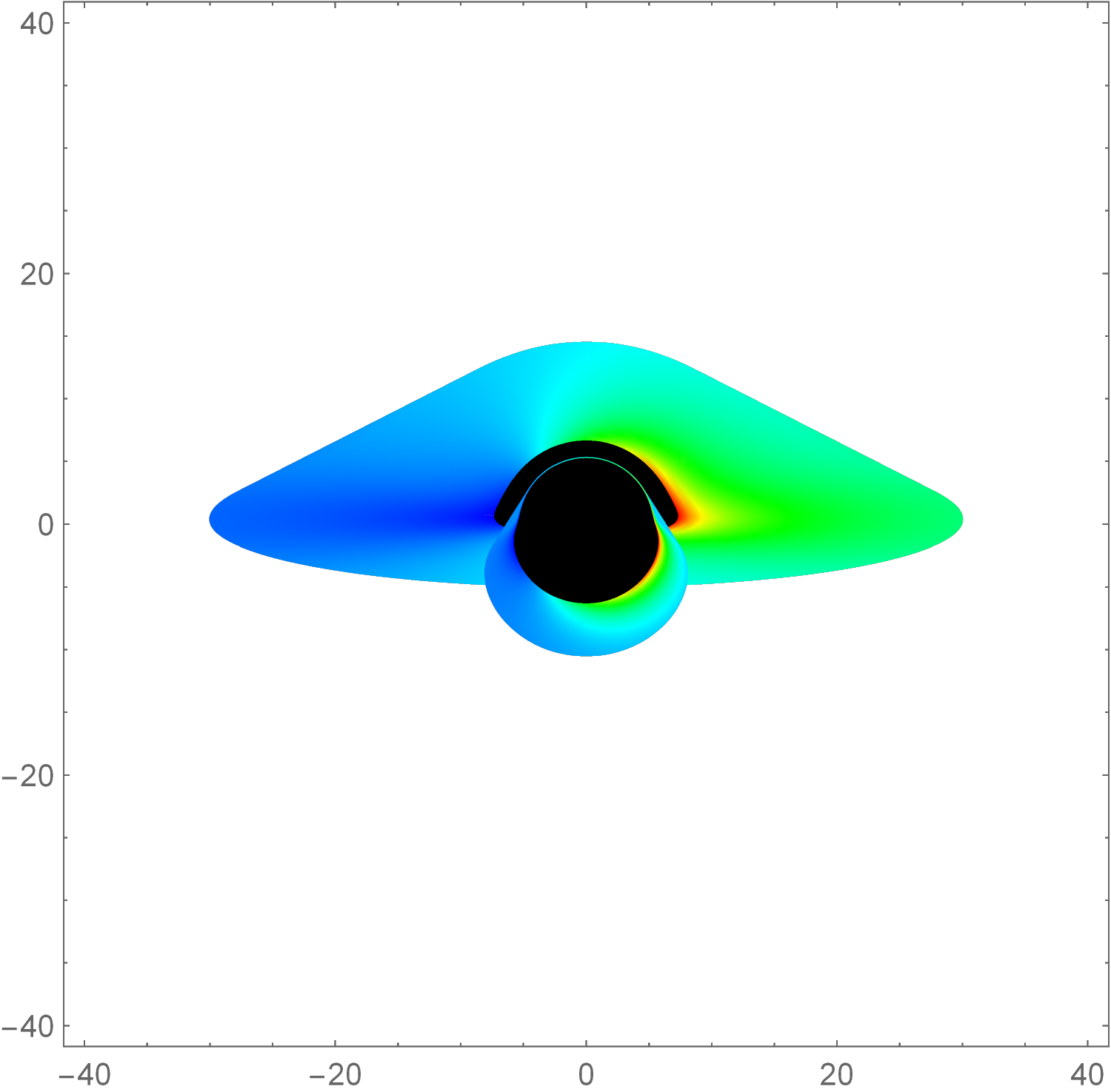}
				\put(1,100){\color{black}\large
					$r_s/M=0.1,\theta=80^{\circ}$}
				\put(-8,48){\color{black}Y'}
				\put(48,-10){\color{black}X'}
			\end{overpic}
			\raisebox{0.2\height}{
				\begin{overpic}[width=0.07\linewidth]{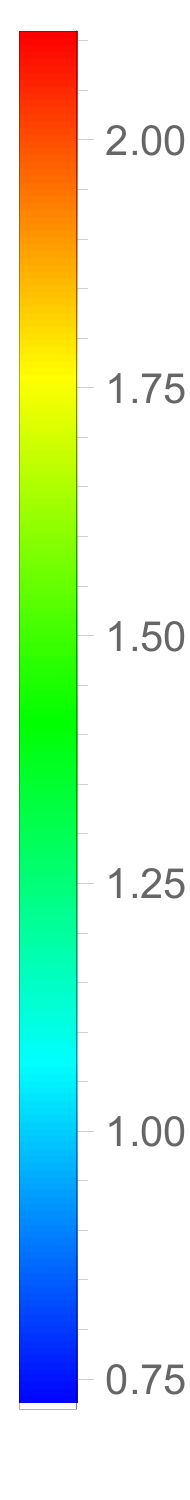}
				\end{overpic}
			}
		\end{minipage}

		\\[-12pt]

\begin{minipage}[t]{0.3\textwidth}
	\centering
	\hspace*{-0.05\linewidth}%
	\begin{overpic}[width=1.1\linewidth]{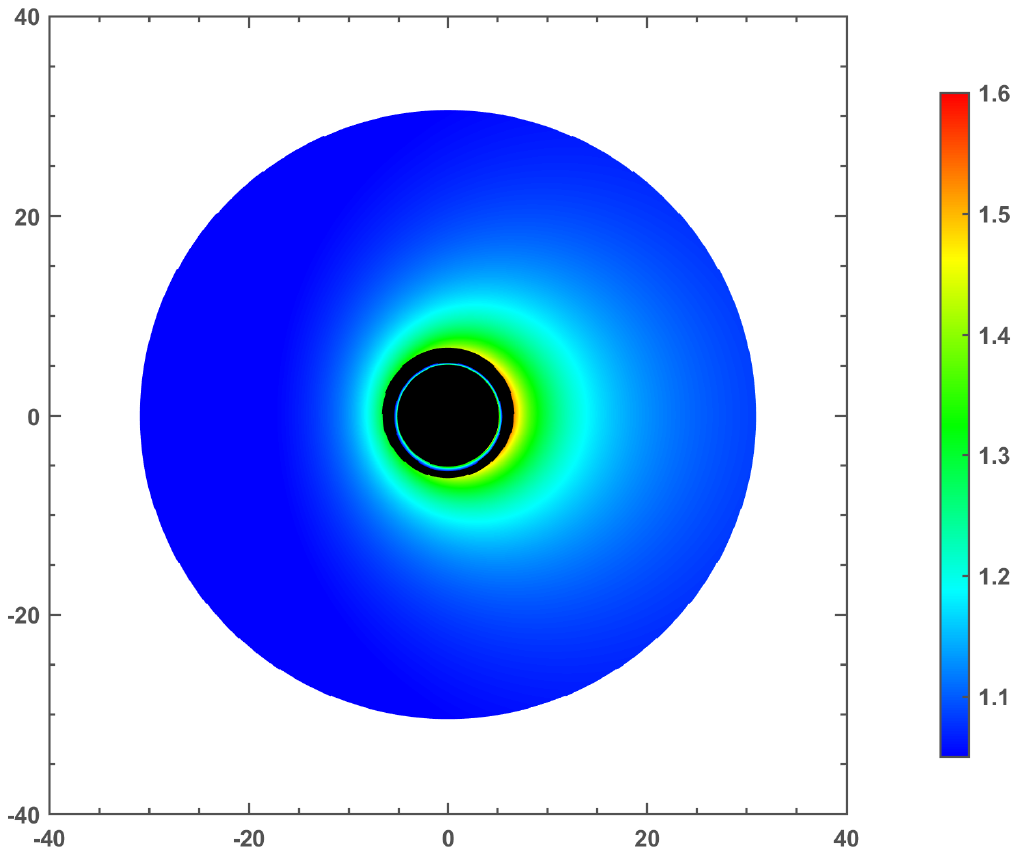}
		\put(10,77){\color{black}\large
			$R_h/M=0.1,\theta=10^{\circ}$}
		\put(5,49){\color{black}Y'}
		\put(32,20){\color{black}X'}
	\end{overpic}
\end{minipage}
&
\begin{minipage}[t]{0.3\textwidth}
	\centering
	\hspace*{-0.05\linewidth}%
	\begin{overpic}[width=1.1\linewidth]{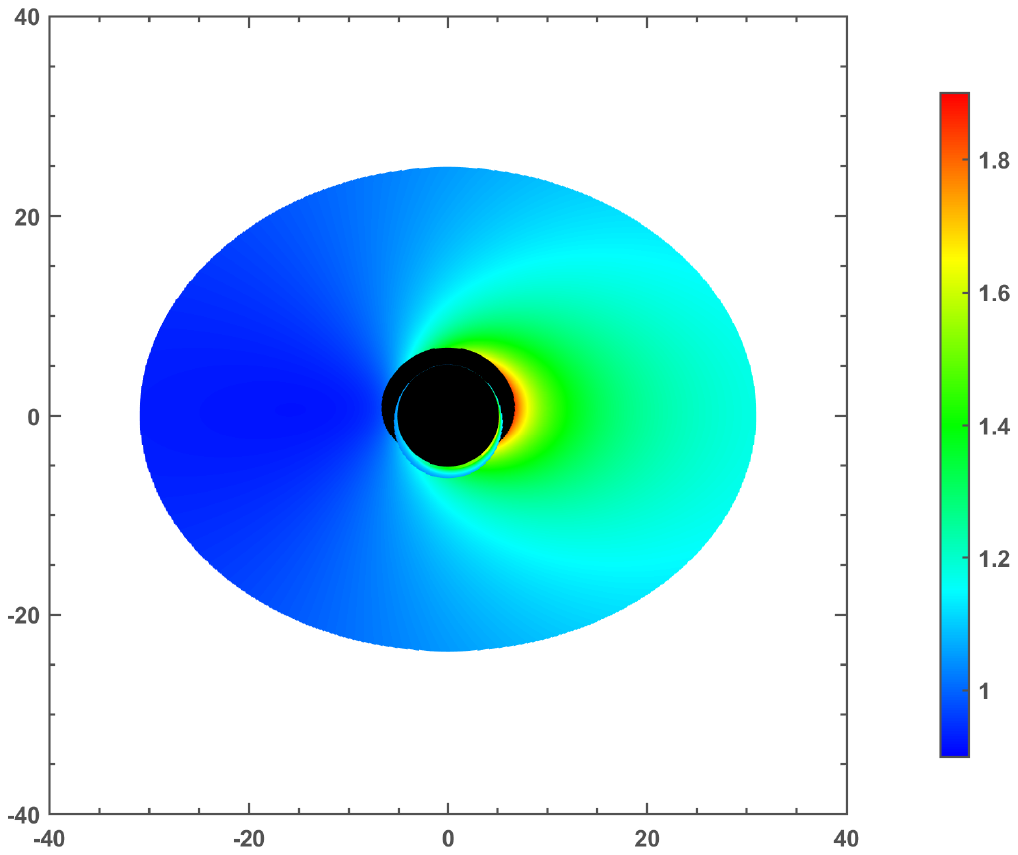}
		\put(10,77){\color{black}\large
			$R_h/M=0.1,\theta=40^{\circ}$}
		\put(5,49){\color{black}Y'}
		\put(32,20){\color{black}X'}
	\end{overpic}
\end{minipage}
&
\begin{minipage}[t]{0.3\textwidth}
	\centering
	\hspace*{-0.05\linewidth}%
	\begin{overpic}[width=1.1\linewidth]{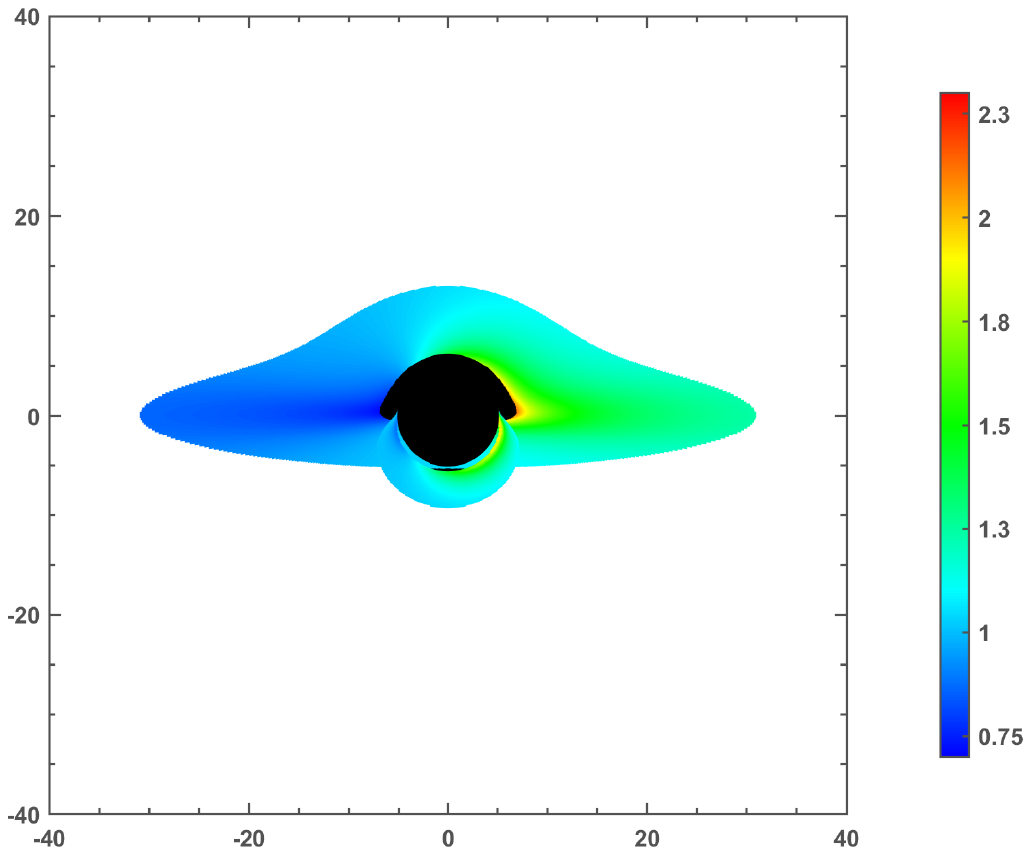}
		\put(10,77){\color{black}\large
			$R_h/M=0.1,\theta=80^{\circ}$}
		\put(5,49){\color{black}Y'}
		\put(32,20){\color{black}X'}
	\end{overpic}
\end{minipage}

	\end{tabular}

	\vspace{-40pt}

	\caption{\label{fig.17}  The first row shows the redshift maps of the BH under the NFW profile at inclination angles of 
    $10^\circ$, $40^\circ$, and $80^\circ$. The second row corresponds to the redshift maps for the Dehnen-type density profile. The third row corresponds to the redshift maps for the Moore density profile. In all panels, the dimensionless central density is taken as $0.1$.}

\end{figure}

Next, taking Sgr A* as the central BH, we estimate the DM densities at \(r = 5M\) and \(r = 100\,\mathrm{pc}\) for the Moore profile. The density at \(5M\) is found to be approximately \(2.89 \times 10^{-9}\,\mathrm{g/cm^3}\), which is weaker than that of the PFDM case, while at \(100\,\mathrm{pc}\) it is about \(3.02 \times 10^{-21}\,\mathrm{g/cm^3}\), which is higher than the PFDM counterpart. This dramatic variation in density over the radial range provides a distinct observational signature for discriminating among different DM models. Similarly, the third row of Fig. \ref{fig.17} shows the redshift rendering of the accretion disk for the Moore DM BH, which is also consistent with the PFDM case. Hence, the redshift maps do not provide a discriminating signature among different DM BHs scenarios. This behavior is likely a universal effect dominated by the viewing inclination. However, the physical origin of this behavior has been explicitly clarified in the preceding section.

\section{Conclusion}
The study begins by examining the spacetime structure of PFDM-Bardeen BH. This analysis identifies the parameter ranges that ensure the existence of regular horizons and clarifies how the DM parameter and magnetic charge reshape the underlying geometry. 
From EHT shadow constraints we derive a rough prediction for the PFDM density near the BH: for Sgr A*, the allowed parameter range \(b/M \sim 10^{-3}\)–\(10^{-2}\) yields \(\rho_{\mathrm{DM}} \sim 0.27\)–\(2.67\,\mathrm{g/cm^{3}}\) at the shadow scale \(R_{\mathrm{sh}}\sim5M\), dropping to \(\sim10^{-24}\)–\(10^{-25}\,\mathrm{g/cm^{3}}\) at 100 pc.
Subsequently, we investigated photon dynamics around the BH, where null trajectories are classified into direct, lensing, and photon ring branches. The results show that the DM parameter has a pronounced effect on the photon sphere, shadow radius, transfer function, and intensity profiles generally enlarging the shadow and suppressing the observed flux. In contrast, variations in \(g\) produce only minor corrections across all observables. The magnetic charge \(g\) contributes negligibly because on the scales probed (\(r\sim5M\)) we have \(r\gg g\); the Bardeen term expands as \(-2M/r+\mathcal{O}(g^2/r^3)\), so \(g\) enters only as a strongly suppressed \(\mathcal{O}(g^2/r^3)\) correction. In contrast, the PFDM term \(-(b/r)\ln(r/b)\) directly modifies the geometry at these radii. Quantitatively, increasing \(b/M\) from 0 to 0.08 enlarges the photon sphere by \(\sim14\%\), while increasing \(g/M\) from 0 to 0.3 changes it by only \(1.7\%\). Hence all observational signatures are dominated by \(b\), with \(g\) playing a negligible role.

The final part of the analysis explores the imaging of thin accretion disks, including primary and secondary images, radiative flux distributions, and redshift maps. The morphology of the disk images is strongly influenced by the viewing inclination, with edge-on perspectives revealing distinct disk contours. We find that increasing the parameter $b$ suppresses the observed flux. This phenomenon may be explained by the fact that increasing $b$ alters the velocity and density distributions of the accreting matter, thereby affecting the photon escape process and modifying the flux distribution observed at infinity. 
Moreover, redshift dominates at low observation angles, while blueshift appears only at sufficiently high inclination. This behavior could stem from two effects. First, the static, spherically symmetric PFDM–Bardeen spacetime lacks frame-dragging, so no substantial Doppler enhancement occurs on the disk side moving toward the observer. Second, the PFDM term shifts the ISCO and emission region outward, deepening the gravitational potential and strengthening the gravitational redshift. Additionally, we compare the DM densities among the PFDM, NFW, Dehnen‑type and Moore models. Near the shadow radius, the PFDM model yields the highest density, followed by the Moore model, then the NFW model, and finally the Dehnen‑type model. This density ordering can serve as a criterion for distinguishing four DM BHs. Regarding the redshift behavior, all four types exhibit the same trend: the blueshift increases with the inclination angle. Even though the redshift profiles do not allow us to distinguish among the four BHs, should future observations detect significant blueshifted emission at low inclination angles from the accretion disk of Sgr A* or M87*, the predictions of all four DM models would be severely challenged.

\begin{acknowledgments}
This study was supported by the National Natural Science Foundation of China (Grant Nos. 12333008 and 12305070) and the Basic Research Program of Shanxi Province (Grant Nos. 202303021222018 and 202503021211204).
\end{acknowledgments}

\bibliographystyle{unsrt}
\bibliography{PFDM}
\end{document}